\documentclass[twoside,12pt]{article}
\usepackage{epsfig}
\usepackage{amsbsy} 
\def\Journal#1#2#3#4{{#1} {#2} (#4) #3 }

\def\PRO{{\em Prog. Theor. Phys.}}

\def\PRL{\em Phys. Rev. Lett.}
\def\PREV{\em Phys. Rev.}

\def\ANNP{\em Ann. Phys. (N.Y.)}

\newcommand{\be}{\begin{equation}}
\newcommand{\ee}{\end{equation}}
\newcommand{\bea}{\begin{eqnarray}}
\newcommand{\eea}{\end{eqnarray}}

\begin{document}

\title{ \vspace{1cm} Nuclear Chiral Dynamics and Thermodynamics}
\author{Jeremy W. Holt$^1$, Norbert Kaiser$^2$, Wolfram Weise$^{2,3}$\\
\\
$^1$Physics Department, University of Washington, Seattle, WA 98195-1550, USA \\
$^2$Physik-Department, Technische Universit\"at M\"unchen, D-85747 Garching, Germany\\
$^3$ECT$^*$, Villa Tambosi, I-38123 Villazzano (Trento), Italy\\}
\maketitle
\begin{abstract} This presentation reviews an approach to nuclear many-body systems 
based on the spontaneously broken chiral symmetry of low-energy QCD. In the low-energy 
limit, for energies and momenta small compared to a characteristic symmetry breaking 
scale of order 1 GeV, QCD is realized as an effective field theory of Goldstone bosons 
(pions) coupled to heavy fermionic sources (nucleons). Nuclear forces at long and 
intermediate distance scales result from a systematic hierarchy of one- and two-pion 
exchange processes  in combination with Pauli blocking effects in the nuclear medium. 
Short distance dynamics, not resolved at the wavelengths corresponding to typical 
nuclear Fermi momenta, are introduced as contact interactions between nucleons. Apart 
from a set of low-energy constants associated with these contact terms, the parameters 
of this theory are entirely determined by pion properties and low-energy pion-nucleon 
scattering observables. This framework (in-medium chiral perturbation theory) can 
provide a realistic description of both isospin-symmetric nuclear matter and neutron 
matter, with emphasis on the isospin-dependence determined by the underlying chiral 
NN interaction. The importance of three-body forces is emphasized, and the role of 
explicit $\Delta(1232)$-isobar degrees of freedom is investigated in detail. Nuclear 
chiral thermodynamics is developed and a calculation of the nuclear phase 
diagram is performed. This includes a successful description of the first-order phase 
transition from a nuclear Fermi liquid to an interacting Fermi gas and the coexistence 
of these phases below a critical temperature $T_c$. Density functional methods for finite 
nuclei based on this approach are also discussed. Effective interactions, their density dependence 
and connections to Landau Fermi liquid theory are outlined. Finally, the density and 
temperature dependence of the chiral (quark) condensate is investigated. 
\end{abstract}
\section{Introduction}
\begin{figure}
      \centerline{\includegraphics[width=10cm] {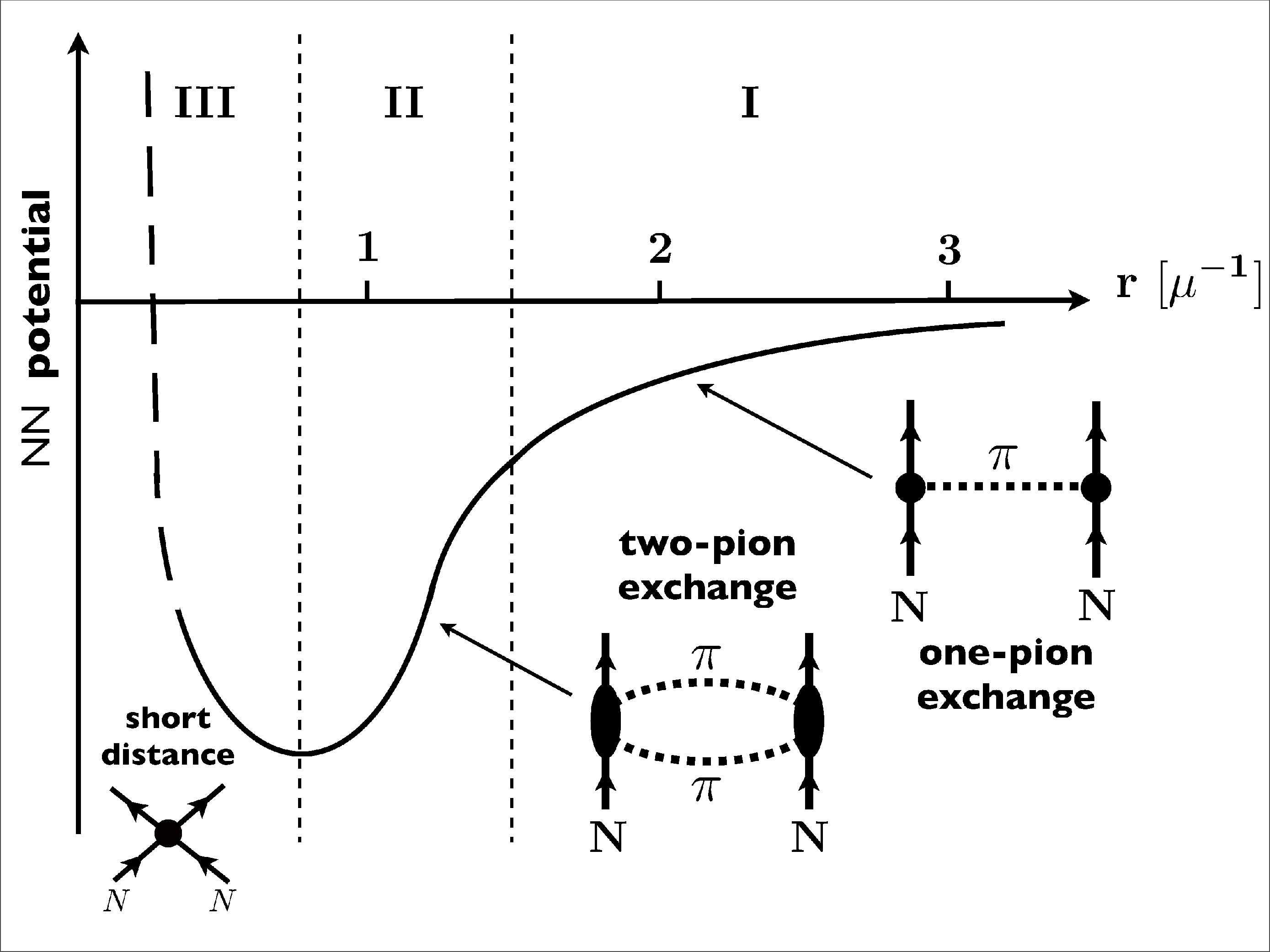}}
   \caption{Schematic picture displaying the hierarchy of scales governing the 
nucleon-nucleon potential (adapted from Taketani  \cite{Tak56}). The distance $r$ is 
given in units of the pion Compton wavelength $\mu^{-1} \simeq 1.4$ fm.}
   \label{fig:1}
 \end{figure}

Almost eight decades ago, Yukawa's pioneering article \cite{Yuk35} introduced the 
framework for a systematic approach to the nucleon-nucleon interaction, based on the 
exchange of a boson identified later as the pion. In the footsteps of Yukawa's original 
work the next generation of Japanese theorists already made impressive efforts to 
proceed from the long-range part of the interaction to shorter distances between 
nucleons.  A cornerstone of these developments was the visionary conceptual design by 
Taketani et al. \cite{TNS51} of an inward-bound hierarchy of scales governing the 
nucleon-nucleon potential, sketched in Fig.\ \ref{fig:1}. The long distance region I is 
determined by one-pion exchange. It continues inward to the region II of intermediate 
distances dominated by two-pion exchange. The basic idea was to construct the NN 
potential in regions I and II by explicit calculation of $\pi$ and $2\pi$ exchange 
processes, whereas the detailed behaviour of the interaction in the short distance 
region III remains unresolved at the low-energy scales characteristic of nuclear 
physics. This short distance part is then given a suitably parametrized form. The 
parameters are fixed by comparison with scattering data.

Taketani's picture, although at the time of its appearance not at all universally 
accepted by the international community of theorists, turned out to be immensely 
useful. Today this strategy is the one pursued by modern effective field theory 
approaches. It is amazing how far this program had already been developed in the late 
fifties of the twentieth century. One example is the pioneering calculation of the 
two-pion exchange potential \cite{Tak52,KMO57} (using dispersion relation techniques) 
and early knowledge \cite{FM50} of the resonant pion-nucleon amplitude which 
anticipated the $\Delta$-isobar models of later decades \cite{BW75}.

Today's theory of the strong interaction is Quantum Chromodynamics (QCD). 
There exist two limiting situations in which QCD is accessible with ``controlled" 
approximations. At momentum scales exceeding several GeV (corresponding to short 
distances, $r < 0.1$ fm), QCD is a theory of weakly interacting quarks and gluons. At 
low momentum scales considerably smaller than 1 GeV (corresponding to long distances, 
$r > 1$ fm), QCD is governed by color confinement and a non-trivial vacuum: the ground 
state of QCD hosts strong condensates of quark-antiquark pairs and gluons. Confinement 
is accompanied by the dynamical (spontaneous) breaking of Chiral Symmetry, a global 
symmetry of QCD that is exact in the limit of massless quarks. Spontaneous chiral 
symmetry breaking in turn implies the existence of pseudoscalar Nambu-Goldstone bosons. 
For two quark flavours $(N_f = 2)$ with (almost) massless $u$ and $d$ quarks, these 
Goldstone bosons are identified with the isospin triplet of pions. Low-energy QCD is 
thus realized as a (chiral) effective field theory in which these Goldstone bosons are 
the active, light degrees of freedom. In the low-energy, long-wavelength limit, 
Goldstone bosons interact weakly with one another or with any massive hadrons. In this 
limit perturbative methods are applicable and a systematic expansion in powers of a 
``small" parameter can be performed (chiral perturbation theory \cite{Wei66/67,GL84}).

The chiral effective field theory (ChEFT) just sketched provides the interface between 
QCD and nuclear physics. This theory has become the framework for a successful 
description of the nucleon-nucleon interaction \cite{evgenireview, hammerreview, 
machleidtreview}, and it is the starting point of a systematic approach to the nuclear 
many-body problem and its thermodynamics at densities and temperatures well within the 
confined (hadronic) phase of QCD. This review is written to serve as an introduction 
and state-of-the-art overview of this active area of research. It does not aim for
completeness, however, and should be read in combination with complementary overviews, 
such as refs.\ \cite{evgenireview,hammerreview,machleidtreview}. 

\section{Low-Energy QCD and Chiral Symmetry}

Before turning to the more detailed presentation of the ChEFT approach to nuclear 
many-body systems, it is useful and instructive to briefly recall how the special role 
of the pion emerges in the context of chiral symmetry in QCD, through the 
Nambu-Goldstone mechanism of spontaneous symmetry breaking. In other words: how does 
Yukawa's pion figure in the framework of QCD?

\subsection{\it Chiral Symmetry and the Pion}

Historically, the foundations for understanding the pion as a Nambu-Goldstone boson 
\cite{G61,NJL61} were initiated in the 1960's, culminating in the current algebra 
approaches \cite{AD68} (combined with the PCAC relation for the pion) of the pre-QCD 
era. A most inspiring work from this early period is the one by Nambu and 
Jona-Lasinio \cite{NJL61} (NJL). Just as the BCS theory provided an understanding of the 
basic mechanism behind superconductivity, the NJL model helped clarify the dynamics 
that drives spontaneous chiral symmetry breaking and the formation of pions as 
Goldstone bosons. 

Consider as a starting point the color current of quarks, ${\bf J}_\mu^a = \bar{q}
\gamma_\mu {\bf t}^a q$, where ${\bf t}^a$ ($a = 1, ... ,8$) are the generators of the 
$SU(3)_c$ color gauge group and $q$ denotes the quark fields with $4N_c N_f$ components 
representing their spin, color and flavor degrees freedom $(N_c=3)$. This current 
couples to the gluon fields. Assume that the distance over which color propagates is 
restricted to a short correlation length $l_c$. Then the interaction experienced by 
low-momentum quarks can be schematically viewed as a local coupling between their color 
currents:
\begin{equation}
{\cal L}_{\rm int} = -G_c\,{\bf J}_\mu^a(x)\,{\bf J}^\mu_a(x)\,\, ,
\label{eq:Lint1}
\end{equation}
where $G_c \sim \bar{g}^2\, l_c^2$ is an effective coupling strength of dimension 
{\it length}$^2$ which encodes the QCD coupling, $g$, averaged over the relevant distance 
scales, in combination with the squared correlation length, $l_c^2$. 

Now adopt the local interaction Eg.(\ref{eq:Lint1}) and write the following model 
Lagrangian for the quark fields $q(x)$:
\begin{equation}
{\cal L} = \bar{q}(x)(i\gamma^\mu\partial_\mu - m_q)q(x) + {\cal L}_{\rm int}(\bar{q},q)
\,.\label{eq:NJL}
\end{equation}
In essence, by ``integrating out" gluon degrees of freedom and absorbing them in the 
four-fermion interaction ${\cal L}_{\rm int}$, the local $SU(3)_c$ gauge symmetry of QCD 
is now replaced by a global one only. Confinement is obviously lost, but all other 
symmetries of QCD are maintained. The mass matrix $m_q$ incorporates small ``bare" 
quark masses. In the limit $m_q \rightarrow 0$, the Lagrangian (\ref{eq:NJL}) has a 
chiral symmetry of left- and right-handed quarks, $SU(N_f)_L\times SU(N_f)_R$, that it 
shares with the original QCD Lagrangian for $N_f$ massless quark flavors.

A Fierz transform of the color current-current interaction (\ref{eq:Lint1}) produces a 
set of exchange terms acting in quark-antiquark channels. For the $N_f = 2$ case: 
\begin{equation}
{\cal L}_{\rm int} \rightarrow {G\over 2}\left[(\bar{q}q)^2 + (\bar{q}\,i\gamma_5\,
\vec\tau \,q)^2\right] + ... \,\, , \label{eq:Lint2}
\end{equation}
with the isospin Pauli matrices $\vec \tau = (\tau_1,\tau_2,\tau_3)$. For brevity we 
have not shown a series of terms with combinations of vector and axial vector currents, 
both in color singlet and color octet channels. The constant $G$ is proportional to 
the color coupling strength $G_c$. The ratio of these two constants is uniquely 
determined by the number of colors and flavors, $N_c$ and $N_f$.

The steps just outlined can be viewed as a contemporary way of introducing the 
time-honored NJL model \cite{NJL61}. This model has been further developed and 
applied \cite{VW91,HK94} to a variety of problems in hadron physics. The virtue of 
this schematic model is its simplicity in illustrating the basic mechanism of 
spontaneous chiral symmetry breaking, as follows. In the mean-field (Hartree) 
approximation the equation of motion derived from the Lagrangian (\ref{eq:NJL}) leads 
to a gap equation
\begin{equation}
M_q = m_q - G\langle0|\bar{q}q|0\rangle\,\, ,
\end{equation}
which links the dynamical generation of a constituent quark mass $M_q$ to the 
appearance of the chiral quark condensate
\begin{equation}
\langle 0|\bar{q}q|0\rangle= -{\rm tr}\lim_{\,x\rightarrow 0}\langle0| {\cal T}q(0)
\bar{q}(x)|0\rangle = -2iN_fN_c\int {d^4p\over (2\pi)^4}{M_q\,\theta(\Lambda -
|\vec{p}\,|)\over p^2 - M_q^2 + i\epsilon}\,\, .
\end{equation}
This condensate plays the role of an order parameter of spontaneous chiral symmetry 
breaking. Starting from $m_q = 0$ a non-zero quark mass $M_q$ develops dynamically, 
together with a non-vanishing chiral condensate $\langle 0|\bar{q}q|0\rangle$, once 
$G$ exceeds a critical value of order $G_{\rm crit} \simeq 10$ GeV$^{-2}$. The procedure 
requires a momentum cutoff $\Lambda \simeq 2M_q$ beyond which the interaction is 
``turned off". Note that the strong non-perturbative interactions, by polarizing the 
vacuum and turning it into a condensate of quark-antiquark pairs, transmute an 
initially pointlike quark with its small bare mass $m_q$ into a massive quasi-particle 
with a size of order $(2M_q)^{-1}$.

\subsection{\it The Pseudoscalar Meson Spectrum}

The NJL model demonstrates lucidly the appearance of chiral Nambu-Goldstone bosons. 
Solving Bethe-Salpeter equations in the color singlet quark-antiquark channels 
generates the lightest mesons as quark-antiquark excitations of the correlated QCD 
ground state with its condensate structure. Several such calculations have been 
performed in the past with $N_f = 3$ quark flavors \cite{VW91,HK94,KLVW90}. Such a 
model has an unwanted $U(3)_L\times U(3)_R$ symmetry to start with, but due to the axial 
$U(1)_A$ anomaly of QCD this symmetry\footnote{The axial $U(1)$-transformations of 
quarks, $ q \to \exp(i\alpha \gamma_5)q $, constitute only a symmetry of classical 
chromodynamics, but not of its quantum version.} should be reduced to $SU(3)_L\times 
SU(3)_R\times 
U(1)_V$. In the NJL model, instanton driven interactions are incorporated in the form 
of a flavor determinant \cite{tH76} $\det[\bar{q}_i(1 \pm \gamma_5)q_j]$. This 
interaction involves all three flavors $u, d, s$ simultaneously in a genuine 
three-body (contact) term. 

 \begin{figure}
       \centerline{\includegraphics[width=8.3cm] {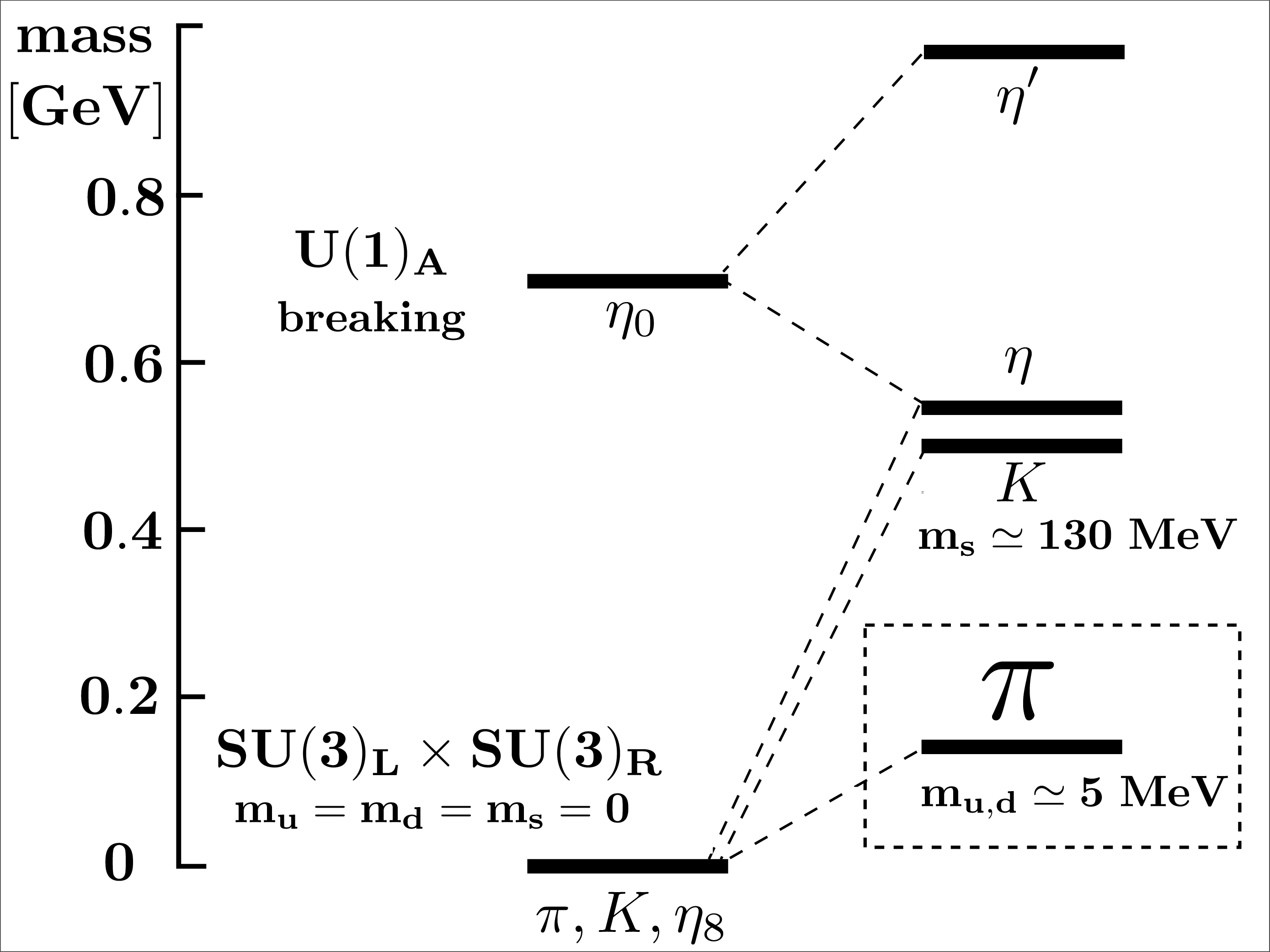}}
\caption{Symmetry breaking pattern in the pseudoscalar meson nonet calculated in the 
three-flavor NJL model \cite{KLVW90} .}
   \label{fig:2}
 \end{figure}

The symmetry breaking pattern resulting from such a calculation is apparent in the 
pseudoscalar meson spectrum of Fig.\ \ref{fig:2}. Starting from massless $u, d$ and 
$s$ quarks, the pseudoscalar octet emerges as a set of massless Goldstone bosons of 
spontaneously broken $SU(3)_L\times SU(3)_R$, while the broken $U(1)_A$ drives the 
singlet $\eta_0$ away from the Goldstone boson sector. Finite quark masses shift the 
pseudoscalar ($J^\pi = 0^-$) nonet into its empirically observed position, including 
$\eta_0$-$\eta_8$ mixing. 

The very special nature of the pion as a Nambu-Goldstone boson is manifest in the famous 
Gell-Mann-Oakes-Renner relation \cite{GOR68} derived from current algebra and PCAC:
\begin{equation}
m^2_{\pi}\,f_{\pi}^2 = - {1\over 2}(m_u + m_d) \langle 0| \bar{q} q |0\rangle + 
{\cal O}(m^2_{u,d} \ln m_{u,d}) .
\label{eq:GOR}
\end{equation}
It involves the pion decay constant, $f_\pi = 92.4$\,MeV, defined by the matrix 
element which connects the pion-state with the QCD vacuum via the isovector axial 
vector current, $A^\mu_a = \bar{q}\gamma^\mu\gamma_5 {\tau_a \over 2} q$:
\begin{equation}
\langle 0 | A^{\mu}_a (0) | \pi_b (p) \rangle = i \delta_{ab}\, p^{\mu} f _\pi\, .
\label{eq:fpi}
\end{equation}
Just like the chiral condensate $\langle 0| \bar{q} q |0\rangle$, the pion decay 
constant $f_\pi$ is a measure of spontaneous chiral symmetry breaking expressed in 
terms of a characteristic scale $\Lambda_\chi \sim 4\pi f_\pi \sim 1$ GeV. The non-zero 
pion mass, $m_\pi \simeq 138$\,MeV\,$\ll \Lambda_\chi$, reflects the explicit symmetry 
breaking by the small quark masses, with $m^2_{\pi} \sim m_q$. One should note that the 
quark masses $m_{u,d}$ and the condensate $\langle 0|\bar{q} q |0\rangle$ are both 
scale dependent quantities. Only their product is scale independent, i.e. invariant 
under the renormalization group. At a renormalization scale of about 2 GeV, a typical 
average quark mass ${1\over 2}(m_u + m_d) \simeq 3.5$ MeV implies $\langle 0 |\bar{q} q
|0 \rangle \simeq -(0.36$\,GeV$)^3$.

\subsection{\it Scales and Symmetry Breaking Patterns}

The quark masses are the only parameters that set primary scales in QCD. Their 
classification into sectors of ``light" and ``heavy" quarks determines very different 
physics phenomena. While the heavy quarks (i.e. the $t$, $b$ and - within limits - 
the $c$ quarks) offer a natural ``small parameter" in terms of their reciprocal masses, 
such that non-relativistic approximations (expansions of observables in powers of 
$1/m_{t,b,c}$) tend to work increasingly well with increasing quark mass,
the sector of the light quarks (i.e. the $u$, $d$ quarks and - to some extent - the 
$s$ quark) is governed by quite different principles and rules. Evidently, the quark 
masses themselves are now ``small parameters", to be compared with a characteristic 
``large" scale of dynamical origin. This large scale is  visible as a characteristic 
mass gap of about 1 GeV which separates the QCD vacuum from almost all of its 
excitations, with the exception of the pseudoscalar meson octet of pions, kaons and 
the eta meson. This mass gap is in turn comparable to $4\pi f_\pi$, the scale 
associated with spontaneous chiral symmetry breaking in QCD. 

\section{Chiral Effective Field Theory}

Low-energy QCD is the physics of systems of light quarks at energy and momentum scales 
smaller than the 1 GeV mass gap observed in the hadron spectrum. 
This 1 GeV scale set by $4 \pi f_{\pi} $ offers a natural separation between ``light" 
and ``heavy" (or, correspondingly, ``fast" and ``slow") degrees of freedom. The basic 
idea of an effective field theory is to
introduce the active light particles as collective degrees of freedom,  while the
heavy particles are frozen and treated as (almost) static sources. The dynamics
is described by an effective Lagrangian which incorporates all relevant
symmetries of the underlying fundamental theory. In QCD, confinement and spontaneous 
chiral symmetry breaking implies that the ``fast" degrees of freedom are the 
Nambu-Goldstone bosons. With Yukawa's pion in mind, we restrict ourselves to $N_f = 2$. 

\subsection{\it The Nambu-Goldstone Boson Sector}

We first briefly summarize the steps \cite{Wei66/67,GL84} required in the pure meson 
sector (baryon number $B= 0$) and later for the pion-nucleon sector ($B= 1$). A chiral 
field is introduced as 
\begin{equation}
U(x) = \exp\bigg({i \over f_\pi}\, \vec \tau \cdot \vec \pi(x) \bigg) \in SU(2)~,
\end{equation}
with the Goldstone pion fields $\vec \pi(x)$ divided by the pion decay constant $f_\pi$ 
in the chiral limit ($m_\pi \rightarrow 0$). The QCD Lagrangian is replaced by an 
effective Lagrangian which involves the chiral field $U(x)$ and its derivatives:
\begin{equation}
{\cal L}_{QCD} \to {\cal L}_{eff} (U, \partial^\mu U, ...) .
\label{eq:Leff}
\end{equation}
Goldstone bosons interact only when they carry non-zero four-momenta, so the low-energy
expansion of (\ref{eq:Leff}) is an ordering in powers of $\partial^{\mu} U$. Lorentz
invariance permits only even numbers of derivatives. One writes a series
\begin{equation}
{\cal L}_{\rm eff} = {\cal L}_{\pi\pi}^{(2)} + {\cal L}_{\pi\pi}^{(4)} + ...\,,
\end{equation}
where the leading term (the non-linear sigma model) involves just two derivatives:
\begin{equation}
{\cal L}^{(2)}_{\pi\pi} = {f_\pi^2 \over 4} {\rm tr} ( \partial^{\mu} U \partial_{\mu} 
U^{\dagger} )\, .\end{equation}
At fourth order, the terms permitted by symmetries are (apart from further
contributions involving the light quark mass $m_q$ and external fields, not shown here):
\begin{equation}
{\cal L}^{(4)}_{\pi\pi} = {\ell_1 \over 4} \Big[ {\rm tr}(\partial^{\mu} U \partial_{\mu} 
U^{\dagger})\Big]^2 + {\ell_2 \over 4}  {\rm tr} (\partial_{\mu} U \partial_{\nu} U^{\dagger}) 
\, {\rm tr} (\partial^{\mu} U \partial^{\nu} U^{\dagger}) +\dots \, ,\end{equation}
The constants $\ell_1, \ell_2$ (following canonical notations \cite{GL84}) absorb 
divergences of loops, and their finite scale-dependent parts must be determined 
from experiment. 

The symmetry breaking mass term is small, so that it can be handled
perturbatively, together with the power series in small momenta. The leading
contribution introduces a term linear in the quark mass matrix $m_q = {\rm diag}
(m_u,m_d)$:
\begin{equation}
{\cal L}^{(2)}_{\pi\pi} = {f_\pi^2 \over 4} {\rm tr}(\partial_{\mu} U \partial^{\mu} 
U^{\dagger}) + {f_\pi^2 \over 4} m_\pi^2 \, {\rm tr}(U + U^{\dagger}) \,,
\end{equation}
with $m_\pi^2 \sim (m_u+m_d)$. The fourth order term ${\cal L}^{(4)}_{\pi\pi}$ also 
receives explicit chiral symmetry breaking contributions (proportional to $m_q$ and 
$m_q^2$) with additional low-energy constants $\ell_{3,4}$.

To the extent that the effective Lagrangian includes all terms allowed by symmetries 
of QCD, the chiral effective field theory is the low-energy equivalent \cite{Wei79,L94} 
of the original QCD Lagrangian. Given the effective Lagrangian, the framework for 
systematic perturbative calculations of (on-shell) $S$-matrix elements involving 
Goldstone bosons, Chiral Perturbation Theory (ChPT), is then defined by the following 
rules:\\  Collect all Feynman diagrams generated by ${\cal L}_{eff}$. Classify
all terms according to powers of a small quantity $p$
which stands generically for three-momenta or energies 
of Goldstone bosons, or for the pion mass $m_{\pi}$. The small
expansion parameter is $q/(4\pi f_{\pi})$. Loops are evaluated in
dimensional regularization and get renormalized by appropriate chiral counter terms. 

\subsection{\it The Chiral Pion-Nucleon Effective Lagrangian}

The prominent role played by the pion as a Goldstone boson of spontaneously
broken chiral symmetry has its impact on the low-energy structure
and dynamics of nucleons as well \cite{TW01}. When probing the nucleon with 
long-wavelength electroweak fields, a substantial part of the response comes from 
the pion cloud, the ``soft'' surface of the nucleon. 
The calculational framework for this, baryon chiral perturbation theory 
\cite{EM96,BKM95} has been applied quite successfully to a variety of low-energy 
processes (such as low-energy pion-nucleon scattering, threshold pion photo- and 
electroproduction and Compton scattering on the nucleon). 

Consider now the sector with baryon number $B = 1$ and the physics of the pion-nucleon 
system. The nucleon is represented by an isospin-$1/2$ doublet, Dirac spinor 
$\Psi= \Big(\!\!\begin{array}{c} p \\ n\end{array}\!\!\Big)$ of proton and neutron. The free Lagrangian
\begin{equation}
{\cal L}_N^{\rm free} = \bar{\Psi}(i\gamma_{\mu}\partial^{\mu} - M_0)\Psi
\end{equation}
includes the nucleon mass in the chiral limit, $M_0$. One should note that the nucleon, 
unlike the pion, has a large mass of the same order as the chiral symmetry breaking 
scale $4\pi f_\pi$, which survives in the limit of vanishing quark masses, 
$m_{u,d}\to 0$. 

The previous pure meson Lagrangian ${\cal L}_{eff}$
is now replaced by ${\cal L}_{eff}(U,\partial^{\mu} U, \Psi, ...)$ which also
includes the nucleon field. The additional term involving the nucleon, denoted
by ${\cal L}_{\pi N}$, is expanded again in powers of derivatives (external
momenta) of the Goldstone boson field and of the quark masses: 
\begin{equation}
{\cal L}_{\pi N} = {\cal L}_{\pi N}^{(1)} + {\cal L}_{\pi N}^{(2)} + \dots
\end{equation}
In the leading term, ${\cal L}_{\pi N}^{(1)}$ there is a replacement of $\partial^{\mu}$ 
by a chiral covariant derivative which introduces a vector current coupling between 
the pions and the nucleon. Secondly, there is an axial vector coupling. This structure 
of the $\pi N$ effective Lagrangian is again dictated by chiral symmetry. We have
\begin{equation}
{\cal L}_{\pi N}^{(1)} =  \bar{\Psi}\Big[i\gamma_{\mu}(\partial^{\mu} +\Gamma^{\mu})- M_0 + 
g_A \gamma_{\mu}\gamma_5\, u^{\mu}\Big]\Psi \,,\label{eq:LeffN}
\end{equation}
with vector and axial vector quantities involving the Goldstone boson (pion)
fields via $\xi = \sqrt{U}$ in the form:
\begin{eqnarray}
\Gamma^{\mu} & = & {1\over 2}[\xi^{\dagger},\partial^{\mu}\xi] = {i\over 4f_\pi^2}
 \, \vec \tau \cdot (\vec \pi \times \partial^{\mu}\vec \pi) + ...~~, \\
u^{\mu} & = & {i\over 2}\{\xi^{\dagger},\partial^{\mu}\xi\}= - {1\over 2f_\pi}\, \vec \tau
\cdot \partial^{\mu}\vec \pi + ...~~,
\end{eqnarray}
where the last steps result when expanding $\Gamma^{\mu}$ and $u^{\mu}$ to 
leading order in the pion fields. So far, the only parameters that enter are the 
nucleon mass $M_0$, the pion decay constant $f_\pi$, and the nucleon axial vector 
coupling constant $g_A$, all three taken in the chiral limit.

At next-to-leading order, ${\cal L}_{\pi N}^{(2)}$, the chiral symmetry breaking 
quark mass term enters. It has the effect of shifting the nucleon mass from
its value in the chiral limit to the physical one. The nucleon sigma term
\begin{equation}
\sigma_N = m_q\frac{\partial M_N}{\partial m_q} =
\langle N | m_q(\bar{u}u + \bar{d}d) |N\rangle
\end{equation}
measures the contribution of the non-vanishing quark mass, $m_q =
\frac{1}{2}(m_u + m_d)$, to the nucleon mass $M_N$. Its empirical value is in the range
$\sigma_N \simeq (45 \pm 8)$ MeV and has been deduced \cite{GLS91} by a sophisticated 
extrapolation of low-energy  pion-nucleon data using dispersion relation techniques.
Up to this point, the $\pi N$ effective Lagrangian, expanded to second order
in the pion field, has the form
\begin{eqnarray}
{\cal L}_{eff}^{N} & = & \bar{\Psi}(i\gamma_{\mu}\partial^{\mu} - M_N)\Psi - 
{g_A \over 2f_{\pi}} \bar{\Psi}\gamma_{\mu}\gamma_5\vec\tau\,\Psi \cdot\partial^{\mu}
\vec \pi  \nonumber \\ && -{1 \over 4f_{\pi}^2} \bar{\Psi}\gamma_{\mu} \vec \tau\,\Psi
\cdot (\vec \pi\times \partial^{\mu} \vec \pi\,) +{\sigma_N\over  2f_\pi^2}\,\bar{\Psi}
\Psi\,\vec \pi^{\,2} + ...~~, \end{eqnarray}
where we have not shown a series of additional terms involving 
$(\partial^{\mu} \vec\pi)^2$ arising from the complete Lagrangian ${\cal L}_{\pi N}^{(2)}$.
These terms come with further low-energy constants $c_{3,4}$ encoding physics at smaller 
distances or higher energies. These constants need to be fitted to experimental data, 
e.g. from pion-nucleon scattering. 

The ``effectiveness" of such an effective field theory relies on the proper 
identification of the active low-energy degrees of freedom. Pion-nucleon scattering is 
known to be dominated by the $p$-wave  $\Delta(1232)$ resonance with spin and isospin 
3/2. The excitation energy of this resonance, given by the mass difference 
$\Delta=M_\Delta - M_N \simeq 293\,$MeV, is not large, although it does not vanish in 
the chiral limit. If the physics of the $\Delta(1232)$ is absorbed in low-energy 
constants of an effective theory that works with pions and nucleons only (as done 
in heavy-baryon ChPT), the limits of applicability of such a theory is clearly 
narrowed down to an energy-momentum range small compared to $\Delta$. The $B=1$ chiral 
effective Lagrangian is therefore often extended \cite{HHK97} by incorporating the 
$\Delta(1232)$ isobar as an explicit degree of freedom.

\section{Chiral Nuclear Interactions}
 \begin{figure}
      \centerline{\includegraphics[width=18cm] {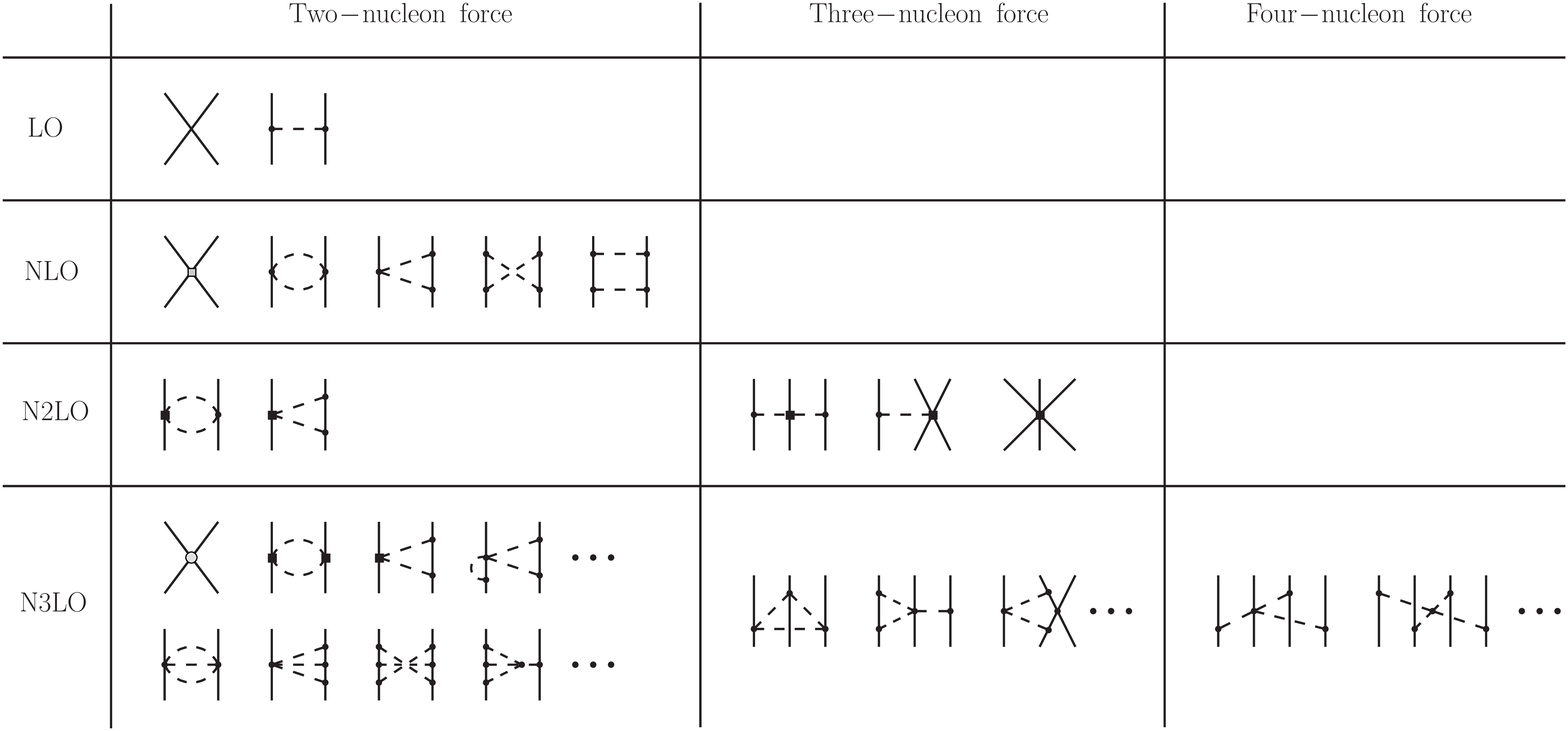}}
\caption{Hierarchical organization of nuclear forces in chiral effective field theory.}
\label{fignforce}
 \end{figure}
\subsection{\it Two-nucleon interaction}
In this section, we briefly review the structure of the chiral nucleon-nucleon potential 
in momentum-space \cite{evgenireview,hammerreview,machleidtreview}. It amounts to a series 
of contributions from explicit multi-pion exchanges and short-distance contact-terms 
(shown in Fig.\ \ref{fignforce}) controlled by chiral symmetry and an expansion in 
small external momenta. The leading-order contribution consists of the well-known one-pion 
exchange piece and two contact-terms operating in the spin-singlet and spin-triplet $S$-waves:
\begin{equation} V_{NN}^{(\rm LO)} = -{g_A^2\over 4 f_\pi^2} \, {\vec \sigma_1 \cdot \vec q\,\,
\vec \sigma_2 \cdot \vec q \over m_\pi^2+q^2}\,\vec \tau_1\cdot \vec \tau_2 + C_S + C_T \,
\vec \sigma_1 \cdot\vec \sigma_2 \,.\end{equation}
Here,  $\vec \sigma_{1,2}$ and  $\vec \tau_{1,2}$ denote usual spin- and isospin operators of 
the two nucleons and $\vec q$ is the momentum transfer between both nucleons. At 
next-to-leading order the two-pion exchange diagrams generated by the vertices of the 
chiral $\pi N$-Lagrangian ${\cal L}_{\pi N}^{(1)}$ come into play. The corresponding potential 
reads \cite{nnpap1}:
\begin{eqnarray} V_{NN}^{(\rm 2\pi-NLO)} &=& {1\over 384 \pi^2 f_\pi^4} \bigg\{ 4m_\pi^2
(1+4g_A^2-5g_A^4) +q^2(1+10g_A^2-23g_A^4) - {48g_A^4 m_\pi^4 \over 4m_\pi^2+q^2} \bigg\} L(q)\,\, 
\vec \tau_1\cdot \vec \tau_2 \nonumber \\ && +{3g_A^4 \over 64\pi^2 f_\pi^4}  L(q) \,
( \vec \sigma_1 \cdot\vec \sigma_2 \, \vec q^{\,2} -  \vec \sigma_1 \cdot \vec q\,\,
\vec \sigma_2 \cdot \vec q\,)\,, \end{eqnarray}
with the logarithmic loop function
\begin{equation} L(q) = {\sqrt{4m_\pi^2+q^2} \over q} \ln{ q+\sqrt{4m_\pi^2+q^2} \over 2m_\pi} 
\,.\end{equation} 
Notably, there are only contributions to the isovector central channel and the isoscalar 
spin-spin and tensor channel of the NN-interaction. Additional polynomial pieces generated 
by the pion-loops have been left out in the above expression. Without loss of 
information, these can be absorbed into the most general contact-term at NLO:
\begin{eqnarray} V_{NN}^{(\rm ct-NLO)} &=& C_1 \, \vec q^{\,2} +  C_2 \, \vec k^{\,2} +  (C_3 \, 
\vec q^{\,2} +  C_4 \, \vec k^{\,2})\, \vec \sigma_1 \cdot\vec \sigma_2 + {i \over 2 }C_5 \,
( \vec \sigma_1 +\vec \sigma_2) \cdot (\vec q \times \vec k\,) \nonumber \\
&& + C_6\,\vec \sigma_1 \cdot \vec q\,\,\vec \sigma_2 \cdot \vec q\, +  C_7\,\vec \sigma_1 
\cdot \vec k\,\,\vec \sigma_2 \cdot \vec k\, , \end{eqnarray}
with $\vec k = (\vec p + \vec p\,')/2$ the half-sum of initial and final nucleon momenta.
The low-energy contants $C_1, \dots, C_7$ are adjustable parameters to be determined in 
fits to empirical NN-phase shifts. Note that $C_5$ parametrizes the strength of the 
short-distance spin-orbit NN-interaction. In fact it dominates over the finite-range 
contributions to the spin-orbit interaction which arise at higher orders. The low-energy constant 
$C_5$ will reappear in the discussion of the nuclear 
energy density functional, where the same coupling determines the (self-consistent) 
single-particle spin-orbit potential in finite nuclei proportional to the density 
gradient.       

The most important pieces from chiral two-pion exchange which generate attraction in the 
isoscalar central channel and reduce the too strong $1\pi$-exchange isovector tensor force 
are still absent at this level. These arise first from subleading $2\pi$-exchange 
through the chiral $\pi\pi NN$ contact couplings $c_{1,3,4}$ in ${\cal L}_{\pi N}^{(2)}$  or 
from the inclusion of the $\Delta(1232)$-isobar as an explicit degree of freedom. In the first 
approach one gets \cite{nnpap1}:
\begin{eqnarray} V_{NN}^{(\rm 2\pi-N^2LO)} &=& {3g_A^2 \over 16\pi f_\pi^4} \Big[ c_3\, 
q^2+2m_\pi^2(c_3-2c_1) \Big] (2m_\pi^2+q^2) A(q) \nonumber \\ && + {g_A^2 c_4\over 32\pi 
f_\pi^4} (4m_\pi^2+q^2) A(q)\, ( \vec \sigma_1 \cdot\vec \sigma_2 \, \vec q^{\,2} -  
\vec \sigma_1 \cdot \vec q\,\,\vec \sigma_2 \cdot \vec q\,)\,\vec \tau_1\cdot \vec \tau_2 
\,,\end{eqnarray}
with the loop function
\begin{equation} A(q) = {1\over 2 q} \arctan{q \over 2m_\pi} \,.\end{equation}
At the same order there are additional relativistic $1/M_N$-corrections to $2\pi$-exchange. 
Their explicit form depends on the precise definition of the nucleon-nucleon potential 
$V_{NN}$, which by itself is not an observable. Covariant perturbation theory \cite{nnpap1} and 
the method of unitary transformations \cite{evgenireview} thus lead to slightly different 
expressions for these small $1/M_N$-corrections. As the state of the art, the 
chiral NN-potential has 
been constructed up to order N$^3$LO and it includes two-loop $2\pi$-exchange, 
$3\pi$-exchange and contact-terms quartic in the momenta parametrized by 15 additional low-energy 
constants $D_1, \dots, D_{15}$. When inserted into the Lippmann-Schwinger equation (in order to 
solve for the unitary $S$-matrix or the $T$-matrix) the chiral NN-potential is multiplied by 
an exponential regulator function with a cutoff scale $\Lambda =500 - 700\,$MeV in order to 
restrict the potential to the low-momentum region where chiral perturbation theory is applicable. 
Furthermore, methods of spectral function regularization \cite{evgenireview} have been employed 
in order to eliminate the high-momentum region in the pion-loop integrals directly. In this case 
the loop functions $L(q)$ and $A(q)$ receive an additional dependence on a regulator scale 
$\tilde \Lambda$.  

At order N$^3$LO the chiral NN-potential reaches the quality of a 
``high-precision'' potential in reproducing empirical NN-phase shifts and deuteron properties.
At the same time it provides the appropriate two-body interaction constrained by chiral 
symmetry of QCD for nuclear few- and many-body calculations.

\subsection{\it Nuclear Many-Body Forces}

Within the chiral effective field theory framework employing nucleons and pions
as the explicit degrees of freedom, the leading-order contribution to the 
nuclear three-body force arises at order N$^2$LO and consists of three terms.
The two-pion exchange three-nucleon force contains terms proportional to the 
low-energy constants $c_1$, $c_3$, and $c_4$ and has the form
\begin{equation}
V_{3N}^{(2\pi)} = \sum_{i\neq j\neq k} \frac{g_A^2}{8f_\pi^4} 
\frac{\vec{\sigma}_i \cdot \vec{q}_i \, \vec{\sigma}_j \cdot
\vec{q}_j}{(\vec{q_i}^2 + m_\pi^2)(\vec{q_j}^2+m_\pi^2)}
F_{ijk}^{\alpha \beta}\tau_i^\alpha \tau_j^\beta,
\label{3n1}
\end{equation}
where $\vec{q}_i$ denotes the difference between the final and initial momenta of 
nucleon $i$, and the isospin tensor $F_{ijk}^{\alpha \beta}$ is explicitly written
\begin{equation}
F_{ijk}^{\alpha \beta} = \delta^{\alpha \beta}\left (-4c_1m_\pi^2
 + 2c_3 \,\vec{q}_i \cdot \vec{q}_j \right ) + 
c_4 \,\epsilon^{\alpha \beta \gamma} \tau_k^\gamma \,\vec{\sigma}_k
\cdot \left ( \vec{q}_i \times \vec{q}_j \right ),
\label{3n4}
\end{equation}
resulting in $c_1$ and $c_3$ terms proportional to $\vec \tau_i \cdot \vec \tau_j$ and
the $c_4$ term proportional to $\vec \tau_k \cdot (\vec \tau_i \times \vec \tau_j)$. 
The low-energy constants $c_1$, $c_3$, and $c_4$ can be fit to pion-nucleon
\cite{buttiker00} or nucleon-nucleon scattering data \cite{rentmeester03,entem03}.

The one-pion exchange term in the N$^2$LO chiral three-nucleon interaction is
proportional to the low-energy constant $c_D$ and has the form
\begin{equation}
V_{3N}^{(1\pi)} = -\sum_{i\neq j\neq k} \frac{g_A c_D}{8f_\pi^4\, \Lambda} 
\,\frac{\vec{\sigma}_j \cdot \vec{q}_j}{\vec{q_j}^2+m_\pi^2}\, \vec{\sigma}_i \cdot
\vec{q}_j \, {\vec \tau}_i \cdot {\vec \tau}_j \, ,
\label{3n2}
\end{equation}
while the three-nucleon contact interaction introduces the low-energy constant $c_E$:
\begin{equation}
V_{3N}^{(\rm ct)} = \sum_{i\neq j\neq k} \frac{c_E}{2f_\pi^4\, \Lambda}
\,{\vec \tau}_i \cdot {\vec \tau}_j\, ,
\label{3n3}
\end{equation}
where $\Lambda = 700$ MeV sets a natural scale. 
The low-energy constants $c_D$ and $c_E$, associated with the one-pion exchange
and contact three-nucleon forces respectively, ideally are fit to properties of three-body
systems only. The binding energies of $A=3$ nuclei and the $\beta$-decay lifetime of $^3$H
provide largely uncorrelated nuclear observables with which to fit these two low-energy
constants \cite{gazit09}. 

At order N$^3$LO in the chiral power counting, corresponding to ${\cal O}(q^4)$ in powers of momentum,
additional three- and four-nucleon force contributions arise without any additional undetermined low-energy 
constants. The N$^3$LO three-body force can be written schematically as
\be
V_{3N}^{(4)} = V_{2\pi}^{(4)} + V_{2\pi-1\pi}^{(4)}+V_{\rm ring}^{(4)} + V_{1\pi-\rm cont.}^{(4)}
+ V_{2\pi-\rm cont.}^{(4)} + V_{1/m}^{(4)},
\ee
where the individual terms denote specific topologies. The $2\pi$ and $1\pi-{\rm cont.}$ topologies
are present already in the N$^2$LO chiral three-nucleon force, whereas the $2\pi-1\pi$, ring, and 
$2\pi-{\rm cont.}$ topologies enter first at order N$^3$LO. All contributions have been worked out
and presented in refs.\ \cite{ishikawa07,bernard08,bernard11}. 
The N$^3$LO four-nucleon force requires the
evaluation of much fewer diagrams than the N$^3$LO three-body force, and the resulting analytic
expressions (presented in ref.\ \cite{epelbaum06}) are also considerably simpler.
 
\subsection{\it Role of Explicit $\Delta$(1232) Degrees of Freedom}

The standard version of chiral meson-baryon effective 
field theory works with pions and nucleons only, both of which are stable particles with respect to the 
strong interaction. Effects of the $\Delta(1232)$ are encoded in low-energy constants
such as $c_3$ and $c_4$ that are determined either by pion-nucleon scattering data or by fits to 
nucleon-nucleon phase shifts.

\begin{figure}
\centerline{\includegraphics[width=14cm] {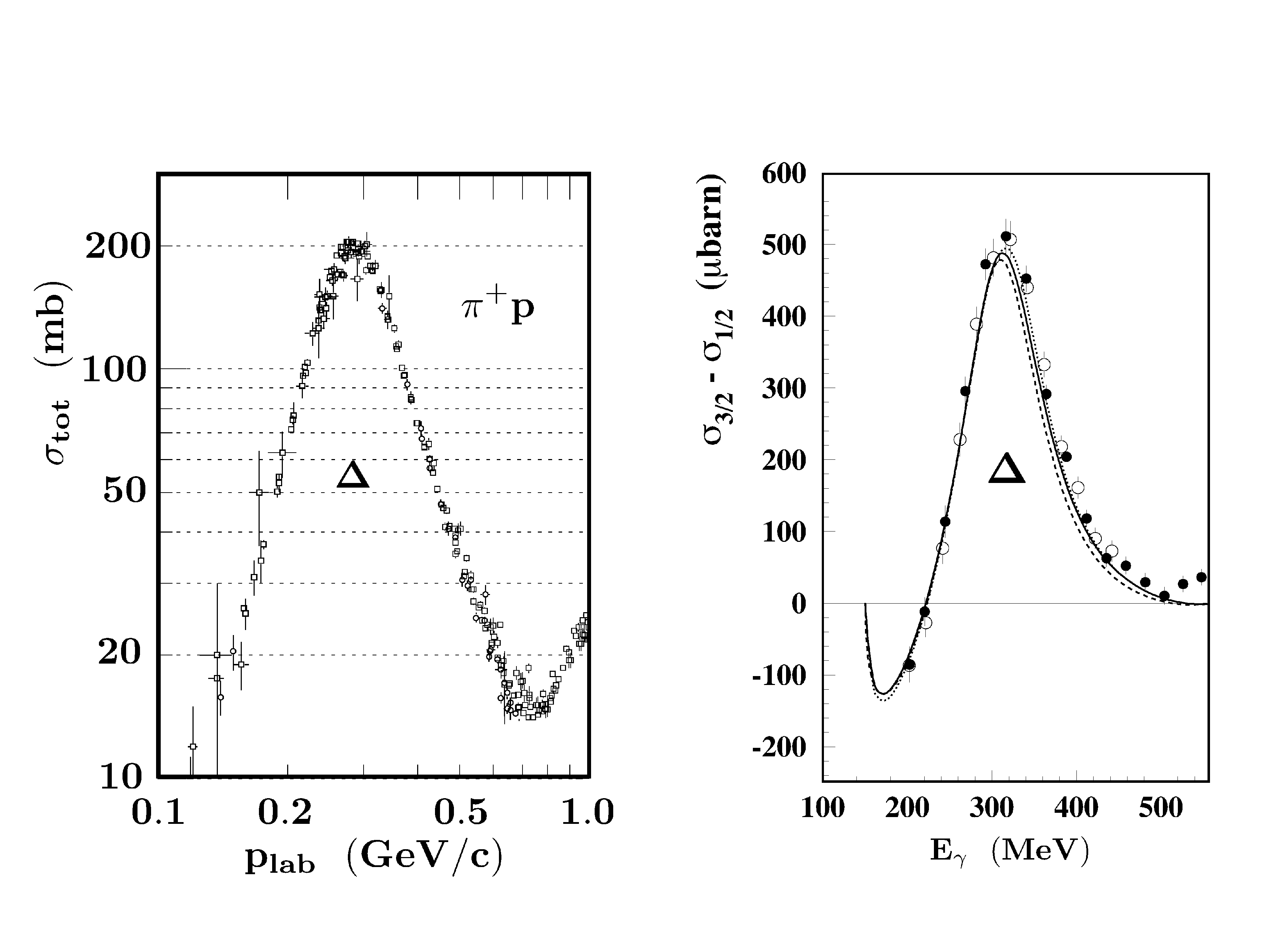}}
\caption{Left: Total $\pi^+p$ cross section in the region of the $\Delta(1232)$ resonance (adapted from \cite{pdg2012}).
Right: Difference of polarized Compton scattering cross sections of the proton, $\sigma_{3/2}$ and $\sigma_{1/2}$, referring to channels with total angular momentum ${3\over 2}$ and ${1\over 2}$, respectively. Data taken from \cite{ahrens2001}. Curves represent dispersion relation and multipole analysis cited in \cite{ahrens2001}.}
\label{Delta}
\end{figure}

On the other hand, the mass difference between the $\Delta(1232)$ isobar and the nucleon is only about $0.3\,{\rm GeV} \simeq 2m_\pi$, yet another ``small" scale compared to the chiral symmetry breaking scale,
$4\pi f_\pi \sim$ 1\,GeV. This suggests incorporating the $\Delta(1232)$ as an additional {\it explicit} degree of 
freedom in the effective field theory. In fact the $\Delta$ isobar is by far the dominant feature in the 
excitation spectrum of the nucleon observed in pion and photon scattering measurements. 
An example is the strong spin-isospin excitation observed in pion-nucleon scattering 
(see Fig.\ \ref{Delta}, left). The $\Delta(1232)$ is also seen prominently in polarized Compton scattering on the proton
(Fig.\ \ref{Delta}, right).

In the low-energy expansion of the spin-independent $\pi N$ scattering amplitude, 
the strong spin-isospin response of the nucleon manifests itself in a large ``axial'' polarizabilty: 
\begin{equation} 
\alpha_A^{(\Delta)} = {g_A^2\over f_\pi^2 (M_\Delta - M_N)} \simeq 5\, {\rm fm}^3~~.
\end{equation}
Here the factor ${g_A^2/f_\pi^2}$ comes from the axial vector coupling of the pion that 
drives the $\pi N\rightarrow\Delta$ transition. The mere magnitude of this polarizability, several times the volume of the
nucleon itself, already illustrates the importance of this effect.

\begin{figure}
\begin{center}
\includegraphics[width=3cm]{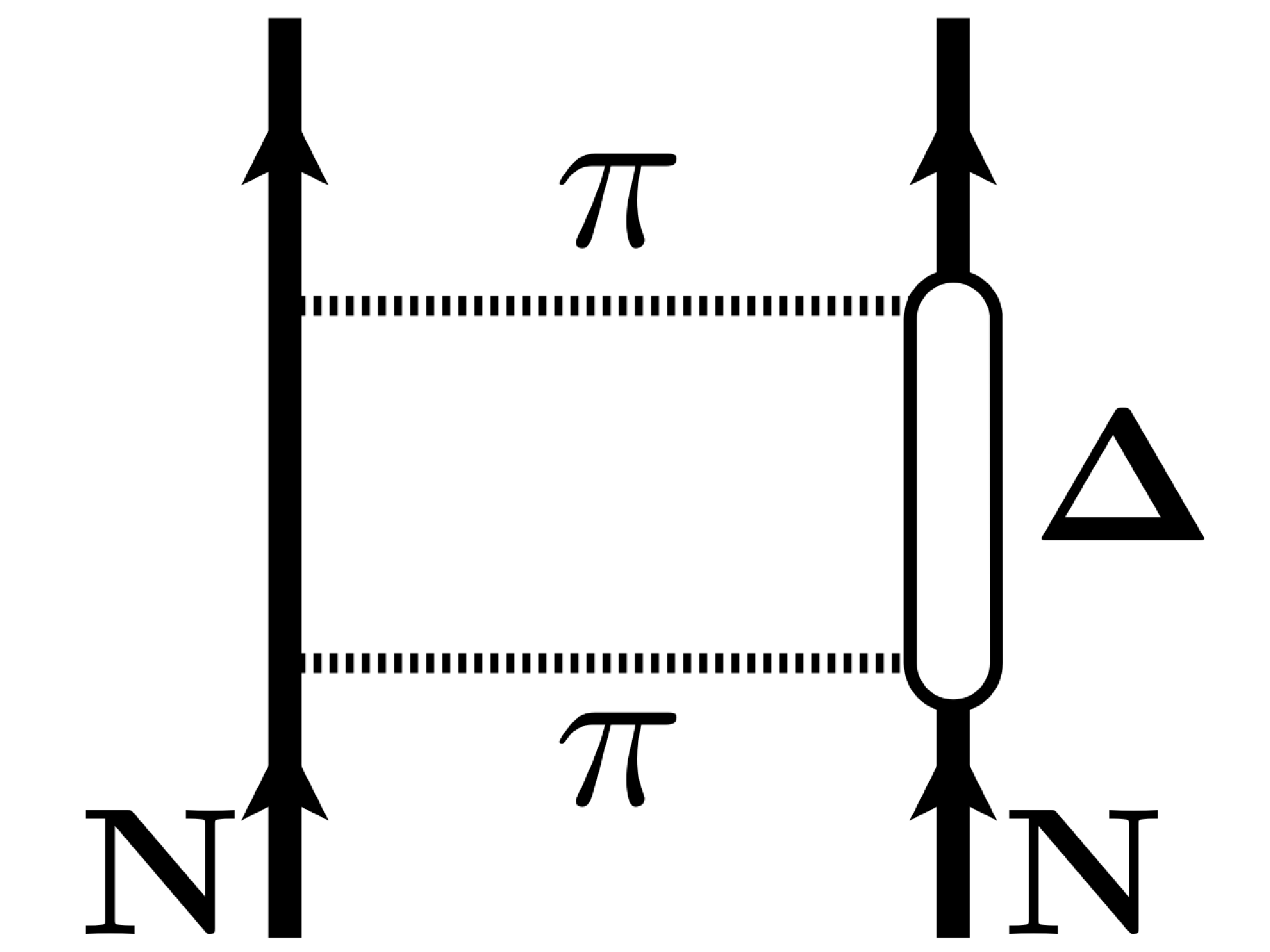}\hspace{.4in}
\includegraphics[height=2.3cm]{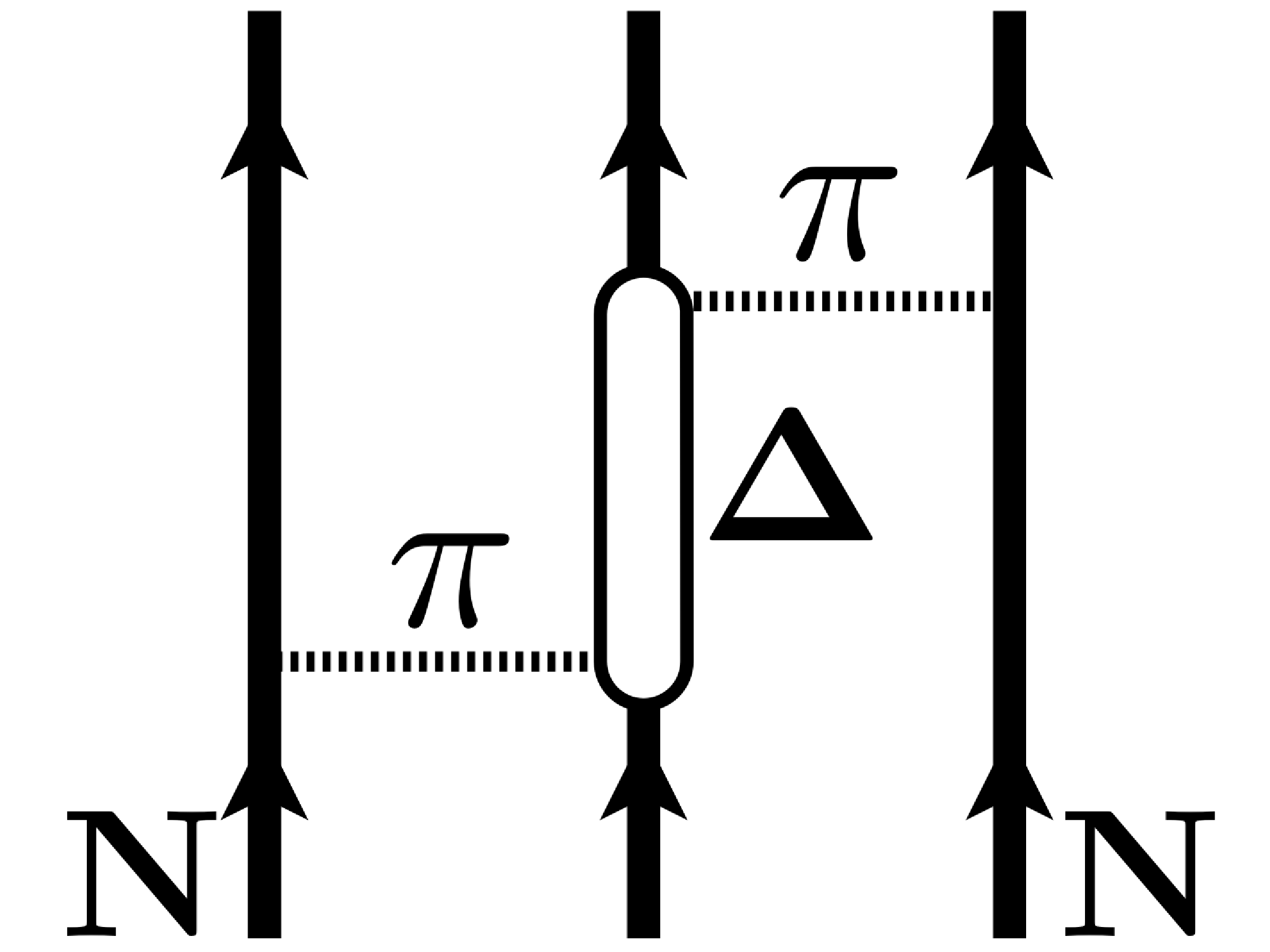}
\end{center}
\vspace{-.5cm}
\caption{Left panel: two-pion exchange involving a virtual $N\rightarrow\Delta\rightarrow N$ transition. Right panel:
three-nucleon interaction generated by the same mechanism.}
\label{2piDelta}
\end{figure}

When implemented in the nucleon-nucleon interaction, the $\Delta$ isobar plays an important role in  
two-pion exchange processes such as the one shown in Fig.\ \ref{2piDelta} (left). This mechanism 
contributes a large part of the attractive isoscalar central part of the NN interaction \cite{fkw2},
the one often parametrized in terms of an ad-hoc ``sigma" boson in phenomenological one-boson exchange 
potentials. A parameter-free calculation of this isoscalar central potential with single and double $\Delta$ excitation
\cite{gerstendorfer} agrees almost perfectly with phenomenological ``$\sigma$" exchange at distances r $>$ 2 fm, 
but not at shorter distances. The detailed behavior of the $2\pi$-exchange isoscalar central potential with 
virtual excitation of a single $\Delta$ is instead more reminiscent of a van der Waals potential: 
\begin{equation} 
V_C^{N\Delta}(r) = - {3g_A^2\,\alpha_A^{(\Delta)}\over (8\pi f_\pi)^2}\,{e^{-2m_\pi r}\over r^6} P(m_\pi r)~~,
\end{equation}
where $P(x)=6+12x+10x^2+4x^3+x^4$ is a fourth-order polynomial in $x = m_\pi r$. The familiar 
$r^{-6}$ dependence of the non-relativistic
van der Waals interaction emerges in the chiral limit, $m_\pi \rightarrow 0$.

When ``integrating out" the $\Delta$ degrees of freedom, the van der Waals -- like two-pion exchange mechanism just described -- gives rise to an important effective three-nucleon interaction, Fig.\\ref{2piDelta} (right), that was actually suggested already more than half a century ago by Fujita and Miyazawa \cite{fujita}. In a many-body framework with explicit $\Delta$ degrees of freedom, such a mechanism automatically occurs as an iterated one-pion exchange process involving the $N\rightarrow\Delta\rightarrow N$ transition. In this case the low-energy constants $c_3$ and $c_4$ in eq.\ (\ref{3n4}), the ones related to $p$-wave pion-nucleon scattering, are readjusted and considerably reduced in magnitude since they have to account only for the remaining non-resonant background, once the $\Delta(1232)$ is treated explicitly. Then important physics of the $\Delta$ such as the terms illustrated in Fig.\ \ref{2piDelta} (left) are actually promoted from N$^2$LO to NLO in the chiral hierarchy of NN interaction terms \cite{evgenireview}, leading to improved convergence. 

These considerations about the role of the $\Delta(1232)$ in the low-energy description of nuclear forces should 
already suffice to answer one of the principal questions at the starting point of the chiral effective field theory approach: 
how much information about the intrinsic structure of the nucleon is actually needed in order for this theory to work efficiently? The $\Delta(1232)$ is indeed by far the dominant feature of nucleon structure in the relevant low-energy 
range. Other, less prominent properties such as baryon resonances appearing at higher energies are conveniently absorbed in the set of low-energy constants.
 
\section{Nuclear Chiral Dynamics}
\subsection{\it In-medium Chiral Perturbation Theory: Nuclear Matter}

The tool to investigate the implications of spontaneous
and explicit chiral symmetry breaking in QCD is chiral perturbation theory \cite{GL84}.
Observables are calculated within the framework of an effective field theory
of interacting pions (Goldstone bosons) and nucleons \cite{BKM95}. The diagrammatic 
expansion in the number of loops has a one-to-one correspondence to an 
expansion in small external momenta and the pion (or light quark) mass. 
In nuclear matter, the relevant momentum scale is the Fermi momentum $k_f$, related 
to the nucleon density by $\rho=2k_f^3/3\pi^2$. At the empirical saturation density $\rho_0 \simeq 
0.16\,$fm$^{-3}$ the Fermi momentum and the pion mass are of comparable magnitude, 
$k_{f0} \simeq 2 m_\pi$. This immediately implies that pions must be included as explicit degrees 
of freedom: their propagation in nuclear matter is resolvable and thus relevant at the densities 
of interest.  With two small scales, $k_f$ and $m_\pi$, at hand the nuclear matter equation of 
state as obtained from in-medium chiral perturbation 
theory  will be given by an expansion in powers of the Fermi momentum.
The expansion coefficients are non-trivial functions of $k_f/m_\pi$, the dimensionless
ratio of the two relevant small scales inherent to the problem.

The only new feature in performing calculations in a nuclear many-body system (as compared to 
scattering processes in the vacuum) is the in-medium nucleon propagator. For a 
non-relativistic nucleon with four-momentum $p^\mu =(p_0, \vec p\,)$ it reads:
\begin{equation} G(p_0, \vec p\,) = { i \over p_0 -\vec p^{\,2}/2M_N + i \epsilon} - 
2\pi \delta(p_0 -\vec p^{\,2}/2M_N) \, \theta(k_f -|\vec p\,|)\,, \end{equation}
where the second term, the so-called ``medium insertion'', accounts for the presence of a filled 
Fermi sea of nucleons. The expression for $G(p_0, \vec p\,)$ is identical to the 
conventional decomposition of the in-medium propagator into particle and hole components used 
in non-relativistic many-body perturbation 
theory. With the decomposition into vacuum propagator and medium insertion, closed 
multi-loop diagrams representing the energy density can then be organized 
systematically in the number of medium insertions. The non-relativistic 
approximation is also compatible with the $k_f$-expansion at leading order. 
 
\begin{figure}
      \centerline{\includegraphics[width=10cm] {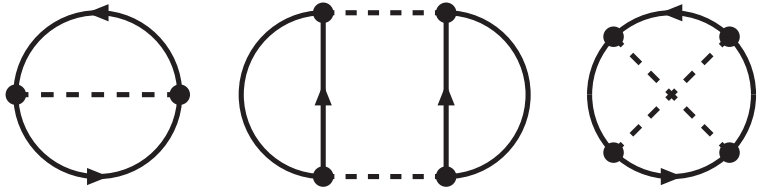}}
\caption{One-pion exchange Fock diagram and iterated one-pion exchange Hartree and Fock diagrams 
contributing to the energy density $\rho \bar E(k_f)$ of isospin-symmetric nuclear matter.}
\label{pionfigs}
 \end{figure}
We outline the leading contributions to the energy per particle $\bar E(k_f)$ of 
isospin-symmetric spin-saturated nuclear matter \cite{kfw1}. The relativistically improved 
kinetic energy is of course well-known:
\begin{equation}  \bar E(k_f)^{\rm (kin)} = {3k_f^2 \over 10M_N}- {3k_f^4 \over 56M_N^3}\,.   
\end{equation}          
Nuclear chiral dynamics up to three loop-order introduces the closed one- and two-ring 
diagrams shown in Fig.\ \ref{pionfigs}. The contribution of the one-pion exchange Fock 
term (represented by the left diagram) to the energy per particle reads:
\begin{equation} \bar E(k_f)^{(1\pi)} = {3g_A^2m_\pi^3 \over(4\pi f_\pi)^2} 
\bigg\{{u^3\over 3} +{1\over 8u} -{3u\over 4}+\arctan 2u -\Big( {3\over
8u}+{1\over 32u^3}\Big) \ln(1+4u^2)\bigg\}\,,\end{equation}
with the dimensionless variable $u= k_f/m_\pi$. Next in the $k_f$-expansion come the 
contributions from iterated (second-order) $1\pi$-exchange. The latter refers to the 
(two-particle) reducible part of the planar $2\pi$-exchange box diagram which gets 
enhanced by a small energy denominator proportional to differences of nucleon kinetic 
energies. The corresponding Hartree term (see middle diagram in Fig.\ \ref{pionfigs}) 
with two medium insertions reads:    
\begin{equation} \bar E(k_f)^{(H2)} = {3g_A^4M_N m_\pi^4 \over 5(8\pi)^3f_\pi^4}\bigg\{
{9\over 2u}-59 u+(60+32u^2) \arctan 2u -\Big( {9\over 8u^3} +{35\over 2u}
\Big) \ln(1+4u^2) \bigg\} \,. \end{equation}  
The analytical evaluation of this contribution is based on the iterated $1\pi$-exchange 
amplitude in the forward direction. For the Fock diagram (see right diagram in 
Fig.\ \ref{pionfigs}) with two medium insertions one encounters, in the same way, the 
iterated $1\pi$-exchange amplitude in the backward direction. The corresponding 
contribution to the energy per particle reads: 
\begin{equation} \bar E(k_f)^{(F2)} = {g_A^4 M_N m_\pi^4 \over (4\pi)^3f_\pi^4}\bigg\{
{u^3\over 2} + \int_0^u \!dx {3x (u-x)^2(2u+x) \over 2u^3(1+2x^2)} \Big[
(1+8x^2+8x^4) \arctan x-(1+4x^2)\arctan2x\Big] \bigg\}\,. \end{equation}
Note that the expressions for these two-body terms do not include the contribution of a linear 
divergence $\int_0^\infty dl\, 1$ of the momentum-space loop integral. Such a linear divergence 
is set to zero in dimensional regularization, but employing a cutoff regularization 
gives a contribution proportional to the momentum cutoff $\Lambda$. There is an interpretational problem with 
results of dimensional regularization. Contributions which are expected to be 
attractive according to second-order perturbation theory arguments show only their finite 
repulsive parts. We restore the attractive component via a term linear in the 
cutoff and the density
\begin{equation}  \bar E(k_f)^{(\Lambda)} = -{10 g_A^4 M_N \over (4\pi f_\pi)^4}\, \Lambda\, k_f^3 
\,, \end{equation}
to which the Hartree and Fock diagrams have contributed in the ratio $4:1$. 

The Pauli blocking corrections due to the nuclear medium are included through diagrams 
with three medium insertions. The corresponding Hartree term has the form:
\begin{equation} \bar E(k_f)^{(H3)} = {9g_A^4 M_N m_\pi^4 \over (4\pi f_\pi)^4 u^3}
\int_0^u \!dx\, x^2\int_{-1}^1 \!dy \bigg[2uxy+(u^2-x^2y^2)\ln{u+x y \over u-x y}\bigg]\bigg\{
{2s^2+s^4\over 1+s^2}-2\ln(1+s^2)\bigg\}\,, \end{equation}
with the abbreviation $s= xy+\sqrt{u^2-x^2+x^2y^2}$. Similarly, the
Fock-diagram in Fig.\ \ref{pionfigs} containing three medium insertions has the form:
\begin{equation} \bar E(k_f)^{(F3)} = {9g_A^4 M_N m_\pi^4 \over (4\pi f_\pi)^4 u^3}
\int_0^u \!dx \bigg\{ {G^2\over 8}+{x^2\over 4}\int_{-1}^1\!dy \int_{-1}^1
\!dz {yz \,\theta(y^2+z^2-1) \over |yz|\sqrt{y^2+z^2-1}}\Big[s^2-\ln(1+s^2)
\Big] \Big[ \ln(1+t^2)-t^2\Big]\bigg\}\,, \end{equation}
with the auxiliary function,
\begin{equation} G = u(1+u^2+x^2) -{1\over 4x}\big[1+(u+x)^2\big] \big[1+
(u-x)^2\big] \ln{1+(u+x)^2\over 1+(u-x)^2 } \,,  \end{equation} 
and $t= xz+\sqrt{u^2-x^2+x^2z^2}$. Let us emphasize that Pauli blocking has been 
treated here exactly, without employing any simplification due to angular averaging.

\begin{figure}
      \centerline{\includegraphics[width=10cm] {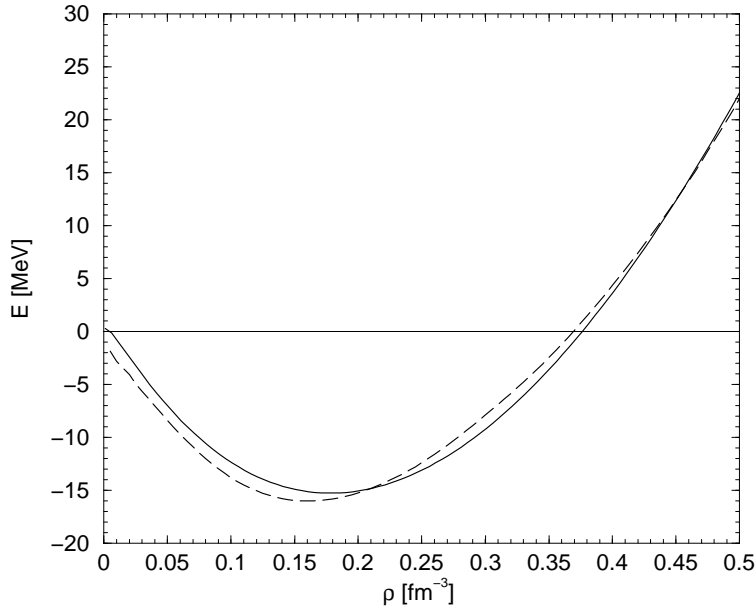}}
\caption{Saturation curve of nuclear matter obtained from $1\pi$-exchange and iterated 
$1\pi$-exchange together with a fine-tuned short-distance term linear in density. The dashed 
line stems from the many-body calculation of ref.\ \cite{friedpand}.}
\label{ebarpion}
\end{figure}

Summing the contributions to the energy per particle $\bar E(k_f)$ listed above and 
treating the cutoff scale $\Lambda$ as an adjustable parameter to incorporate 
(in a global fashion) unresolved short-distance dynamics, one obtains the equation of state 
of nuclear matter as shown in Fig.\ \ref{ebarpion}. With a fine-tuned cutoff scale of  
$\Lambda=611\,$MeV, the saturation minimum lies at $\rho_0 =0.173\,$fm$^{-3}$ and $\bar E(k_{f0}) 
= -15.3\,$MeV. The nuclear matter compressibility, related to the curvature of the saturation 
line at its minimum, comes out as $K = k_{f0}^2 \bar E''(k_{f0}) = 252\,$MeV, in good agreement 
with the presently accepted empirical value of $K = (250\pm 50)\,$MeV 
\cite{blaizot,youngblood,stoitsov}. Although this looks at first sight to be a very 
successful reproduction of nuclear matter bulk properties, 
one should not hide the fact that strong cancellations between individually large terms of 
opposite sign are involved here. For example, the adjusted term linear in density amounts to 
$-177.4\,$MeV at saturation density. The feature that strong cancellations are involved, which 
is also common to other approaches to nuclear matter, does not allow to make definite 
statements about the convergence of the $k_f$-expansion or to give reliable theoretical 
error bars. The appropriate framework to address such questions is a many-body calculation based 
on chiral low-momentum interactions whose long- and short-distance parts are determined together 
by empirical NN phase shifts and not separated any further.       

Let us now give an explanation for how saturation of nuclear matter is achieved in the 
framework of in-medium chiral perturbation theory. For that purpose it is instructive to 
consider the following simple parametrization of the energy per particle \cite{kfw1}:
\begin{equation}  \bar E(k_f) = {3k_f^2 \over 10 M_N} - \alpha {k_f^3 \over M_N^2} + \beta
{k_f^4 \over M_N^3} \,, \end{equation}      
which includes an attractive $k_f^3$-term and a repulsive $k_f^4$-term. This two-parameter 
form has generically a saturation minimum if $\alpha, \beta>0$. Its striking feature is than 
once $\alpha = 5.27$ and $\beta =12.22$ are adjusted to the empirical saturation point $\rho_0 =
0.16\,$fm$^{-3}$, $\bar E_0 = -16\,$\,MeV the compressibility $K \simeq 240$\,MeV comes out 
correctly. Moreover, such a parametrized curve for $\bar E(k_f)$ follows the results of 
sophisticated many-body calculations \cite{friedpand} up to quite high densities 
$\rho \simeq 1\,$fm$^{-3}$. 

In the chiral limit $m_\pi=0$ the leading interaction contributions calculated from $1\pi$- and 
iterated $1\pi$-exchange turn into exactly such a two-parameter form with the coefficient 
$\beta$ of the $k_f^4$-term given by \cite{kfw1}:  
\begin{equation}  \beta = {3 \over 70}\bigg({g_{\pi N}\over 4\pi}\bigg)^4(4\pi^2+237-24\ln 2) 
= 13.6\,, \end{equation}
where $g_{\pi N} =g_A M_N/f_\pi=13.2$ is the strong pion-nucleon coupling constant. This number 
is quite close to $\beta =12.22$ as extracted from a realistic nuclear matter equation of state. 
The mechanism for nuclear matter to saturate can be summarized roughly as follows: while 
pion-exchange at second order generates the necessary attraction, the Pauli-blocking effects 
due to the nuclear medium counteract this attraction in the form of a repulsive 
contribution with a stronger density dependence (a $k_f^4$-term). 

\begin{figure}
      \centerline{\includegraphics[width=10cm] {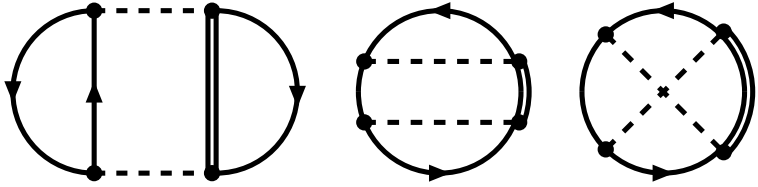}}
\caption{Hartree and Fock three-body diagrams related to $2\pi$-exchange with virtual 
$\Delta(1232)$-isobar excitation. They represent long-range interactions between three 
nucleons in the Fermi sea.}
\label{figdelt}
 \end{figure}

Calculations of nuclear matter in this framework have been  extended further by including 
the (irreducible) two-pion exchange contributions in the medium \cite{fkw2}. A compact form of 
the corresponding Fock term is given in terms of a (subtracted) spectral function representation:
\begin{eqnarray} \bar E(k_f)^{(2\pi F)}&=& {1 \over 8\pi^3} \int_{2m_\pi}^{\infty}
\!\! d\mu\,{\rm Im}(V_C+3W_C+2\mu^2V_T+6\mu^2W_T) \bigg\{ 3\mu k_f -{4k_f^3
\over 3\mu }\nonumber \\ &&+{8k_f^5 \over 5\mu^3 } -{\mu^3 \over 2k_f}-4\mu^2
\arctan{2k_f\over\mu} +{\mu^3 \over 8k_f^3}(12k_f^2+\mu^2) \ln\bigg( 1+{4k_f^2
\over \mu^2} \bigg) \bigg\} \,, \label{specrep} \end{eqnarray} 
where Im$V_C$, Im$W_C$, Im$V_T$ and Im$W_T$ are the imaginary parts (or spectral functions) of 
the isoscalar and isovector central and tensor NN-amplitudes, respectively. At this level also 
the $2\pi$-exchange with excitation of virtual $\Delta(1232)$-isobars comes into play. It is 
known to provide a dominant mechanism for intermediate range attraction in the isoscalar 
central channel \cite{gerstendorfer}. Furthermore, the Pauli-blocking effects to these 
$\Delta(1232)$-induced $2\pi$-exchange mechanisms, as represented by diagrams with three medium 
insertions, are equivalent to contributions from genuine long-range three-body forces. 
The two-ring Hartree diagram in Fig.\ \ref{figdelt} leads to the following contribution to the 
energy per particle \cite{fkw2}:
 \begin{equation} \bar E(k_f)^{(\Delta H3)}={g_A^4 m_\pi^6 \over \Delta(2\pi f_\pi)^4}
\bigg[ {2\over3}u^6 +u^2-3u^4+5u^3 \arctan2u-{1\over 4}(1+9u^2)
\ln(1+4u^2) \bigg] \,, \end{equation}  
with $\Delta = 293\,$MeV the delta-nucleon mass splitting.  On the other hand, the total 
contribution of the one-ring Fock diagrams in Fig.\ \ref{figdelt} reads: 
\begin{equation} \bar E(k_f)^{(\Delta F3)}=-{3g_A^4 m_\pi^6 u^{-3}\over 4\Delta(4\pi
f_\pi)^4 }\int_0^u \!\! dx \Big[ 2G^2_S(x,u)+G^2_T(x,u)\Big] \,, \end{equation}
where we have introduced the two auxiliary functions:
\begin{eqnarray} G_S(x,u) &=& {4ux \over 3}( 2u^2-3) +4x\Big[
\arctan(u+x)+\arctan(u-x)\Big] \nonumber \\ && + (x^2-u^2-1) \ln{1+(u+x)^2
\over  1+(u-x)^2} \,,\end{eqnarray}
\begin{eqnarray} G_T(x,u) &=& {ux\over 6}(8u^2+3x^2)-{u\over
2x} (1+u^2)^2  \nonumber \\ && + {1\over 8} \bigg[ {(1+u^2)^3 \over x^2} -x^4 
+(1-3u^2)(1+u^2-x^2)\bigg] \ln{1+(u+x)^2\over  1+(u-x)^2} \,.\end{eqnarray}
Evidently, the three-body Hartree term is repulsive while the corresponding Fock term is 
weakly attractive.  

The subtraction constants asscociated with the spectral function 
representation eq.\ (\ref{specrep}) give rise to short-distance contributions of the form
\begin{equation}\bar E(k_f)^{(sd)} = B_3 {k_f^3 \over M_N^2}+ B_5 {k_f^5 \over M_N^4}+  B_6 {k_f^6 
\over M_N^5}\,,\end{equation} 
where the last term stems from a three-body contact term. The parameters $B_3,B_5,B_6$ are 
assumed to represent all unresolved short-distance NN-dynamics relevant for isospin-symmetric 
nuclear matter at low and moderate densities. They are adjusted to few empirical bulk properties 
of nuclear matter,  and thereafter the approach is able to make predictions for more detailed 
quantities like the single-particle potential or the thermodynamic properties of nuclear matter 
at finite temperatures. Another benefit of this approach is that the pion-mass dependence of all 
interaction terms is explicitly known. By differentiating the nuclear matter equation of state
with respect to the light quark mass or pion mass one gets access to the density and temperature 
dependent chiral condensate $\langle \bar q q \rangle(\rho,T)$.            

\subsection{\it Resummation Strategies for Neutron Matter at Low Densities}
\label{rsnm}

Dilute degenerate many-fermion systems with large scattering lengths have attracted much
interest in recent years \cite{zwerger}. Experimental advances have allowed the
possibility to tune (magnetically) atomic interactions in ultracold fermionic gases through
so-called Feshbach resonances, enabling the exploration of weakly and strongly interacting 
many-body systems together with transitions from the superconducting BCS phase to the 
Bose-Einstein condensed state. Of special interest is the so-called unitary limit in which the 
two-body interaction is capable of supporting a single bound state at zero energy and the 
$S$-wave scattering length diverges, $a\to \infty$. Under these conditions  
the strongly-interacting many-fermion system is scale invariant. The  
ground-state energy per particle is then determined by a universal number, the 
so-called Bertsch parameter $\xi$, which is equal to the ratio of the energy per particle 
of the strongly interacting system at unitary, $\bar E(k_f)^{(\infty)}$, to that of a free 
Fermi gas, $\bar E(k_f)^{(0)}=3k_f^2/10M$.

Low-density ($\rho_n =k_f^3/3\pi^2 < 0.05\,$fm$^3$) neutron matter is expected to 
behave as a unitary Fermi gas due to the very large 
neutron-neutron $S$-wave scattering length, $a_{nn} \simeq 19\,$fm \cite{chen}.  
The results of a variety of
sophisticated many-body calculations \cite{gandolfi,gezerlis} indicate that at low densities the 
interactions behind the large neutron-neutron $S$-wave scattering length lead to a reduction 
of the free Fermi gas energy by approximately a factor $\xi_{nn} \simeq 1/2$.

Here, we outline the basic steps of a novel resummation technique \cite{resum} which allows to 
sum up the complete series of fermionic in-medium ladder diagrams generated by a contact interaction 
proportional to the scattering length $a$.  The solution of this (restricted) problem is enabled through
a different organization of the many-body calculation. Instead of treating 
particles and holes separately, one keeps them together and takes into account 
effects of the filled Fermi sea  by a ``medium-insertion''. This approach is realized 
by the following identical rewriting of the (non-relativistic) in-medium single-particle propagator:
\begin{eqnarray} G(p_0,\vec p\,) &=& i\bigg({\theta(|\vec p\,|-k_f)\over p_0-
\vec p^{\,2}/2M+i \epsilon }+{\theta(k_f-|\vec p\,|)\over p_0-\vec p^{\,2}/2M-i 
\epsilon}\bigg) \nonumber \\ &=& {i \over p_0-\vec p^{\,2}/2M+i \epsilon } 
-2\pi\, \delta(p_0-\vec p^{\,2}/2M)\,\theta(k_f-|\vec p\,|)\,, \end{eqnarray}
for an internal fermion line with energy $p_0$ and momentum $\vec p$. In the above equation, $M$
denotes the large fermion mass.

\begin{figure}
      \centerline{\includegraphics[width=8.5cm] {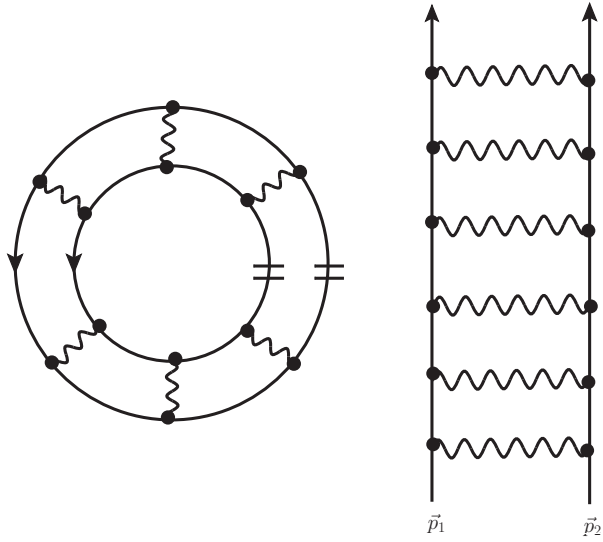}}
\caption{Left: Closed multi-loop diagram representing a contribution to the energy density.
Right: Planar ladder diagram obtained by opening at a minimal pair of adjacent medium-insertions.
Wavy lines symbolize the contact interaction proportional to the scattering length $a$.}
\label{medium1}
\end{figure}

Many-body contributions to the energy per particle come from diagrams with 
at least two medium-insertions. For the closed ladder diagram, shown to the left in
Fig.\ \ref{medium1}, this minimal pair of medium-insertions must be placed on opposing 
positions of the double-ring. By opening them one obtains
a planar ladder diagram containing energy denominators that are all given as differences 
of fermion kinetic energies. It is only for this planar topology that the factors 
of $M$ from the energy denominators balance the $1/M$ factors coming from the 
interaction vertices in such a way that in the non-relativistic limit a finite result remains 
(to any order $a^n$). The open ladder diagram, shown to the right in Fig.\ \ref{medium1}, 
comes in with additional medium-insertions on internal lines in all possible ways. Due 
to the special nature of the momentum-independent contact interaction, all multi-loop 
diagrams factorize (successive loops are independent), which allows their sum to be written
in the form of a power of the in-medium loop.

\begin{figure}
      \centerline{\includegraphics[width=10cm] {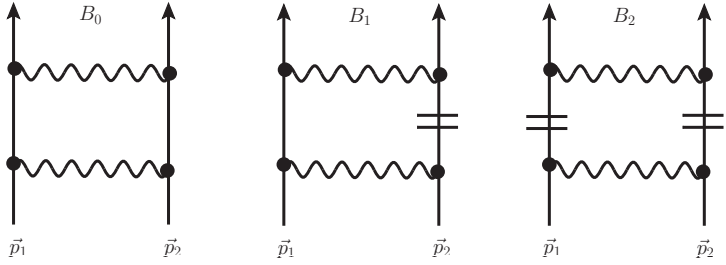}}
\caption{In-medium loop with different numbers of medium insertions. The middle diagram 
has a reflected partner. The external momenta $|\vec p_{1,2}|<k_f$ are inside the 
Fermi sphere.}
\label{mediumloop}
\end{figure}

The in-medium loop is decomposed into contributions from zero, one and two medium insertions: 
$B_0+B_1+B_2$ (see Fig.\ \ref{mediumloop}). The first term $B_0$ is the well-known rescattering 
in vacuum:  
\begin{equation}B_0 = 4\pi a \int\! {d^3 l \over (2\pi)^3} \,{1 \over  
\vec l^{\,2}-\vec q^{\,2}-i \epsilon }= {2a \over \pi } \int\limits_0^\infty\! dl 
\,\bigg( 1 + {\vec q^{\,2} \over  l^2 -\vec q^{\,2}-i \epsilon } \bigg)  = 0 + 
i\, a |\vec q \,|\,,   \end{equation}
where the rule $\int_0^\infty dl\, 1 = 0$ of dimensional regularization has been exploited. In 
a regularization scheme employing a momentum-space cutoff, the emerging scale-dependent 
constant $-2\Lambda/\pi$ can be absorbed into $a^{-1}$, from which one defines the renormalized 
(or physical) scattering length. The expression 
for the contribution $B_1$ to the in-medium loop with one medium insertion reads:  
\begin{equation} B_1 =   -4\pi a \int {d^3 l \over (2\pi)^3} {1 \over  
\vec l^{\,2}-\vec q^{\,2}-i \epsilon } \Big\{ \theta(k_f-|\vec P-\vec l\,|)+ 
\theta(k_f-|\vec P+\vec l\,|)\Big\} \,.\end{equation} 
Solving this integral, one obtains the following result for the real part:
\begin{equation} {\rm Re}\,B_1 = -{a k_f \over \pi} \, R(s,\kappa) \,,
\end{equation}   
where
\begin{equation} R(s,\kappa) = 2 +{1\over 2s}[1-(s+\kappa)^2]\ln{1+s +\kappa
\over |1-s -\kappa|}+{1\over 2s}[1-(s-\kappa)^2]\ln{1+s -\kappa
\over 1-s +\kappa}\,, \end{equation}
is a logarithmic function written in terms of two dimensionless variables $s = |\vec p_1+ \vec p_2|
/2k_f$ and $\kappa= |\vec p_1 -\vec p_2|/2k_f$. An additional constraint, $s^2+\kappa^2<1$, arises
from the fact that both external momenta $\vec p_1$ and $\vec p_2$ lie inside the Fermi sphere.

Finally, there is the contribution $B_2$ with two medium insertions which in fact 
is purely imaginary. Summing up all contributions gives for the total imaginary part 
of the in-medium loop:
\begin{eqnarray} {\rm Im}(B_0+B_1+B_2) &=& 4 \pi a  \int {d^3 l \over (2\pi)^3} \,
\pi\, \delta(\vec l^{\,2}-\vec q^{\,2}) \nonumber \\ && \times \bigg\{\Big[1-
\theta(k_f-|\vec P-\vec l\,|)\Big]\Big[ 1-\theta(k_f-|\vec P+\vec l\,|)\Big]
\nonumber \\ && \quad +\theta(k_f-|\vec P-\vec l\,|)\, \theta(k_f-|\vec P+
\vec l\,|) \bigg\} \,,\end{eqnarray}
where terms with zero, one, and two step-functions have suitably arranged. The 
first term has the form $[1-\theta(...)][1-\theta(...)]$ and gives no contribution to the 
imaginary part since energy conservation prohibits the on-shell scattering of two 
particles within the Fermi sea into the region outside the Fermi sea. 
Consequently, the result for the imaginary part of the in-medium loop takes the form:
\begin{equation} {\rm Im}(B_0+B_1+B_2) = {B_2 \over 2i} = a k_f \, 
I(s,\kappa)\,,  \end{equation}
where $I(s,\kappa)$ is the non-smooth function
\begin{equation}I(s,\kappa)=   \kappa\, \theta(1-s-\kappa) + {1 \over 2s}(1-s^2-\kappa^2)\, 
 \theta(\kappa+s-1)\, .
 \end{equation}
In this expression $\kappa$ is restricted to lie in the interval $0<\kappa<\sqrt{1-s^2}$. 
Putting together the real and imaginary 
parts, the complex-valued in-medium loop is written
\begin{equation} B_0+B_1+B_2=  -{a k_f \over \pi} \,\Big\{ R(s,\kappa)- i \pi \,
I(s,\kappa)\Big\}  \,. \label{imloop}\end{equation}  
We observe that if the contribution from the diagram with two medium-insertions is separated off, 
then the imaginary part of the expression in eq.\ (\ref{imloop}) changes sign:  
\begin{equation} B_0+B_1=  -{a k_f \over \pi} \,\Big\{ R(s,\kappa)+ i \pi \,
I(s,\kappa)\Big\}  \,.\end{equation}
Indeed, this particular property of the in-medium loop will play a critical role in the derivation of
the correct expression for the energy per particle.

\begin{figure}
      \centerline{\includegraphics[width=10cm] {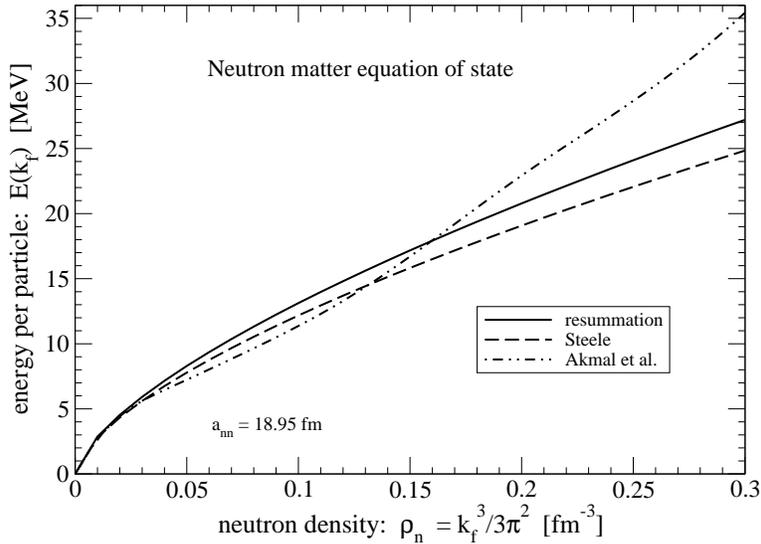}}
\caption{The energy per particle of neutron matter as a function of the neutron density $\rho_n$. 
The dash-dotted line is a reproduction of the results from the variational many-body calculation 
of ref.\ \cite{akmal}.}
\label{neutronmatter}
\end{figure}

In ref.\ \cite{resum} it was shown how the contributions to the energy per particle 
$\bar E(k_f)$ at a given order in $a^n$ can be constructed from the in-medium loop and how the 
resulting series in $a k_f$ can even be summed to all orders. A naive iteration method would 
suggest $(R-i \pi I)^{n-1}$ as 
the integrand for the energy density at order $a^n$, but it is complex-valued. The deficit of 
the naive iteration method lies in the fact that a closed diagram with a repeated double 
medium-insertion has a symmetry factor which is not included in the binomial series expansion 
of $[(R+i \pi I)+(-2i \pi I)]^{n-1}$. A detailed combinatorial analysis \cite{resum} shows that 
the $j$-th power of $-2i \pi I$ coming from the diagrams with repeated double medium-insertions 
has to be re-weighted by a factor $1/(j+1)$. This crucial observation leads to the following 
formula 
\begin{equation} \sum_{j=0}^{n-1}  (R+i \pi I)^{n-1-j}(-2i \pi I)^j {n-1 \choose j}
{1\over j+1} = {1\over 2i \pi I n}\Big\{ (R+i \pi I)^n- (R-i \pi I)^n\Big\} 
\,, \end{equation}   
where the right hand side is now manifestly real for all $n$. In this form the whole subset of 
ladder diagrams can be summed to all orders, since the series $\sum_{n=1}^\infty [-a k_f(R\pm i 
\pi I)/\pi ]^n/n$ is solved easily in terms of a (complex) logarithm. Putting all pieces 
together the final expression for the resummed energy per particle reads \cite{resum}:
\begin{equation} \bar E(k_f)^{(\rm lad)}= -{24k_f^2 \over \pi M} \int\limits_0^1 \!ds\, s^2  
\!\!\int\limits_0^{\sqrt{1-s^2}}  \!\!d\kappa \, 
\kappa \, \arctan { a k_f\, I(s,\kappa) \over 1+ \pi^{-1}ak_f \,R(s,\kappa)} \,, 
\label{ladseries}\end{equation}
where the arctangent function refers to the usual branch with odd parity taking on values in the 
interval $[-\pi/2,\pi/2]$.

The expansion of the resummed energy per particle  $\bar E(k_f)^{(\rm lad)}$ in powers of $a k_f$ 
has been checked against known results from many-body perturbation theory based on the 
traditional particle-hole counting scheme \cite{furnstahl,steele}. Up to the order $a^4$ where 
such results are presently available the agreement is perfect. Moreover, since only double 
integrals are involved, the pertinent coefficients can be calculated with very high 
numerical precision \cite{resum}.  

As an application of eq.\ (\ref{ladseries}) for the exact resummation of in-medium ladder 
diagrams, one can consider the equation of state of pure neutron matter. The very large 
neutron-neutron scattering length $a_{nn} = (18.95 \pm  0.40)\,$fm \cite{chen} necessitates a
non-perturbative treatment of neutron matter at low densities. In Fig.\ \ref{neutronmatter} we 
reproduce the energy per particle of neutron matter as a function of the density in the approximation 
$\bar E(k_f)^{(\rm lad)}+ 3k_f^2/10M$ \cite{resum}. The solid line is obtained from eq.\ (\ref{ladseries}) 
by substituting $a=a_{nn}=18.95$\,fm for the 
scattering length and $M=M_n = 939.57\,$MeV for the large fermion mass. The 
dash-dotted line comes from a sophisticated variational many-body calculation \cite{akmal} 
representative of realistic neutron matter calculations. 
In Fig.\ \ref{neutronmatter} the dashed line corresponds to Steele's 
suggestion \cite{steele} of a simple geometrical series, $\bar E(k_f)^{(St)} = 
-a k_f^3[3M(\pi+2a k_f)]^{-1}$. Up to rather high neutron densities, $\rho_n \simeq 0.2\,$fm$^{-3}$, 
where the dimensionless parameter $a_{nn} k_f \simeq 34$, there is rather good agreement 
among the different calculations. For neutron densities beyond that of saturated nuclear matter, 
repulsive effects from three-body forces (present in the variational calculation of ref.\ \cite{akmal}) 
begin to play a more important role. Including the neutron-neutron $S$-wave effective range $r_{nn} = (2.75 
\pm  0.11)\,$fm can also modify the equation of state, as shown in ref.\ \cite{achimpethick}. 


\subsection{\it In-Medium Effective Nucleon-Nucleon Interaction}

\begin{figure}
\begin{center}
\includegraphics[scale=0.65,clip]{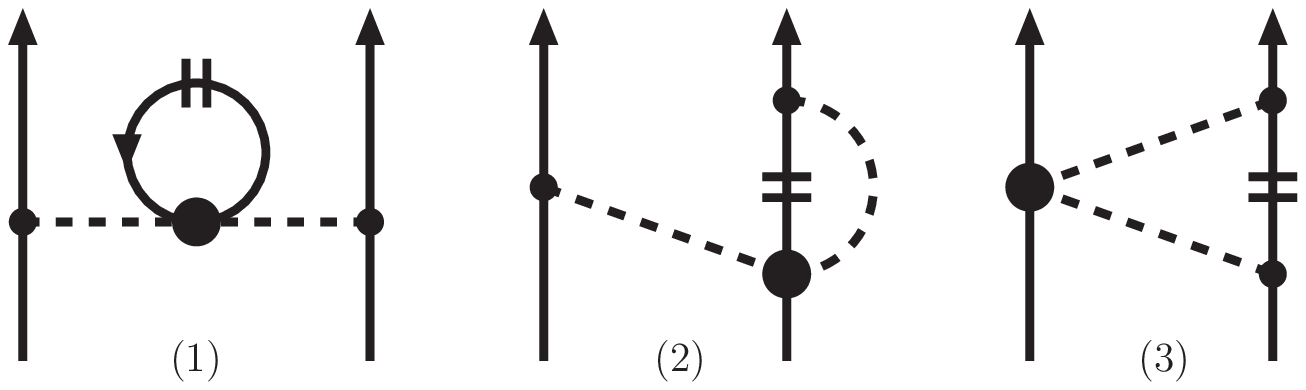}\hspace{.4in}
\includegraphics[scale=0.65,clip]{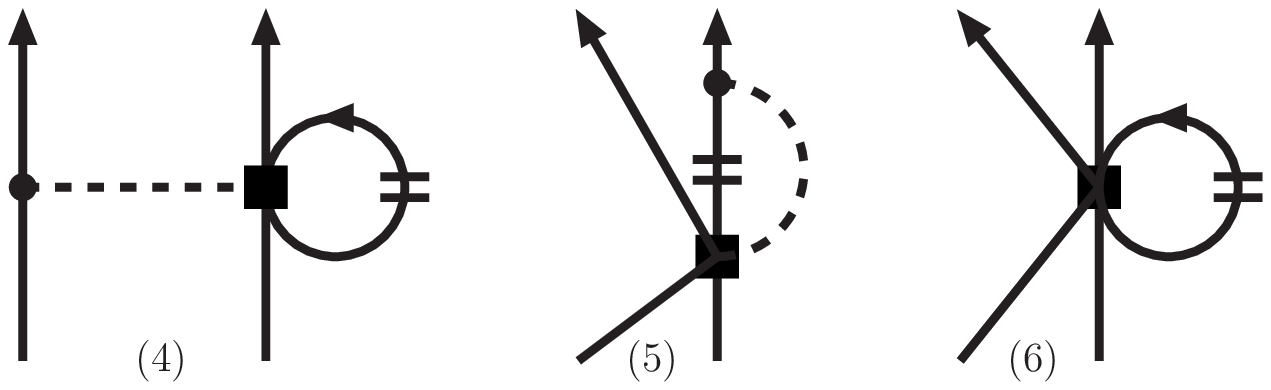}
\end{center}
\vspace{-.5cm}
\caption{Diagrammatic contributions to the quasiparticle interaction in 
neutron matter generated from the two-pion exchange, one-pion exchange,
and contact three-nucleon forces. The short double-line denotes the summation 
over the filled Fermi sea of nucleons. Reflected diagrams of (2), (3), (4), and (5) are not shown.}
\label{mfig1}
\end{figure}

In applications of chiral effective field theory to light nuclei, the N$^2$LO chiral 
three-nucleon force is routinely implemented exactly \cite{evgenireview,navratil07}.
For heavier systems, however, the inclusion of three-nucleon forces is 
computationally demanding and requires resources approximately an order 
of magnitude larger than that for two-nucleon forces alone. An alternative 
strategy has therefore been explored in a number of recent calculations 
\cite{hagen07,holt09,holt10,hebeler10,otsuka10,menendez11} of finite nuclei 
as well as infinite nuclear and neutron matter in which three-nucleon forces are 
replaced by medium-dependent effective two-body interactions. 
The three-body force in second-quantized notation
\be
V_{3N} = \frac{1}{36} \sum_{123456} \langle 1 2 3 | \bar V | 4 5 6 \rangle
\hat a^\dagger_1 \hat a^\dagger_2 \hat a^\dagger_3 \hat a_6 \hat a_5 \hat a_4
\ee
with antisymmetrized matrix elements $\langle 1 2 3 | \bar V | 4 5 6 \rangle$, is normal
ordered with respect to a convenient reference state $|\Omega \rangle$ which yields
\bea
V_{3N} &=& \frac{1}{6} \sum_{ijk}\langle ijk | \bar V_{3N} | ijk \rangle 
+ \frac{1}{2} \sum_{ij1 4} \langle ij1 | \bar V_{3N} | ij4 \rangle :\!\hat a^\dagger_1 \hat a_4\!:
+ \frac{1}{4}\sum_{i1245}\langle i12 | \bar V_{3N} | i 4 5 \rangle :\!\hat a^\dagger_1 \hat a^\dagger_2 
\hat a_5 \hat a_4\!:  \nonumber \\
&& + {1\over 36} \sum_{123456} \langle 1 2 3 | \bar V_{3N} | 4 5 6 \rangle
:\!\hat a^\dagger_1 \hat a^\dagger_2 \hat a^\dagger_3 \hat a_6 \hat a_5 \hat a_4\!:,
\label{nord}
\eea
where the (alphabetic) indices $i,j,k$ represent filled orbitals in the reference state, and $:\,\,:$ denotes 
the normal-ordered product of operators satisfying
\be
:\! \hat a^\dagger_1\dots \hat a_n \!: |\Omega \rangle= 0.
\ee

\begin{figure}
\begin{center}
\includegraphics[angle=270,scale=0.325,clip]{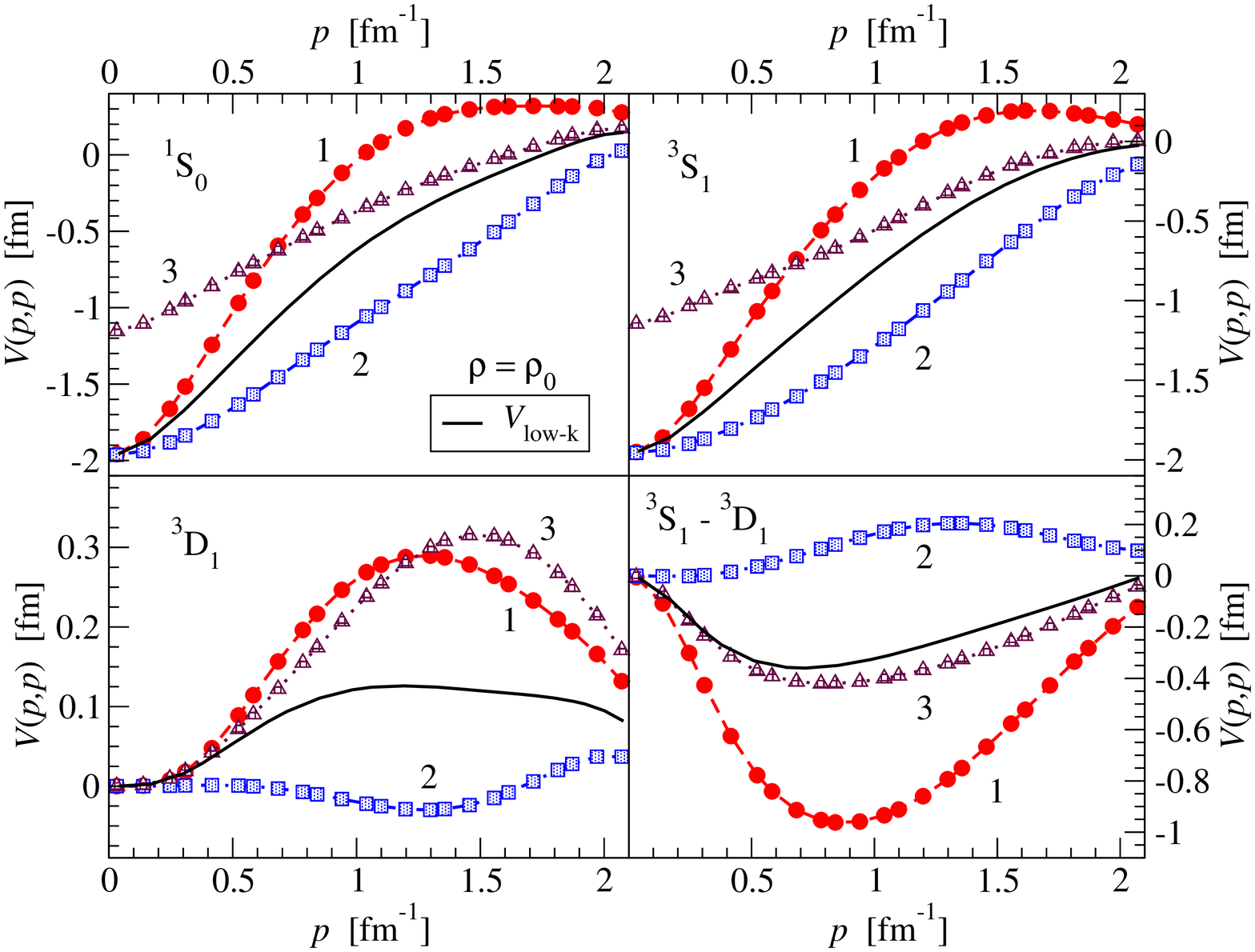}
\includegraphics[angle=270,scale=0.325,clip]{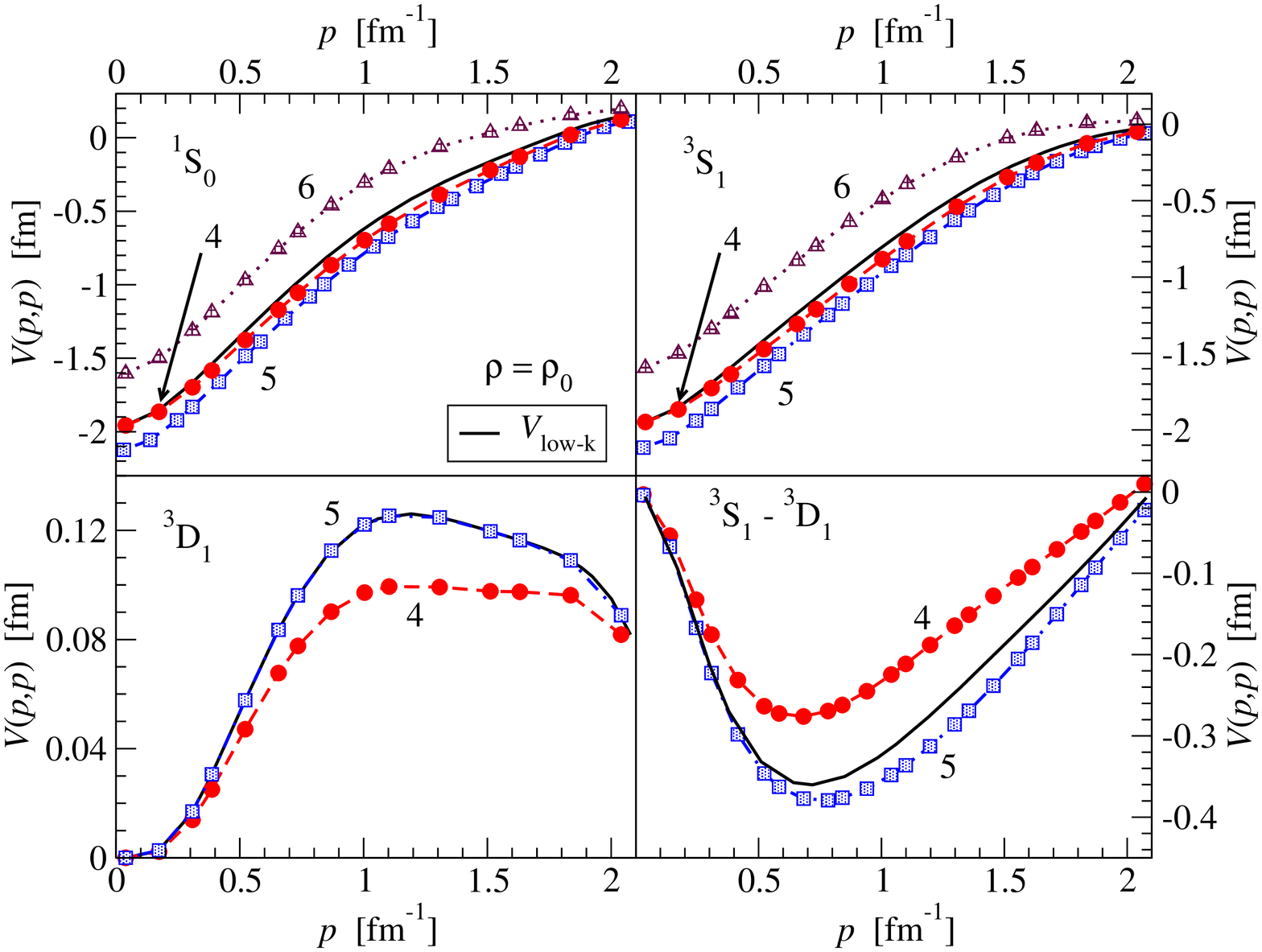}
\end{center}
\vspace{-.5cm}
\caption{Contributions to the on-shell $S$- and $D$-wave matrix elements from the
density-dependent nucleon-nucleon interaction derived in ref.\ \cite{holt09,holt10}. The 
solid curve denotes the matrix elements of a renormalization-group evolved NN potential
and the numbered lines show the modification due to the six classes of diagrams shown
in Fig.\ \ref{mfig1}.}
\label{sdwaves}
\end{figure}

In refs.\ \cite{holt09,holt10} the reference state was chosen to be a filled Fermi sea
of noninteracting nucleons characterized by the density as well as the isospin
asymmetry parameter $\delta_{np} = (\rho_n-\rho_p)/(\rho_n+\rho_p)$. With the
additional kinematical restriction that the scattering takes place in the rest
frame of the medium, the on-shell scattering amplitude (equivalent to the 
normal-ordered two-body part of the full three-nucleon force) has the same 
form as the free-space nucleon-nucleon interaction:
\bea
V_{NN}^{\rm med} &=& V_C + W_C\, \vec \tau_1 \cdot \vec \tau_2 
+ \left [ V_S + W_S \,\vec \tau_1 \cdot \vec \tau_2  \right ] \vec \sigma_1 \cdot \vec \sigma_2 
+\left [ V_T + W_T\, \vec \tau_1 \cdot \vec \tau_2 \right ] 
  \vec \sigma_1 \cdot \vec q \, \vec \sigma_2 \cdot \vec q  \nonumber \\
&+&\left [ V_{LS}+ W_{LS}\, \vec \tau_1 \cdot \vec \tau_2  \right ] 
  i (\vec \sigma_1 + \vec \sigma_2)\cdot ( \vec q \times \vec p\,)
+\left [ V_Q +W_Q\, \vec \tau_1 \cdot \vec \tau_2 \right ] 
  \vec \sigma_1 \cdot (\vec q \times \vec p\,) \vec \sigma_2 \cdot (\vec q \times \vec p\,),
\eea
where the different scalar functions $V$ and $W$ depend on $p$ and $q$, the initial relative 
momentum and the momentum transfer, respectively. The resulting in-medium two-nucleon 
interaction can then be implemented in the many-body method of choice, provided that one 
takes care of the different symmetry factors in the zero- and one-body normal-ordered components 
of a three-nucleon force compared to those from two-body interactions \cite{hebeler10}.

For the N$^2$LO chiral three-nucleon force, there are six topologically distinct
diagrams contributing to the in-medium nucleon-nucleon interaction shown in Fig.\ 
\ref{mfig1}. The contributions labeled (1), (2), and (4) give corrections to one-pion exchange, while
contributions (5) and (6) adjust the strength of the nucleon-nucleon contact interaction. On the other
hand, the Pauli-blocked two-pion exchange diagram, labeled (3) in Fig.\ \ref{mfig1}, gives rise
to central, spin-spin, tensor, spin-orbit, and quadratic spin-orbit contributions. 
In Fig.\ \ref{sdwaves} we reproduce the on-shell momentum-space matrix 
elements of the renormalization-group evolved Idaho N$^3$LO chiral two-nucleon force
together with the modifications arising from the six different components of the density-dependent
nucleon-nucleon interaction in selected relative $S$ and $D$ waves. The values of the 
relevant low-energy constants are $c_1 = -0.76$\,GeV$^{-1}$, $c_3 = -4.78$\,GeV$^{-1}$,
$c_4 = -3.96$\,GeV$^{-1}$, $c_D = -2.06$, and $c_E = -0.625$. The dominant effects
arise from two-pion exchange dynamics, but in all partial waves the pion
self-energy correction $V_{NN}^{\rm med,1}$ and the pion-exchange vertex correction
$V_{NN}^{\rm med,2}$ approximately cancel. This leaves Pauli-blocking effects in 
two-pion exchange, represented by $V_{NN}^{\rm med,3}$, together with additional
repulsion from the chiral three-body contact interaction, encoded in $V_{NN}^{\rm med,6}$,
as the dominant effects from three-body forces. 

\begin{figure}
\begin{center}
\includegraphics[angle=270,scale=0.35,clip]{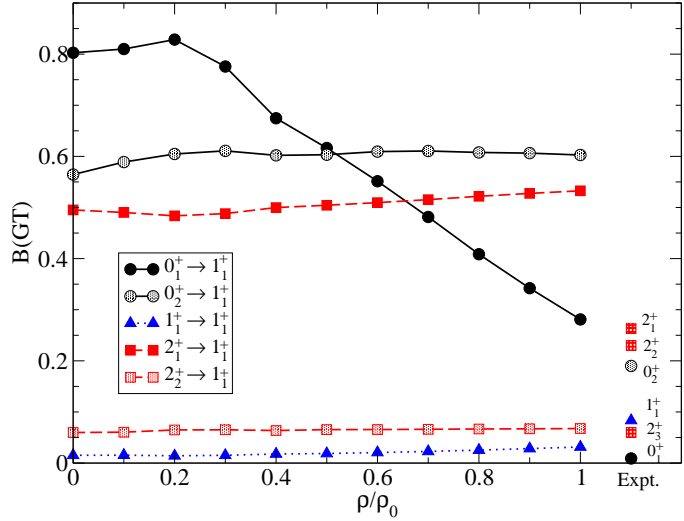}
\end{center}
\vspace{-.5cm}
\caption{Gamow-Teller strengths for transitions from the low-energy even-parity states
of $^{14}$C to the ground state of $^{14}$N as a function of the nuclear density.}
\label{bgtc14}
\end{figure}

\begin{figure}[hbt]
\begin{center}
\includegraphics[scale=0.6,clip]{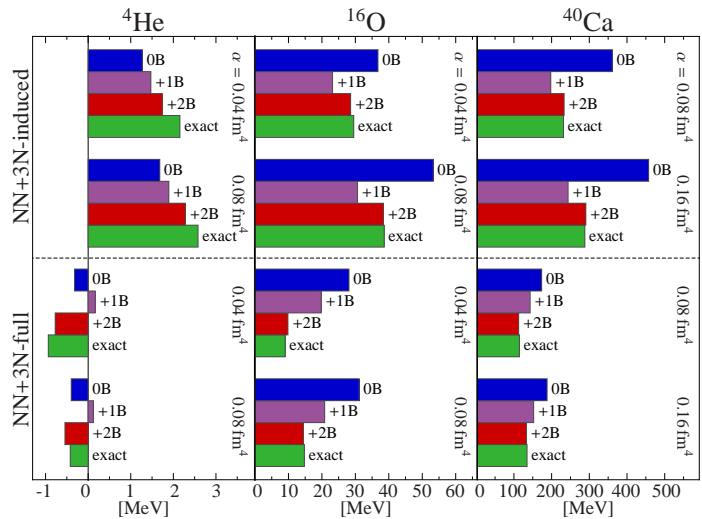}
\end{center}
\vspace{-.5cm}
\caption{Ground-state expectation values of the N$^2$LO chiral three-nucleon force at different 
normal-ordering approximations. The bars are labeled according to the highest $n$-body ($n$B)
contribution included in the calculation. Similarity renormalization group evolved interactions were 
employed for two different values of the flow parameter $\alpha$. Figure reproduced from ref.\ \cite{roth12}.}
\label{roth}
\end{figure}

The normal-ordering approximation, in which the residual three-nucleon force in eq.\ (\ref{nord})
is neglected, has proven to be a useful approximation for calculations in medium-mass
nuclei where exact {\it ab-initio} many-body methods incorporating three-body forces are
computationally challenging. Within this approximation three-nucleon forces were shown to provide the 
microscopic origin of the anomalously-long beta-decay lifetime of $^{14}$C \cite{holt09}, which
subsequent no-core shell model calculations \cite{maris11} have confirmed and clarified. 
In Fig.\ \ref{bgtc14}, we show the Gamow-Teller strengths from low-lying states in $^{14}$C
to the ground state of $^{14}$N computed with density-dependent low-momentum chiral 
nucleon-nucleon interactions. In fact, only the ground state to ground state transition receives 
significant medium modifications that result in a strong enhancement of the $^{14}$C lifetime.
Further work has employed normal-ordered two-body interactions to study three-nucleon force
effects on the neutron drip-line in oxygen isotopes \cite{otsuka10,jdholt13} as well as on the shell structure 
of calcium isotopes \cite{jdholt12,gallant12}. Benchmark calculations of the normal ordering 
approximation in closed shell nuclei have been carried out in the importance-truncated no-core 
shell model, and it was found that beyond the lightest nuclei, normal-ordered Hamiltonians provide 
an accurate substitute for full three-nucleon forces \cite{roth12}. In Fig.\ \ref{roth} we reproduce from 
ref.\ \cite{roth12} the expectation values of the leading-order chiral three-nucleon force at different
levels of the normal-ordering approximation. For both $^{16}$O and $^{40}$Ca, the difference between
the expectation of the normal-ordered 2B approximation, consisting of the first three terms in 
eq.\ (\ref{nord}), and the expectation value of the exact three-body interaction is negligible.

Density-dependent nucleon-nucleon interactions have also been used extensively in calculations of the 
equation of state of isospin-symmetric nuclear matter and neutron matter \cite{hebeler10,hebeler11,tews12,coraggio13}. This has allowed the study of three-nucleon 
forces beyond the Hartree-Fock approximation, but certain topologies are necessarily omitted in such
an approximation. The inclusion of the full second-order contribution 
to the energy per particle from a three-nucleon force has been carried out only in the case of a 
three-body contact interaction \cite{kaiser12}. There it was found that the complete set of 
second-order three-body diagrams gives a contribution to the energy per particle that is roughly 
half that from the density-dependent nucleon-nucleon interaction.

\subsection{\it Nuclear Mean Field}

The average single-particle potential acting on a nucleon in a finite nucleus is a central 
concept in nuclear structure and reaction theory. For negative energy bound states,
the nuclear mean field is associated with the shell model potential, while for positive
energy scattering states it is identified with the optical model potential. The latter is
complex and strongly absorptive, in contrast to the shell model potential which is a
real-valued quantity. While phenomenological optical potentials have been used 
extensively to describe reactions on target nuclei close to the valley of stability,
microscopic optical potentials have no adjustable parameters and therefore provide 
a reliable basis for extending to reactions on exotic, neutron-rich nuclei that will be
studied at the next generation of radioactive beam facilities. Numerous microscopic 
many-body methods have been used to
compute the single-particle potential (self-energy), including the Brueckner-Hartree-Fock
\cite{jeukenne76,grange87,haider88}, Dirac-Brueckner-Hartree-Fock \cite{arnold81,haar87}, 
and Green's function \cite{dickhoff04,waldecker11} methods.
Here we review only the recent calculations \cite{kaiser02,fkw2,holt13b} that have been 
carried out in the framework of in-medium chiral perturbation theory for infinite 
isospin-symmetric nuclear matter.

\begin{figure}[t]
\begin{center}
\includegraphics[height=16cm,angle=270]{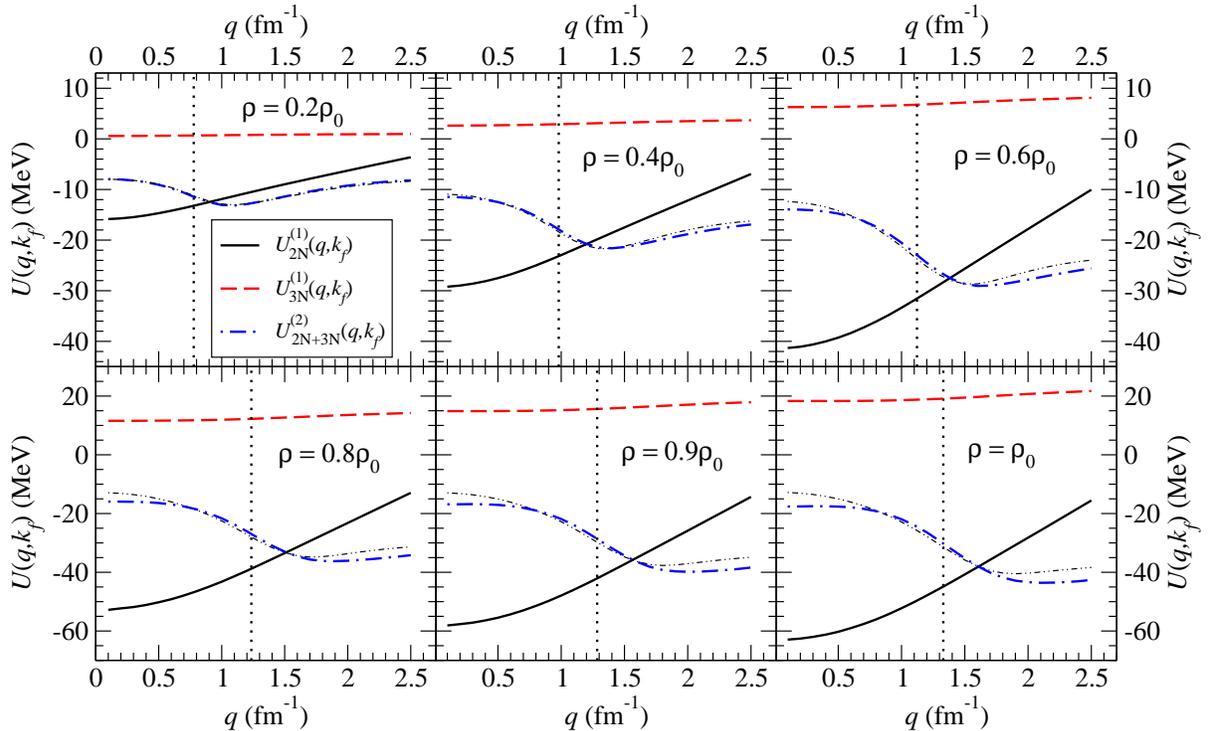}
\end{center}
\vspace{-.5cm}
\caption{Contributions to the real part of the on-shell ($\omega = q^2/(2M_N)$) 
momentum- and density-dependent optical potential from
chiral two- and three-nucleon forces. The solid and dashed-dotted lines are the first- and 
second-order contributions from the two-body potential, while the dashed line is the 
first-order contribution from the chiral three-nucleon force. The vertical dotted line indicates the 
Fermi momentum, and the dashed-double-dotted line is the second-order contribution without 
three-body forces.}
\label{ddopr}
\end{figure}

The starting point is the high-precision Idaho chiral two-nucleon potential with a cutoff
of $\Lambda = 500$\,MeV supplemented with the leading-order chiral three-nucleon force
whose low-energy constants ($c_1=-0.81$\,GeV$^{-1}$, $c_3=-3.2$\,GeV$^{-1}$,
$c_4=5.4$\,GeV$^{-1}$, $c_D=-0.2$, and $c_E=-0.205$) have been fit to nucleon-nucleon
scattering phase shifts \cite{entem03} and the binding energy and lifetime of the triton \cite{gazit09}.
Within microscopic many-body theory, the nuclear mean field is identified with the
nucleon self-energy, $\Sigma(\vec r, \vec r^{\, \prime}, E)= U(\vec r, \vec r^{\, \prime}, E) + i 
W(\vec r, \vec r^{\, \prime}, E)$. For a homogeneous medium
the self-energy can be written as a function of the momentum, energy, and density (or
Fermi momentum) $\Sigma(q,\omega;k_f)$. 
The first two perturbative contributions from two-nucleon forces are given by
\begin{eqnarray}
\Sigma^{(1)}_{2N}(q,\omega;k_f) &=& \sum_{1} \langle \vec q \, \vec h_1 s s_1 t t_1 | \bar V_{2N} | \vec q \,
\vec h_1 s s_1 t t_1 \rangle n_1, \nonumber \\
\Sigma^{(2)}_{2N}(q,\omega;k_f) &=& \frac{1}{2}\sum_{123} \frac{| \langle \vec p_1 \vec p_3 s_1 s_3 t_1 
t_3 | \bar V_{2N} | \vec q \, \vec h_2 s s_2 t t_2 \rangle |^2}{\omega + \epsilon_2 - \epsilon_1
-\epsilon_3 + i \eta} \bar n_1 n_2 \bar n_3 (2\pi)^3 \delta(\vec p_1 + \vec p_3 - \vec q - \vec h_2), \nonumber \\
&+& \frac{1}{2}\sum_{123} \frac{| \langle \vec h_1 \vec h_3 s_1 s_3 t_1 
t_3 | \bar V_{2N} | \vec q \, \vec p_2 s s_2 t t_2 \rangle |^2}{\omega + \epsilon_2 - \epsilon_1
- \epsilon_3 - i \eta} n_1 \bar n_2 n_3 (2\pi)^3 \delta(\vec h_1 + \vec h_3 - \vec q - \vec p_2),
\label{se1}
\end{eqnarray}
where $\bar V_{2N}$ is the antisymmetrized two-body potential, $n_i = \theta(k_f-|\vec k_i|)$ 
is the zero-temperature Fermi distribution, $\bar n_i = 1-n_i$, and the sum is over the 
momentum, spin, and isospin of the intermediate states. The first-order contribution from three-body
forces has the form
\begin{equation}
\Sigma^{(1)}_{3N}(q,\omega;k_f) = \sum_{12} \langle \vec q \, \vec h_1\vec h_2; s s_1s_2; t t_1t_2 |
 \bar V_{3N} | \vec q \, \vec h_1\vec h_2; s s_1s_2; t t_1t_2 \rangle n_1 n_2,
\label{se31}
\end{equation}
where $\bar V_{3N}$ is the antisymmetrized three-body interaction. As discussed in the previous
section, the density-dependent two-nucleon interaction \cite{holt09,holt10} can be used in the last 
equation of (\ref{se1}) to study the effects of three-nucleon forces beyond the Hartree-Fock
approximation.

\begin{figure}
\includegraphics[height=9cm,angle=270]{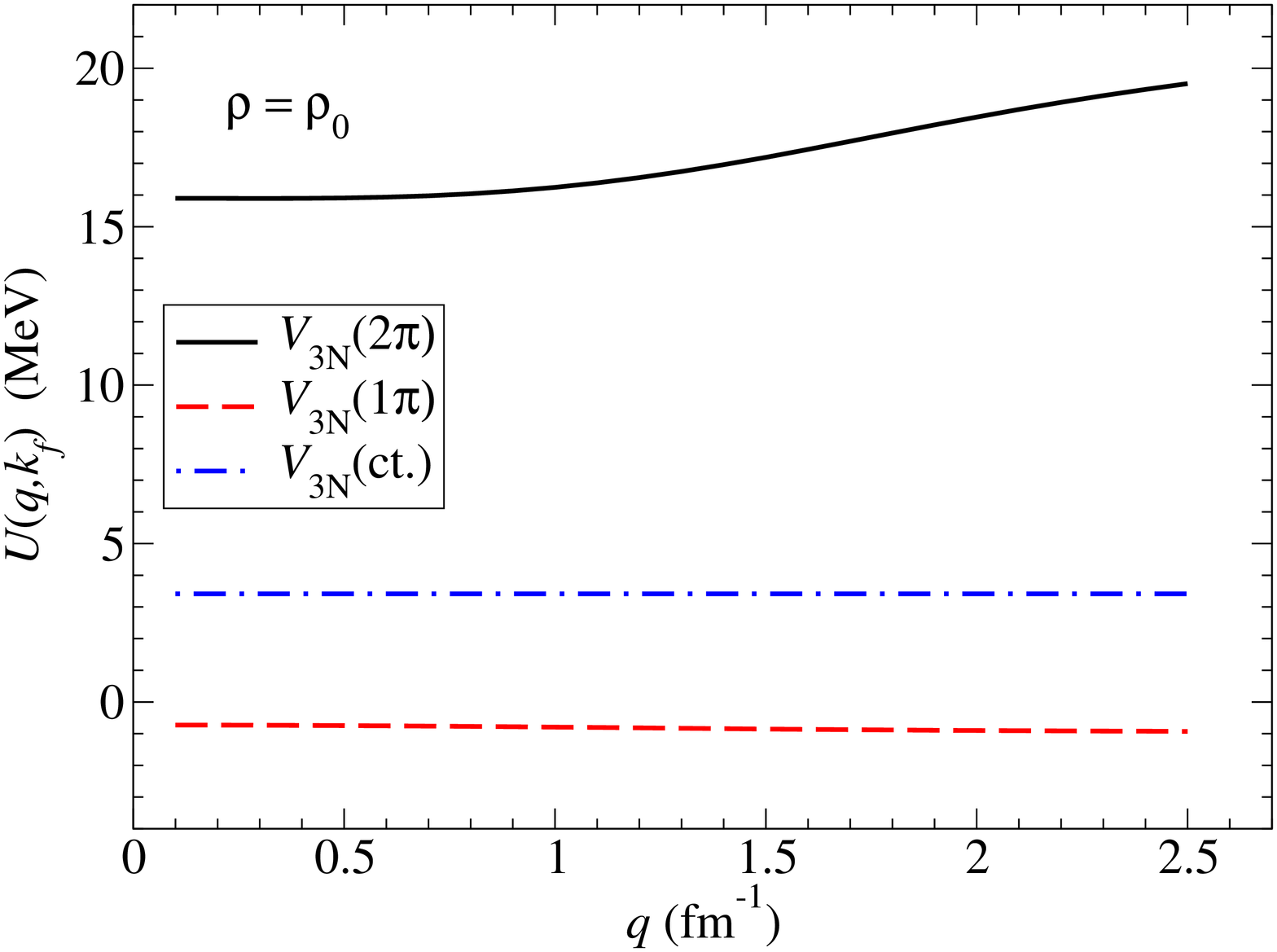}\hspace{.1in}
\includegraphics[height=9cm,angle=270]{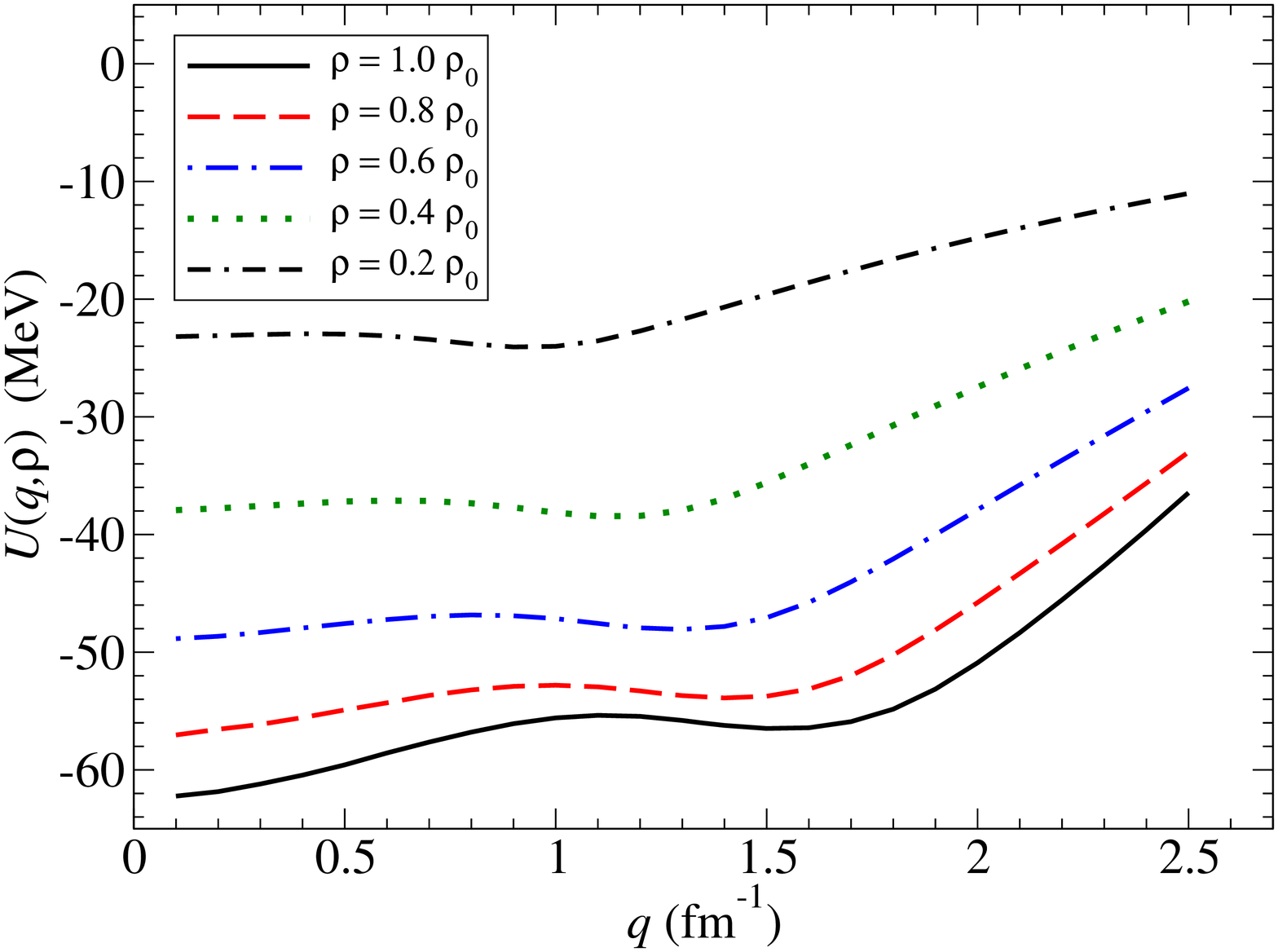}
\caption{Left panel: Three-body Hartree-Fock contributions to the real part of the nuclear optical 
potential. The individual single-particle potentials from the three components of the N$^2$LO chiral 
three-nucleon force are shown separately and plotted as a function of the momentum at nuclear 
matter saturation density.
Right panel: Density dependence of the real part of the momentum-dependent optical potential 
at second order in perturbation theory from chiral two- and three-body forces.}
\label{ddop3}
\end{figure}

In Fig.\ \ref{ddopr} we reproduce results from ref.\ \cite{holt13b} for the real part of the on-shell 
self-energy ($\omega = q^2/(2M_N)$) as a function of the density and momentum. The leading-order
Hartree-Fock contribution from two- and three-body forces are given by the solid lines and dashed
lines, respectively. Both contributions are real and explicitly energy independent, but they have 
qualitatively different properties. The Hartree-Fock contribution from two-body forces is attractive up 
to the maximum momentum plotted in Fig.\ \ref{ddopr} and decreases in magnitude as the 
momentum increases. The N$^2$LO chiral three-nucleon force, on the other hand, gives a repulsive 
contribution to
the single-particle potential that increases nearly linearly with the background density but which
varies only mildly with the momentum of the propagating nucleon. As shown in the left panel 
of Fig.\ \ref{ddop3}, the two-pion exchange component of the N$^2$LO chiral three-nucleon force
provides about 80\% of the total repulsion and is responsible for nearly all of the momentum
dependence of the three-body Hartree-Fock mean field.

A notable feature of the contact and $1\pi$-exchange three-body forces is that they generate
a momentum-independent and a nearly momentum-independent single-particle potential, respectively. 
This implies a strong correlation in the low-energy constants $c_D$ and $c_E$, where variations 
along the line
\be
c_E = \alpha \cdot c_D + {\rm const}
\ee
give nearly equivalent descriptions of the mean field. The constant of proportionality has the 
value $\alpha \simeq 0.21\pm 0.02$ and is weakly dependent on the momentum and density. 
As noted in ref.\ \cite{holt13b}, in the chiral limit the single-particle potential associated with 
the one-pion exchange three-body force has a particularly simple form which then yields the
correlation coefficient $\alpha = g_A/4 \simeq 0.3$.

\begin{figure}
\begin{center}
\includegraphics[height=16cm,angle=270]{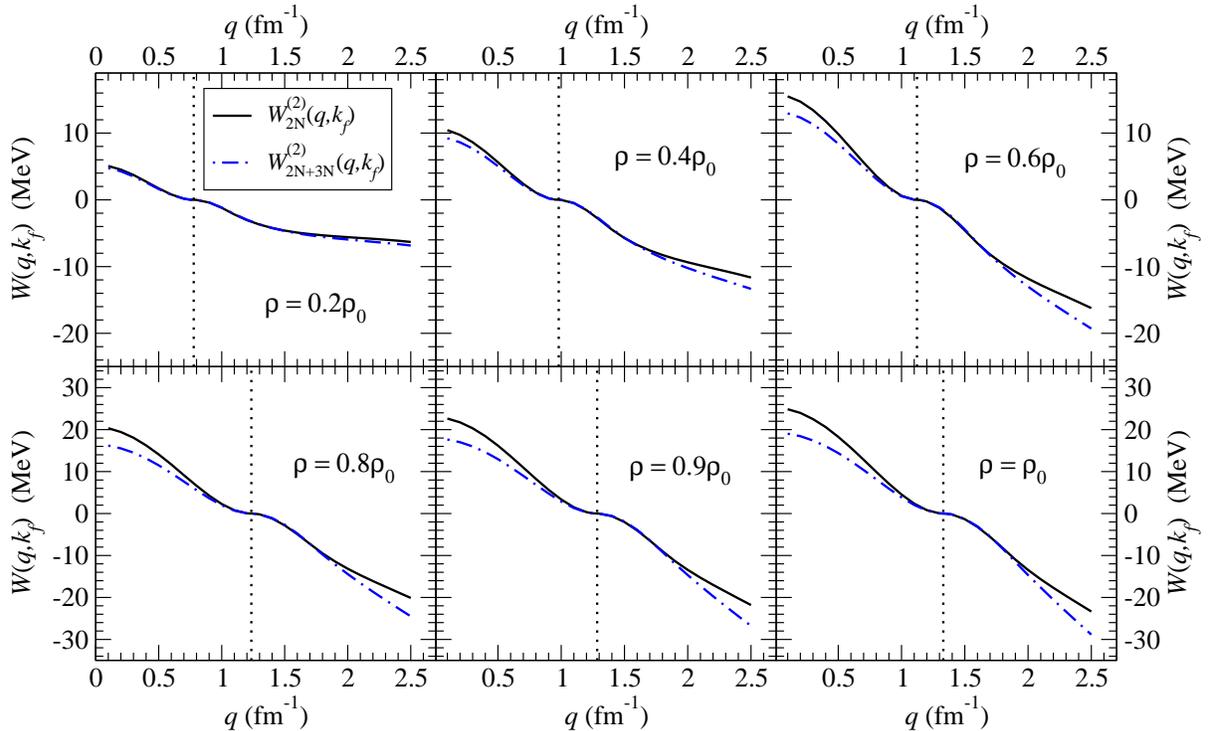}
\end{center}
\vspace{-.5cm}
\caption{The imaginary part of the nuclear optical potential from chiral two- and 
three-body forces at second order in many-body perturbation theory. The vertical 
dotted line indicates the Fermi momentum $k_f$.}
\label{ddopi}
\end{figure}

Second-order contributions to the single-particle potential 
from two- and three-body forces are shown as the dashed-dotted
lines in Fig.\ \ref{ddopr}. The momentum dependence of the second-order terms peaks close 
to the Fermi surface, and when combined with the Hartree-Fock mean field from two- and
three-body forces the resulting momentum dependence is quite small below $k_f$. The result 
is a nucleon 
effective mass at the Fermi surface that is close to the free-space mass. In 
Fig.\ \ref{ddopr} we show with the dashed-double-dotted line also the second-order contribution
without three-nucleon forces. Second-order effects from three-body forces appear to be rather 
small and result in additional attraction at both low and high momenta. 
In the right panel of Fig.\ \ref{ddop3} we show the total real part of the nucleon self-energy as a 
function of density and momentum. For low to moderate densities the single-particle potential
is nearly momentum independent below $k_f$. The overall well depth for scattering at zero incident 
energy ($q=k_f$) is approximately $-57$\,MeV at nuclear matter saturation density, which is within 
about 10\% of the depth, $-52$\,MeV, of phenomenological optical potentials \cite{koning03}.

In Fig.\ \ref{ddopi} we show the imaginary part of the single-particle potential as a function
of the density and momentum. At positive energies (measured with respect to the Fermi energy) 
the imaginary part of the potential is negative and strongly absorptive for particles propagating with
large momenta. We see that the imaginary part vanishes quadratically close to the Fermi surface 
for both two- and three-nucleon force contributions, in agreement with Luttinger's theorem 
\cite{luttinger61}. With two-body forces alone the imaginary part is approximately 
inversion-symmetric about the Fermi momentum, $W(q,k_f) \simeq -W(2k_f-q,k_f)$, but three-nucleon
forces lower the imaginary part at momenta well below and above $k_f$. For intermediate-energy
scattering, $E \sim 100$\,MeV, the phenomenological strength of the imaginary part of the 
optical potential is approximately $|W| \simeq 10$\,MeV \cite{koning03}, which is much less than
the value $\sim 30$\,MeV obtained in the second-order calculation with two- and three-body forces.
Effective mass corrections \cite{negele81,fantoni81}, self-consistent single-particle energies
in the denominators of eq.\ (\ref{se1}), and an inclusion of an energy gap at the Fermi surface are 
all expected to reduce significantly the magnitude of the imaginary part of the calculated optical
potential. Future comparisons to rare-isotope reactions will require also the inclusion of isospin 
asymmetry and the extension to finite nuclear systems.

\section{Density Functional Methods and Finite Nuclei}

In this section we review how chiral low-momentum interactions can be employed
for the description of properties of finite nuclei \cite{dmeimprov,microefun,stoitsov2010,
platter,efun,isofun}. We leave aside the highly 
sophisticated ab-initio methods which have been developed over the past decades in order to 
solve (numerically) the nuclear $A$-body problem with a given two- and three-nucleon 
interaction. Here we concentrate on the nuclear energy density functional approach, which remains
the many-body method of choice to compute the properties of medium-mass and 
heavy nuclei \cite{reinhard,stone}. Non-relativistic Skyrme functionals \cite{sly,pearson} 
with a small number of adjustable parameters as well as relativistic mean-field models 
\cite{serot,ring} have been employed in such self-consistent mean-field calculations to 
describe a wide range of nuclear properties. An alternative and complementary 
scheme \cite{lesinski,drut,platter} focuses less on fitting to experimental data 
and attempts instead to constrain the analytical form of the functional as well as the 
values of its couplings from a microscopic approach based on many-body perturbation 
theory with realistic two- and three-nucleon forces. The use of low-momentum nucleon-nucleon
potentials \cite{vlowkreview,vlowk}, rather than conventional hard-core interactions, is 
essential in this regard, because the former exhibit much improved convergence properties
compared to the latter in perturbative solutions to the nuclear many-body problem. Indeed, 
when evolved low-momentum interactions are employed at second-order in perturbation theory,
including also three-body forces, a good description of the bulk correlations in infinite nuclear 
matter \cite{achim,hebeler11} and in doubly-magic nuclei \cite{roth} can be achieved.

Within the framework of many-body perturbation theory, contributions to the energy are given
in terms of density-matrices convoluted with finite-range interaction kernels. The resulting
expression for the energy is then highly non-local in both space and time. In order for mean-field
calculations with such functionals to be numerically feasible in heavy open-shell nuclei, it is 
essential to develop approximations for these functionals in which only local densities and 
currents enter. For this purpose the density-matrix expansion is highly useful as it provides a
means to remove the non-local character of the exchange contribution to the
energy by expanding it in the form a generalized Skyrme functional with couplings that depend
on the density. For many years such an approach has been based on the density-matrix 
expansion of Negele and Vautherin \cite{negele}, but recently
Gebremariam, Duguet and Bogner \cite{dmeimprov} have suggested an improved 
version that specifically addresses the challenges that arise for spin-unsaturated nuclei. 
In fact, the phase-space averaging techniques they employed were shown to allow for a 
consistent expansion of both the spin-independent part as well as the spin-dependent part of 
the nuclear density-matrix. The improved features of the phase-space averaged density-matrix
expansion have been studied \cite{dmeimprov} via the Fock energy densities of schematic 
finite-range central, spin-orbit, and tensor interactions for a large group of semi-magic nuclei. 

Making use of these new developments a microscopically constrained nuclear energy 
density functional derived from the chiral NN potential at order N$^2$LO has been presented 
in ref.\ \cite{microefun}. There it has been suggested that the density-dependent couplings 
associated with pion-exchange should be added to a standard Skyrme 
functional with re-adjusted parameters. In a subsequent study \cite{stoitsov}, it has 
been shown that this new energy density functional yields numerically stable 
results and that it exhibits a small yet systematic reduction of the $\chi^2$ deviation 
in comparison to traditional Skyrme functionals without any explicit pion-exchange dynamics.
   
In this section we report on a derivation of the nuclear energy density functional with improved 
(chiral) two- and three-nucleon interactions \cite{efun,isofun}. For the 
two-body interaction the N$^3$LO chiral NN-potential is used, which reaches 
at this order in the chiral expansion the quality of a high-precision nucleon-nucleon potential 
in reproducing empirical NN scattering phase shifts and properties of the deuteron. 
The N$^3$LO chiral potential contains long-range one- and two-pion exchange interactions
as well as a short-distance part that is parameterized in terms of 24 low-energy constants. 
The latter contact potential is written in 
momentum space and provides the most general contribution up to fourth power in nucleon
momenta. This high-precision nucleon-nucleon potential is supplemented with the 
N$^2$LO chiral three-nucleon force. In addition to the equation of state
of infinite homogeneous nuclear matter, the energy density functional includes strength 
functions associated with the $(\vec \nabla \rho)^2$ surface term and the spin-orbit coupling 
term. In phenomenological Skyrme parametrizations these strength functions are treated as 
constants, whereas in a microscopic approach based on realistic nuclear interactions the 
finite-range character of the pion-exchange terms gives rise to 
specific density dependences for these strength functions.

\subsection{\it Density-Matrix Expansion and Energy Density Functional}

The construction of an explicit nuclear energy density functional starts from 
the density-matrix given as a sum over the energy eigenfunctions $\Psi_\alpha(\vec r
\,)$ associated with occupied orbitals of the non-relativistic many-body Fermi system. 
Gebremariam, Duguet and Bogner \cite{dmeimprov} have shown that it can be expanded 
in center-of-mass and relative coordinates, $\vec r$ and $\vec a$, in the following way   
\begin{eqnarray} \sum_{\alpha}\Psi_\alpha( \vec r -\vec a/2)\Psi_\alpha^
\dagger(\vec r +\vec a/2) &=& {3 \rho\over a k_f}\, j_1(a k_f)-{a \over 2k_f} 
\,j_1(a k_f) \bigg[ \tau - {3\over 5} \rho k_f^2 - {1\over 4} \vec \nabla^2 
\rho \bigg] \nonumber \\ && + {3i \over 2a k_f} \,j_1(a k_f)\, \vec \sigma
\cdot (\vec a \times \vec J\,) + \dots\,,  \label{dme}\end{eqnarray}
where $j_1(x) = (\sin x - x \cos x)/x^2$ is a spherical Bessel function. The right-hand
side of eq.\ (\ref{dme}) is written in terms of the nucleon density $\rho(\vec r\,) 
=2k_f^3(\vec r\,)/3\pi^2 =  \sum_\alpha \Psi^\dagger_\alpha(\vec r\,) \Psi_\alpha(\vec r\,)$, 
the kinetic energy density $\tau(\vec r\,) =  \sum_\alpha \vec \nabla 
\Psi^\dagger_\alpha (\vec r\,) \cdot \vec \nabla \Psi_\alpha(\vec r\,)$, and the  
spin-orbit density $ \vec J(\vec r\,) = i \sum_\alpha \vec \Psi^\dagger_\alpha(\vec r\,) 
\vec \sigma \times \vec \nabla \Psi_\alpha(\vec r\,)$.

Nuclear interactions derived within the framework of chiral effective field theory are 
generally given in momentum space. Therefore the Fourier transform of the expanded 
density-matrix eq.\ (\ref{dme})
with respect to the coordinates $\vec a$ and $\vec r$ provides the appropriate tool for an 
efficient calculation of the nuclear energy density functional. This Fourier transform:
\begin{eqnarray} \Gamma(\vec p,\vec q\,)& =& \int d^3 r \, e^{-i \vec q \cdot
\vec r}\,\bigg\{ \theta(k_f-|\vec p\,|) +{\pi^2 \over 4k_f^4}\Big[k_f\,\delta'
(k_f-|\vec p\,|)-2 \delta(k_f-|\vec p\,|)\Big] \nonumber \\ && \times \bigg( 
\tau - {3\over 5} \rho k_f^2 - {1\over 4} \vec \nabla^2 \rho \bigg) -{3\pi^2 
\over 4k_f^4}\,\delta(k_f-|\vec p\,|) \, \vec \sigma \cdot (\vec p \times 
\vec J\,)  \bigg\}\,, \label{gamma} \end{eqnarray}
generalizes the concept of ``medium-insertion'' to inhomogeneous many-nucleon systems 
characterized by the time-reversal-even fields $\rho(\vec r\,)$,  $\tau(\vec r\,)$ and 
$ \vec J(\vec r\,)$. At a practical level $\Gamma(\vec p,\vec q\,)$ extends the 
step-function-like momentum distribution $\theta(k_f-|\vec p\,|)$ for infinite nuclear matter 
to inhomogeneous many-nucleon systems. Note that the 
delta-function $\delta(k_f-|\vec p\,|)$ in eq.\ (\ref{gamma}) gives weight to the 
nucleon-nucleon interactions in the vicinity of the local
Fermi momentum $|\vec p\,|=k_f(\vec r\,)$ only. 

Up to second order in spatial gradients, characterizing deviations from homogeneity, the 
energy density functional appropriate for $N=Z$ even-even nuclei has the form
\begin{eqnarray} {\cal E}[\rho,\tau,\vec J\,] &=& \rho\,\bar E(\rho)+\bigg[\tau-
{3\over 5} \rho k_f^2\bigg] \bigg[{1\over 2M_N}-{k_f^2 \over 4M_N^3}+F_\tau(\rho)
\bigg] \nonumber \\ && + (\vec \nabla \rho)^2\, F_\nabla(\rho)+  \vec \nabla 
\rho \cdot\vec J\, F_{so}(\rho)+ \vec J\,^2 \, F_J(\rho)\,.\label{edf} \end{eqnarray}
In this equation, $\bar E(\rho)$ represents the energy per particle of symmetric nuclear 
matter evaluated at the local nucleon density. The strength function 
$F_\tau(\rho)$ is associated with an effective density-dependent nucleon mass 
$M^*(\rho)$, and it is related to the nuclear single-particle potential $U(p,k_f)$:
\begin{equation} F_\tau(\rho) = {1 \over 2k_f} {\partial U(p,k_f) \over \partial p}
\Big|_{p=k_f} = -{k_f \over 3\pi^2} f_1(k_f)\,,\end{equation}
where $\rho = 2k_f^3/3\pi^2$. The second equality gives the relation to the 
``spin and isospin independent'' p-wave Fermi-liquid parameter $f_1(k_f)$. A new  
feature of the improved density-matrix expansion is that it leads to the same 
concept of effective mass as established in Fermi-liquid theory for quasi-particles on 
the Fermi surface. For the original density-matrix expansion of Negele and Vautherin 
\cite{negele} this close relationship does not hold in general, e.g.\ when $U(p,k_f)$ 
deviates from a simple quadratic $p$-dependence.
   
The strength function $F_\nabla(\rho)$ associated with the $(\vec \nabla \rho)^2$ surface 
term can be decomposed into two terms:
\begin{equation} F_\nabla(\rho) = {1\over 4}\, {\partial F_\tau(\rho)
\over  \partial \rho} +F_d(\rho) \,,
\end{equation}
where $F_d(\rho)$ is composed of all contributions for which the $(\vec \nabla \rho
)^2$-factor originates directly from the momentum dependence of the interactions (expanded  
up to order $\vec q^{\,2}$ and combined with a Fourier transformation). It is worth noting that 
only the nuclear matter component $\theta(k_f-|\vec p\,|)$ of the density-matrix expansion enters
 into the derivation of the strength function $F_d(\rho)$. The next-to-last term in eq.\ (\ref{edf}), 
namely $\vec \nabla \rho \cdot\vec J\, F_{so}(\rho)$ with its associated strength function
$F_{so}(\rho)$, gives the spin-orbit interaction in nuclei. The last term, $\vec J\,^2 \, F_J(\rho)$, 
in the expression for energy density is the quadratic spin-orbit term. It gives rise to an additional 
spin-orbit single-particle potential proportional to $\vec J$, whereas the ordinary spin-orbit potential is provided by 
the density-gradient $\vec \nabla \rho$.    

\subsection{\it Two- and Three-Body Contributions}

Here, the two- and three-body contributions to the various density-dependent 
strength functions entering the nuclear energy density functional ${\cal E}[
\rho,\tau,\vec J\,]$ are presented. One would prefer to calculate the two-body 
contributions with perturbative nucleon-nucleon potentials, of which the low-momentum 
NN interaction $V_{\rm low-k}$ \cite{vlowk} is the prototypical example. 
Low-momentum two-body potentials are however non-local and given in terms of partial-wave 
matrix elements, which makes their implementation in the density-matrix expansion rather
difficult. An explicit representation of the momentum-space nucleon-nucleon potential 
(without non-localities) in terms 
of spin- and isospin-operators is much more suitable for this purpose. As a convenient 
substitute for $V_{\rm low-k}$ one can use the chiral NN-potential N$^3$LOW developed in 
refs.\ \cite{machleidtreview,n3low} by imposing a sharp cutoff at the scale $\Lambda= 414\,$MeV.  
This value of the cutoff coincides with the resolution scale below which evolved low-momentum 
NN potentials become nearly model-independent and exhibit desirable convergence properties 
in many-body perturbation theory calculations. The finite-range part of the N$^3$LOW chiral 
NN-potential arises from one- and two-pion exchange processes that have the general form:
\begin{eqnarray} V_{NN}^{(\pi)} &=& V_C(q) + \vec \tau_1 \cdot \vec \tau_2\, W_C(q) +
\big[V_S(q) + \vec \tau_1 \cdot \vec \tau_2\, W_S(q)\big] \, \vec \sigma_1 \cdot 
\vec\sigma_2 \nonumber \\ && +\big[V_T(q) + \vec \tau_1 \cdot \vec\tau_2\,
W_T(q) \big]\,\vec \sigma_1 \cdot \vec q \,\, \vec\sigma_2 \cdot \vec q  
\nonumber \\ && +\big[V_{SO}(q) + \vec \tau_1 \cdot \vec\tau_2\,W_{SO}(q)\big]\, 
i (\vec \sigma_1+\vec\sigma_2)\cdot (\vec q \times \vec p\,) \,, \end{eqnarray}
where $\vec q$ is the momentum transfer and $\vec p$ a single nucleon momentum. 
A special and convenient feature of $V_{NN}^{(\pi)}$ is that it is a local potential: all 
occurring functions $V_C(q),\dots, W_{SO}(q)$ depend only on the momentum transfer $q$, 
and quadratic spin-orbit components $\sim \vec \sigma_1\cdot (\vec q \times \vec 
p\,)\,\vec \sigma_2\cdot (\vec q \times \vec p\,)$ are absent. 

In the Hartree-Fock approximation the finite-range part of the nucleon-nucleon potential arising from
explicit pion exchange leads to the following two-body contributions to the energy density functional:  
\begin{eqnarray} \bar E(\rho) &=& {\rho\over 2} V_C(0) - {3\rho\over 2} \int_0^1
\!\!dx\, x^2(1-x)^2(2+x) \Big[V_C(q)+3W_C(q) \nonumber \\ && \qquad\qquad\quad
+3V_S(q)+9W_S(q) +q^2V_T(q) +3q^2W_T(q)\Big] \,, \end{eqnarray}
\begin{equation} F_\tau(\rho) = {k_f\over 2\pi^2}\int_0^1\!\!dx(x-2x^3)\Big[
V_C(q)+3W_C(q) +3V_S(q)+9W_S(q) +q^2V_T(q) +3q^2W_T(q)\Big] \,, \end{equation}
\begin{equation} F_d(\rho) = {1\over 4} V_C''(0)\,,  \label{fdeq}
\end{equation} 
\begin{equation} F_{so}(\rho) = {1\over 2}V_{SO}(0)+ \int_0^1\!\!dx\,x^3\Big[
V_{SO}(q)+3W_{SO}(q)\Big] \,, \end{equation}
\begin{equation} F_J(\rho) = {3\over 8k_f^2} \int_0^1\!\!dx\Big\{(2x^3-x)\Big[
V_C(q)+3W_C(q) -V_S(q)-3W_S(q)\Big] -x^3 q^2\Big[V_T(q) +3W_T(q)\Big]\Big\}
\,, \end{equation} 
where $q = 2x k_f$. The above expressions are obtained from the density 
matrix-expansion employing the product of two medium-insertions 
$\Gamma(\vec p_1,\vec q\,)\,\Gamma(\vec p_2, -\vec q\,)$). In eq.\ (\ref{fdeq}) 
the double-prime denotes the second derivative, and it 
is worth noting that $F_J(\rho)$ stays finite in the low-density limit $k_f \to 0$. 

In addition to the finite-range parts of the NN potential there are two-body 
contributions from the zero-range contact forces. At order 
N$^3$LO the corresponding expressions in momentum-space include constant, quadratic, 
and quartic terms in momenta \cite{evgeni}. 
The density-dependent strength functions arising from the contact NN potential read: 
\begin{equation} \bar E(\rho) = {3\rho\over 8}(C_S-C_T)+{3 \rho k_f^2 \over 20} 
(C_2-C_1-3C_3-C_6)+{9\rho k_f^4\over 140}(D_2-4D_1-12D_5-4D_{11})\,,
\end{equation} 
\begin{equation} F_\tau(\rho) = {\rho \over 4} (C_2-C_1-3C_3-C_6)+{\rho k_f^2\over
 4}(D_2-4D_1-12D_5-4D_{11})\,,
\end{equation} 
\begin{equation} F_d(\rho) = {1\over 32} (16C_1-C_2-3C_4-C_7)+{k_f^2\over 48}
(9D_3+6D_4-9D_7-6D_8-3D_{12}-3D_{13}-2D_{15})\,,
\end{equation} 
\begin{equation}
F_{so}(\rho) = {3\over 8} C_5+{k_f^2\over 6}(2D_9+D_{10})\,,
\end{equation}
\begin{equation} F_J(\rho) = {1\over 16} (2C_1-2C_3-2C_4-4C_6+C_7)+{k_f^2\over 32}
(16D_1-16D_5-4D_6-24D_{11}+D_{14})\,.
\end{equation} 
The 24 low-energy constants $C_{S,T}$, $C_j$ and $D_j$ are fit (with a sharp cutoff 
regulator at $\Lambda = 414\,$MeV) to empirical nucleon-nucleon scattering phase shifts 
and properties of the deuteron \cite{n3low}. It is worth mentioning that in the first-order Hartree-Fock 
approximation it is not necessary to include a regulator function since the 
nuclear interactions are probed only at small momenta $|\vec p_{1,2}|\leq k_f\leq 285\,$MeV. 
The regulator function as well as its associated cutoff scale $\Lambda$ become 
relevant only at second order (and higher).

In the next step, the three-body contributions to the nuclear energy density 
functional ${\cal E}[\rho,\tau,\vec J\,]$ are included. The leading order chiral 
three-nucleon interaction consists of a contact piece 
(with low-energy constant $c_E$), a $1\pi$-exchange part (proportional to $c_D$), and a 
$2\pi$-exchange component (with low-energy constants $c_1$, $c_3$ and $c_4$). In order simplify
the treatment of the three-body correlations in inhomogeneous nuclear systems, it has been assumed 
\cite{efun} that the relevant product of density-matrices can be represented in the factorized form 
$\Gamma(\vec p_1,\vec q_1)\,\Gamma(\vec p_2,\vec q_2)\,
\Gamma(\vec p_3,-\vec q_1-\vec q_2)$ in momentum 
space. This factorization ansatz respects the correct infinite nuclear matter limit, 
but it involves approximations compared to more 
sophisticated treatments explored in ref.\ \cite{platter}. In fact, the present approach 
is similar to the method ``DME-I'' considered in ref.\ \cite{platter}.

\begin{figure}
\begin{center}
\includegraphics[scale=0.90,clip]{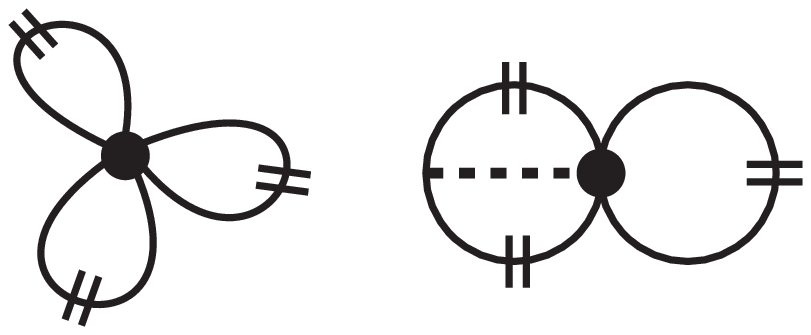}\hspace{.1in}
\includegraphics[scale=0.90,clip]{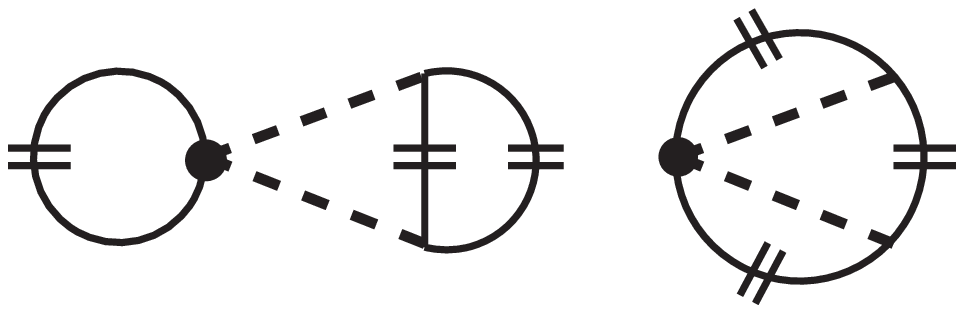}
\end{center}
\vspace{-.6cm}
\caption{Three-body diagrams related to the contact ($c_E$), $1\pi$-exchange ($c_D$), 
and $2\pi$-exchange ($c_1,c_2,c_3$) 
components of the chiral three-nucleon interaction. The short double-line symbolizes the 
medium insertion $\Gamma(\vec p, \vec q\,)$ for inhomogeneous nuclear matter.}
\label{efunfig1}
\end{figure}


The three-body contact-term proportional to $c_E$ (see Fig.\ \ref{efunfig1}) gives rise to 
a contribution to the energy per particle:
\begin{equation} \bar E(\rho) = -{c_E k_f^6 \over 12 \pi^4 f_\pi^4\Lambda}\,,
\end{equation}
that is quadratic in the density $\rho=2k_f^3/3\pi^2$. Obviously, all other strength functions 
$F_{\tau,d,so,J}(\rho)$ receive vanishing contributions due to the momentum-independence of the 
chiral contact interaction. The $1\pi$-exchange component of the chiral 3N-interaction involves the 
low-energy constant $c_D$. With three (inhomogeneous) medium insertions in the corresponding 
three-body diagram (shown in Fig.\ \ref{efunfig1}) one finds the following expressions for the 
strength functions:  
\begin{equation} \bar E(\rho)={g_A c_D m_\pi^6\over(2\pi f_\pi)^4\Lambda}
\bigg\{{u^6 \over3}-{3u^4\over 4}+{u^2\over 8} +u^3 \arctan 2u -{1+12u^2 \over
  32}\ln(1+4u^2) \bigg\}\,,\end{equation}  
\begin{equation} F_\tau(\rho)={2g_A c_D m_\pi^4\over(4\pi f_\pi)^4\Lambda}
\Big\{(1+2u^2)\ln(1+4u^2)-4u^2 \Big\}\,,\end{equation}
\begin{equation} F_d(\rho)={g_A c_D m_\pi\over(4f_\pi)^4 \pi^2\Lambda}
\bigg\{{1\over 2u}\ln(1+4u^2)-{2u\over 1+4u^2} \bigg\}\,,\end{equation}
\begin{equation} F_J(\rho)={3g_A c_D m_\pi\over(4f_\pi)^4 \pi^2\Lambda}
\bigg\{ 2u-{1\over u}+{1\over 4u^3}\ln(1+4u^2)\bigg\}\,, \end{equation}
where $u=k_f/m_\pi$. We observe that there is no contribution to the spin-orbit strength 
$F_{so}(\rho)$, since neither the contact-vertex nor the $1\pi$-exchange component 
produces a spin-orbit interaction.  

In the third diagram of Fig.\ \ref{efunfig1}, we show the three-body contribution arising 
from the $2\pi$-exchange Hartree term. With three (inhomogeneous) medium 
insertions, the corresponding energy density coupling strengths are written \cite{efun}:
\begin{eqnarray}  \bar E(\rho)&=&{g_A^2 m_\pi^6\over(2\pi f_\pi)^4}\bigg\{(12c_1
-10c_3) u^3\arctan 2u -{4\over 3} c_3 u^6 +6(c_3-c_1)u^4 \nonumber \\ && +(3c_1
-2c_3)u^2 +\bigg[{1\over 4}(2c_3-3c_1)+{3u^2\over 2}(3c_3-4c_1)\bigg]\ln(1+4u^2)
\bigg\}\,,  \end{eqnarray}
\begin{eqnarray}  F_\tau(\rho)&=&{g_A^2 m_\pi^4\over(2\pi   f_\pi)^4}\bigg\{
(5c_3-6c_1)u^2 +{(c_3-2c_1)u^2 \over 1+4u^2}\nonumber \\ && +\bigg[
2c_1-{3\over 2}c_3+2(c_1-c_3)u^2\bigg]\ln(1+4u^2) \bigg\}\,,  \end{eqnarray}
\begin{eqnarray}  F_d(\rho)&=&{g_A^2 m_\pi\over  (8\pi)^2 f_\pi^4}\bigg\{(10c_1-
23c_3)\arctan 2u+ 16c_3 u  \nonumber \\ && +{7c_3-5c_1 \over u}\ln(1+4u^2) 
+ {6c_3 u+16(2c_3-c_1)u^3\over 3(1+4u^2)^2} \bigg\}\,,  \end{eqnarray}
\begin{equation} F_{so}(\rho)  = {3g_A^2 m_\pi \over (8\pi)^2 f_\pi^4}
\bigg\{{2\over u}(4c_1-3c_3) -4c_3 u+\bigg[{4\over u}(c_3-c_1)+{3c_3-4c_1
    \over 2u^3}\bigg] \ln(1+4u^2) \bigg\}\,, \label{3bso}\end{equation}
\begin{equation} F_J(\rho)  = {3g_A^2 m_\pi \over (8\pi)^2 f_\pi^4}\bigg\{
{3c_3-4c_1 \over u} -2c_3 u+{4u(2c_1-c_3)\over 1+4u^2}+{4c_1-3c_3 \over 4u^3}
\ln(1+4u^2) \bigg\}\,, \end{equation}
which depend only on the isoscalar coupling constants $c_1$ and $c_3$. Note that
the expression for the spin-orbit strength in eq.\ (\ref{3bso}) provides the dominant part 
of the three-body contribution to $F_{so}(\rho)$, as suggested originally by Fujita and 
Miyazawa \cite{fujita} in the context of $\Delta(1232)$-resonance excitation.
In the present chiral effective field theory treatment without explicit $\Delta$ resonances, 
the two-step process $\pi N\to \Delta \to \pi N$ is written as
an equivalent $\pi\pi NN$ contact vertex proportional to $c_3$. 

Finally, we discuss the three-body $2\pi$-exchange Fock contribution, which
is shown as the fourth diagram of Fig.\ \ref{efunfig1}. This term consists of a 
single closed nucleon ring that generates for isospin-symmetric nuclear matter 
non-vanishing contributions from both the isoscalar and isovector $\pi\pi NN$ 
contact vertices, the latter proportional to the low-energy constant $c_4$. 
In contrast to the three-body $2\pi$-exchange Hartree diagram, the Fock contribution
involves integrals (over three Fermi spheres) that cannot be computed in closed 
analytical form. The resulting rather lengthy expressions for 
the various strength functions can be found in section 4.4.\ of ref.\ \cite{efun}.

\begin{figure}
\begin{center}
\includegraphics[scale=0.4,clip]{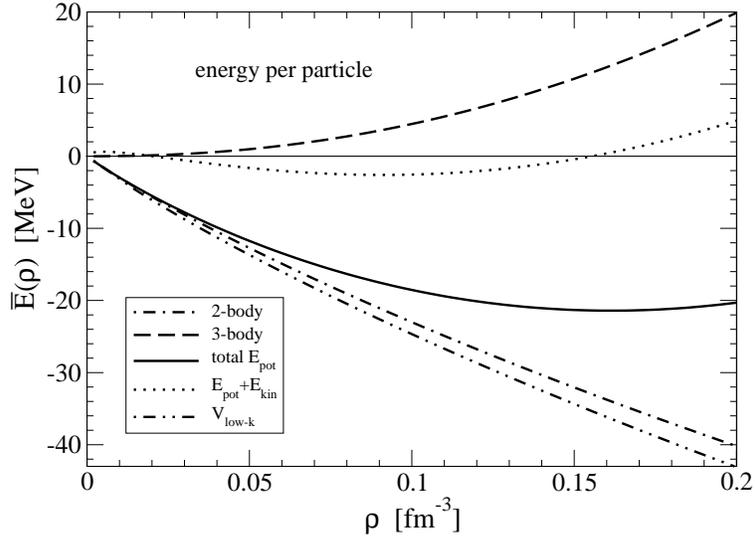}
\end{center}
\vspace{-.8cm}
\caption{The energy per particle $\bar E(\rho)$ of isospin-symmetric nuclear matter derived from chiral nuclear interactions.}\label{ebar23}
\end{figure}

Let us now discuss the results for the nuclear energy density functional obtained in the 
first-order Hartree-Fock approximation. In Fig.\ \ref{ebar23} we show the contributions to the 
energy per particle $\bar E(\rho)$ from chiral two- and three-body forces for densities up to 
$\rho = 0.2\,$fm$^{-3}$. The dash-dotted line denotes the attractive two-body contributions, while
the dashed line gives the repulsive three-body contributions. The results from the N$^3$LOW chiral
nucleon-nucleon interaction are compared to those from the universal low-momentum 
NN potential $V_{\rm low-k}$ obtained by integrating over
its diagonal (on-shell) partial-wave matrix elements. Indeed, the 
treatment of the two-body interaction via the chiral potential N$^3$LOW gives a fairly accurate
reproduction of these results. The full line in Fig.\ \ref{ebar23} shows the sum of the two- and 
three-body contributions, which exhibits a first tendency for the saturation of nuclear matter. 
After including the kinetic energy $\bar E_{\rm kin}(\rho) = 3k_f^2/10M_N-3k_f^4/56M_N^3$,
however, the resulting saturation energy is still much too low. This particular feature of the 
Hartree-Fock approximation has been observed in similar studies \cite{achim,hebeler11}. A much
improved description of the nuclear matter equation of state is achieved by 
treating the two-body (and three-body) interaction at least to second order.

\begin{figure}
\begin{center}
\includegraphics[scale=0.4,clip]{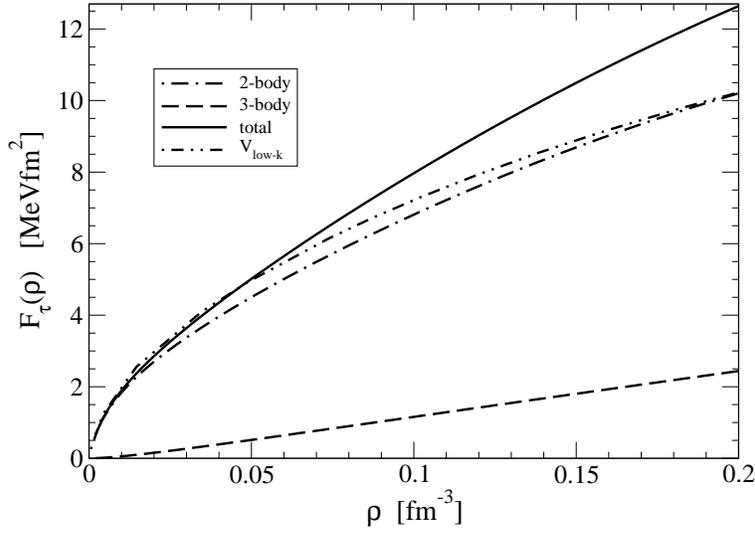}
\end{center}
\vspace{-.6cm}
\caption{Contributions to the strength function $F_\tau(\rho)$ as a function of the nuclear 
density $\rho$.}\label{ftau}
\end{figure}

\begin{figure}
\begin{center}
\includegraphics[scale=0.4,clip]{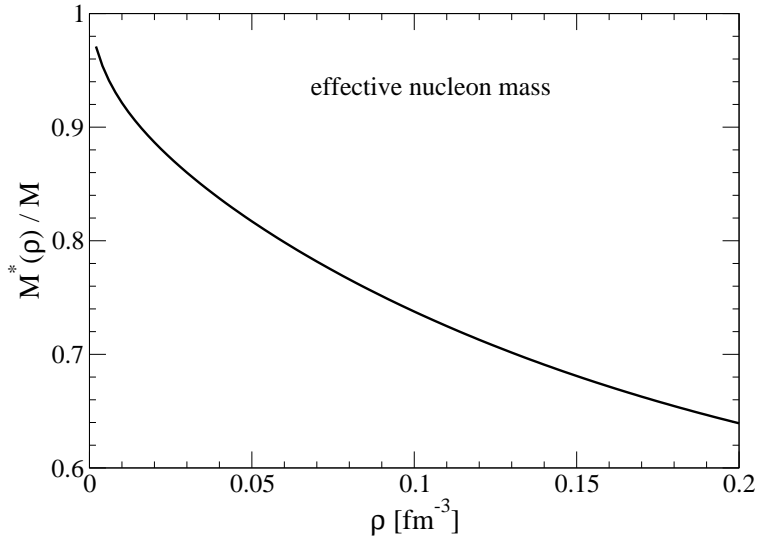}
\end{center}
\vspace{-.8cm}
\caption{Ratio of the effective nucleon mass $M^*(\rho)$ to the free nucleon mass $M$ as 
function of $\rho$.}\label{mstar23}
\end{figure}

\begin{figure}
\begin{center}
\includegraphics[scale=0.4,clip]{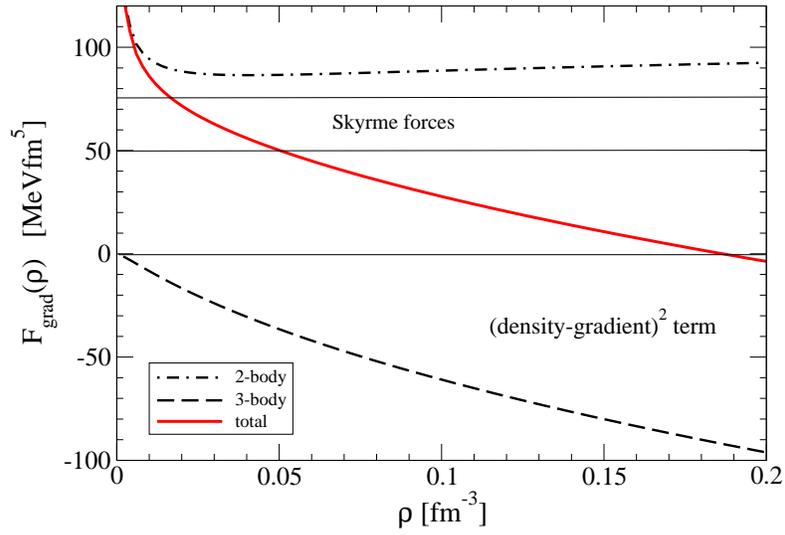}
\end{center}
\vspace{-.8cm}
\caption{The strength function $F_\nabla(\rho)$ of the surface term 
$(\vec \nabla \rho)^2$ versus the nuclear density $\rho$.}\label{fgrad23}
\end{figure}
     
\begin{figure}
\begin{center}
\includegraphics[scale=0.4,clip]{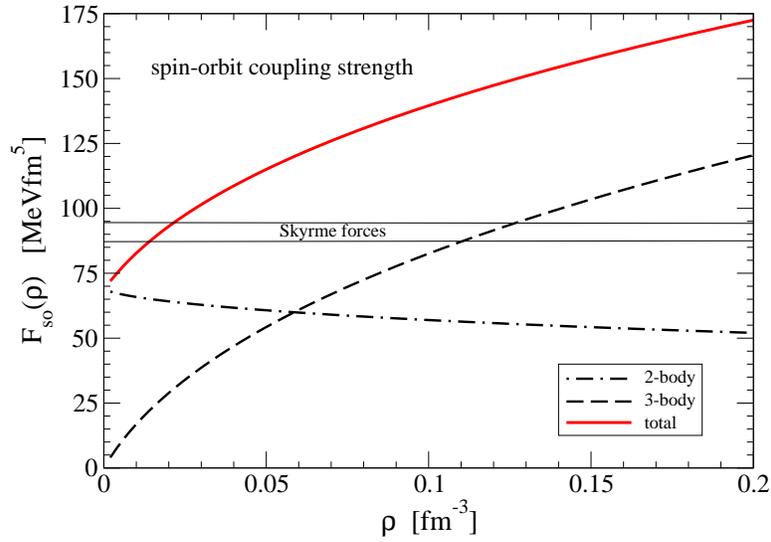}
\end{center}
\vspace{-.8cm}
\caption{Spin-orbit strength function $F_{so}(\rho)$ as a function of the density $\rho$.}\label{fso23}
\end{figure}

\begin{figure}
\begin{center}
\includegraphics[scale=0.4,clip]{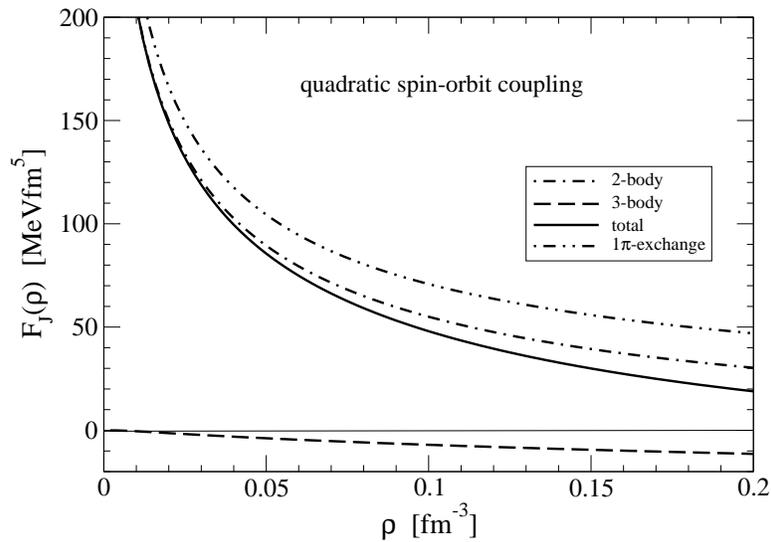}
\end{center}
\vspace{-.8cm}
\caption{Strength function $F_J(\rho)$ of the squared spin-orbit density 
$\vec J^{\,2}$ versus $\rho$.}\label{fj23}
\end{figure}

In Fig.\ \ref{ftau} we reproduce the contributions to the strength function $F_\tau(\rho)$
associated with the surface gradient term in the energy density functional. We note that 
the two-body parts derived from $V_{\rm low-k}$ and the chiral N$^3$LOW potential are
rather similar. The three-body part, which increases nearly linearly with the density, 
comes out relatively small. At nuclear matter saturation density $\rho_0=0.16\,$fm$^{-3}$ 
it gives rise to a correction of about $20\%$ compared to the two-body contribution. 
The bracketed term that multiplies the kinetic energy density 
$\tau(\vec r\,)$ in the expression for the nuclear energy density functional 
in eq.\ (\ref{edf}) can be interpreted 
as the reciprocal of a density-dependent effective nucleon mass:
\begin{equation} 
M^*(\rho) = M_N\bigg[ 1 -{k_f^2 \over 2M_N^2}+2M_N\, F_\tau(\rho)\bigg]^{-1}
\,. \end{equation}
In the improved density matrix expansion it is identical to the so-called 
``Landau'' effective mass introduced in Fermi-liquid theory, and
it is derived in the same way from the derivative of the single-particle potential $U(p,k_f)$ with
respect to the momentum $p$ at the Fermi surface $p = k_f$. 
The relativistic increase in the mass is accounted for with the small correction term 
$-k_f^2/2M_N^2$. In Fig.\ \ref{mstar23} we show the ratio of effective nucleon mass to the 
free nucleon mass $M^*(\rho)/M_N$ as a function of the density $\rho$. In the Hartree-Fock
approximation, the effective nucleon mass is reduced in comparison to the free-space
mass, and the ratio reaches the value  $M^*(\rho_0) \simeq 0.67M_N$ at nuclear 
matter saturation density $\rho_0=0.16\,$fm$^{-3}$. This is consistent with the range 
$0.7 < M^*(\rho_0)/M_N<1$ determined from phenomenological Skyrme forces 
\cite{sly,pearson}. On the other hand, second-order corrections from two-body forces 
significantly enhance the effective mass \cite{fermiliq}.

The strength function $F_\nabla(\rho)$ of the surface gradient term $(\vec \nabla \rho)^2$ 
is plotted in Fig.\ \ref{fgrad23} as a function of the nuclear density. 
At very low densities this coupling strength exhibits a pronounced increase in the 
two-body contribution (dash-dotted line) that has its origin in $1\pi$-exchange. For larger
densities, the two-body contributions are nearly independent of the density.
The three-body contribution (dashed line) is sizable and negative. It reduces the two-body 
contribution such that the total value of $F_\nabla(\rho)$ (full line in Fig.\ \ref{fgrad23}) 
decreases with increasing density $\rho$. To compare with phenomenology we also show
the band (of constant $F_\nabla(\rho)$-values) spanned by parameterized Skyrme 
forces \cite{sly,pearson}. The Hartree-Fock result from realistic chiral two- and three-body
forces is somewhat too small at densities close to $\rho_0/2 =0.08\,$fm$^{-3}$, where the 
surface energy in finite nuclei gains much of its weight. Iterated $1\pi$-exchange has been
considered in ref.\ \cite{efunold} from which it may be deduced that a treatment of the 
two-body interaction to second order in perturbation theory will increase the values of 
$F_\nabla(\rho)$.

Of special interest is the coupling strength $F_{so}(\rho)$ multiplying the spin-orbit 
term $\vec \nabla \rho\cdot \vec J$. Contributions from two- and three-body terms, 
as well as their total sum, are shown in Fig.\ \ref{fso23}. The coupling strength from 
two-body forces is dominated by the low-energy constant $3C_5/8$, as suggested also
by the weak dependence of the 
dash-dotted line on the density $\rho$. Three-body forces induce a density-dependent 
spin-orbit interaction that considerably enhances the contribution from the N$^3$LOW
two-body force. The most significant contribution arises in the 
Hartree term (\ref{3bso}) proportional to the low-energy constant $c_3=-4.78\,$GeV$^{-1}$. 
With this particular value of $c_3$, the three-body spin-orbit strength is considerably larger 
than that proposed by Fujita and Miyazawa \cite{fujita}, where the $\Delta(1232)$-excitation 
mechanism would correspond to $c_3^{(\Delta)} = -g_A^2/2
\Delta\simeq-2.9\,$GeV$^{-1}$ ($\Delta = 293\,$MeV is the delta-nucleon mass splitting).  
At densities around half that of saturated nuclear matter, $\rho_0/2 = 0.08\,$fm$^{-3}$, 
the total Hartree-Fock contribution exceeds the empirical 
spin-orbit coupling strength $F_{so}^{\rm (emp)}(\rho) \simeq 90\,$MeV\,fm$^{5}$ \cite{sly,
pearson} by nearly $50\%$. A compensating effect is therefore required, and it has been 
suggested in ref.\ \cite{efunold} that the $1\pi$-exchange tensor force at second-order 
generates a spin-orbit coupling of the ``wrong-sign''. It is then reasonable to assume that 
the low-momentum two-nucleon tensor potential, when treated to second-order in
perturbation theory, will reduce the strength of the spin-orbit coupling $F_{so}(\rho)$ 
to a value close to that suggested by phenomenology.

To conclude the description of the nuclear energy density functional in $N=Z$ even-even
nuclei, we show in Fig.\ \ref{fj23} the strength function $F_J(\rho)$ of the squared 
spin-orbit coupling. In contrast to all of the previous coupling strengths, $F_J(\rho)$ receives 
only a very small contribution from three-body forces. One observes that the two-body 
contribution is, however, strongly density-dependent, and at low densities it reaches quite 
large values. This strong density dependence originates primarily from the
$1\pi$-exchange contribution, which for comparison we reproduce separately in 
Fig.\ \ref{fj23} (dashed-double-dotted line). At this point it 
should be emphasized that the $\vec J^{\,2}$-term in the energy density functional 
represents non-local Fock contributions from tensor forces, etc. It is therefore not 
surprising that there exists an outstanding $1\pi$-exchange 
contribution to the strength function $F_J(\rho)$.       
 
\subsection{\it Isovector Part of the Nuclear Energy Density Functional}

The previous calculation of the nuclear energy density functional from chiral two- and 
three-nucleon forces can be extended in a straightforward manner to the isovector terms 
\cite{isofun} relevant for nuclei with different proton and neutron densities. The additional 
isovector terms 
play an important role in the description of long isotopic chains of stable nuclei and for neutron-rich 
systems far from stability. By construction the density-matrix expansion of Gebremariam, Duguet 
and Bogner \cite{dmeimprov} applies separately to proton and neutron orbitals. It is 
therefore easily adapted to the situation of isospin-asymmetric nuclear systems with
 different proton and neutron 
local densities $\rho_{p,n}, \tau_{p,n}, \vec J_{p,n}$. After Fourier transforming to momentum 
space one obtains the medium insertion: 
\begin{eqnarray} \Gamma_{\rm iv}(\vec p,\vec q\,)& =& \int\!d^3 r \, e^{-i \vec q \cdot
\vec r}\,\bigg\{ {{\boldsymbol 1}+{\boldsymbol\tau}_3\over 2}\,\theta(k_p-
|\vec p\,|) + {{\boldsymbol 1}-{\boldsymbol\tau}_3\over 2}\,
\theta(k_n-|\vec p\,|)  \nonumber \\ && +{\pi^2 \over 4k_f^4}\Big[k_f\,\delta'
(k_f-|\vec p\,|)-2 \delta(k_f-|\vec p\,|)\Big] \bigg[ \tau_p - \tau_n-\bigg(
k_f^2+{\vec \nabla^2 \over 4}\bigg)\nonumber \\ && \times (\rho_p-\rho_n)\bigg] 
{\boldsymbol\tau}_3-{3\pi^2 \over 4k_f^4}\,\delta(k_f-|\vec p\,|) \, (\vec 
\sigma \times\vec p\,) \cdot( \vec J_p- \vec J_n) {\boldsymbol \tau}_3+\dots 
\bigg\}\,,  \end{eqnarray}
for the inhomogeneous isospin-asymmetric many-nucleon system. In the above
expression,  
${\boldsymbol\tau}_3$ denotes the third Pauli isospin matrix, and we have included 
only the relevant terms proportional to differences of proton and neutron 
densities: $\rho_p - \rho_n$, $\tau_p - \tau_n$, $\vec J_p- \vec J_n$. The two local 
Fermi momenta are related to the (particle) densities in the 
usual way: $\rho_p=k_p^3/3\pi^2$, $\rho_n=k_n^3/3\pi^2$, $\rho=\rho_p+\rho_n=2k_f^3 
/3\pi^2$. When working to quadratic order in deviations from isospin-symmetric nuclear
systems, it is sufficient to employ an average Fermi momentum $k_f$ 
in the factors multiplying $\tau_p - \tau_n$ and $\vec J_p- \vec J_n$. 

Up to second order in proton-neutron density differences and spatial gradients, the isovector 
part of the nuclear energy density functional reads:
\begin{eqnarray} &&{\cal E}_{\rm iv}[\rho_p,\rho_n,\tau_p,\tau_n,\vec J_p,\vec J_n] 
= {1\over \rho}(\rho_p-\rho_n)^2\,\tilde A(\rho)+ {1\over \rho}(\tau_p-\tau_n)
(\rho_p-\rho_n)\, G_\tau(\rho) \nonumber \\ && + (\vec \nabla \rho_p-\vec \nabla 
\rho_n)^2\, G_\nabla(\rho)+  (\vec \nabla \rho_p- \vec \nabla \rho_n)\cdot(\vec J_p-
\vec J_n)\, G_{so}(\rho)+ (\vec J_p-\vec J_n)^2 \, G_J(\rho)\,.\label{isoedf}\end{eqnarray}
Here, $\tilde A(\rho)$ is the interaction part of the isospin-asymmetry energy of 
homogeneous nuclear matter. The non-interacting (or kinetic energy) contribution 
$A_{\rm kin}(\rho) = k_f^2/6M_N$ to the isospin-asymmetry energy is included in the 
energy density functional through the kinetic energy density term, $ {\cal E}_{\rm kin}
=(\tau_p+\tau_n)/2M_N$. The strength function $G_\nabla(\rho)$ multiplying the isovector 
surface term $(\vec \nabla \rho_p-\vec \nabla \rho_n)^2$ can be decomposed as
\begin{equation}
G_\nabla(\rho) = {1\over 4\rho}\, G_\tau(\rho)+G_d(\rho) \,,
\end{equation}
where $G_d(\rho)$ denotes all those contributions for which the $(\vec \nabla 
\rho_p-\vec \nabla \rho_n)^2$ factor originates from the momentum 
dependence of the interactions in an expansion up to order $\vec q^{\,2}$. Performing 
a Fourier transformation converts this factor $\vec q^{\,2}$ into 
$(\vec \nabla k_p-\vec \nabla k_n)^2 \simeq (\vec \nabla \rho_p-\vec \nabla \rho_n)^2 
(\pi/k_f)^4$. The next-to-last term $(\vec \nabla \rho_p- \vec \nabla \rho_n)\cdot
(\vec J_p-\vec J_n)\, G_{so}(\rho)$ in eq.\ (\ref{isoedf}) is the isovector spin-orbit 
interaction in nuclei. Depending on the sign and size of its strength function 
$G_{so}(\rho)$, the spin-orbit potentials for protons and neutrons receive different
weightings from the gradients of the local proton and neutron densities.   

The two- and three-body contributions to the isovector strength functions $\tilde A(\rho), 
G_\tau(\rho), G_d(\rho), G_{so}(\rho)$ and $G_J(\rho)$ follow the same pattern as outlined 
in the previous section for the isoscalar strength functions. The corresponding analytical 
expressions can be found in ref.\ \cite{isofun}. An interesting observation is that the 
Fujita-Mayazawa mechanism for induced spin-orbit couplings from three-body forces is 
not operative in the isovector channel, since the $2\pi$-exchange three-body Hartree diagram 
leads to a vanishing contribution to $G_{so}(\rho)$. 

\begin{figure}
\begin{center}
\includegraphics[scale=0.41,clip]{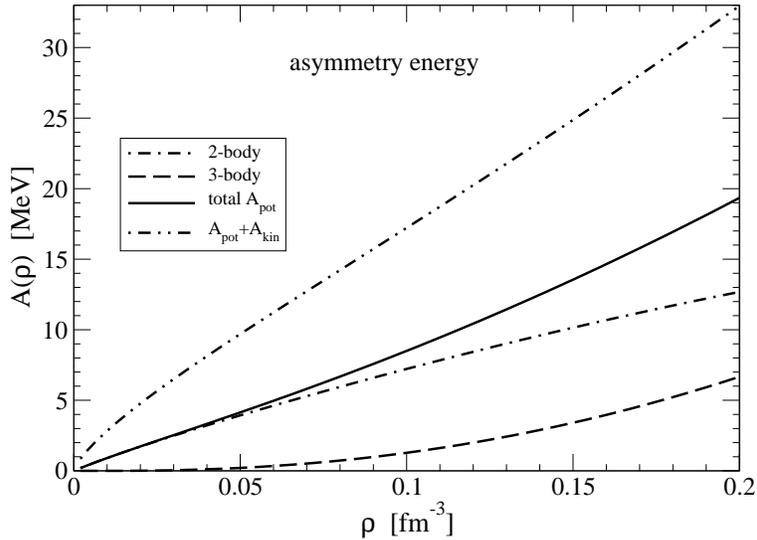}
\end{center}
\vspace{-.9cm}
\caption{Contributions to the asymmetry energy $A(\rho)$ of nuclear matter.}\label{asyme}
\end{figure}

 \begin{figure}
\begin{center}
\includegraphics[scale=0.41,clip]{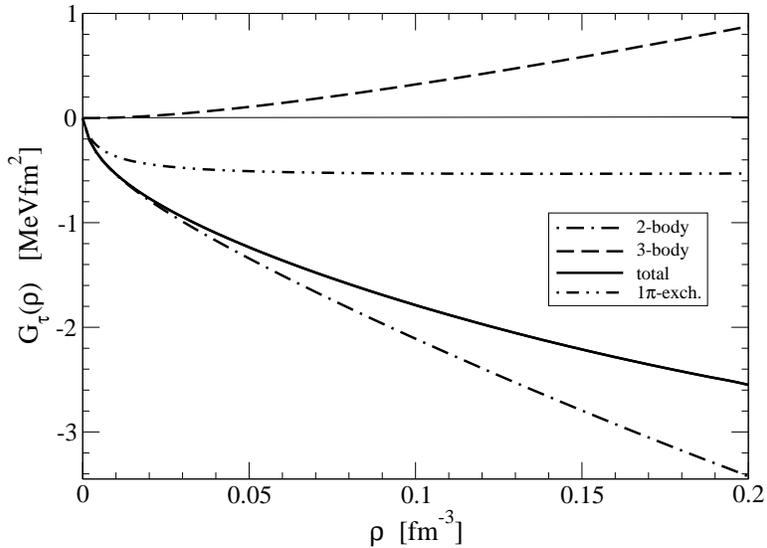}
\end{center}
\vspace{-.9cm}
\caption{Contributions to the strength function $G_\tau(\rho)$ versus the nuclear 
density $\rho$.}\label{gtau}
\end{figure}
           
\begin{figure}
\begin{center}
\includegraphics[scale=0.41,clip]{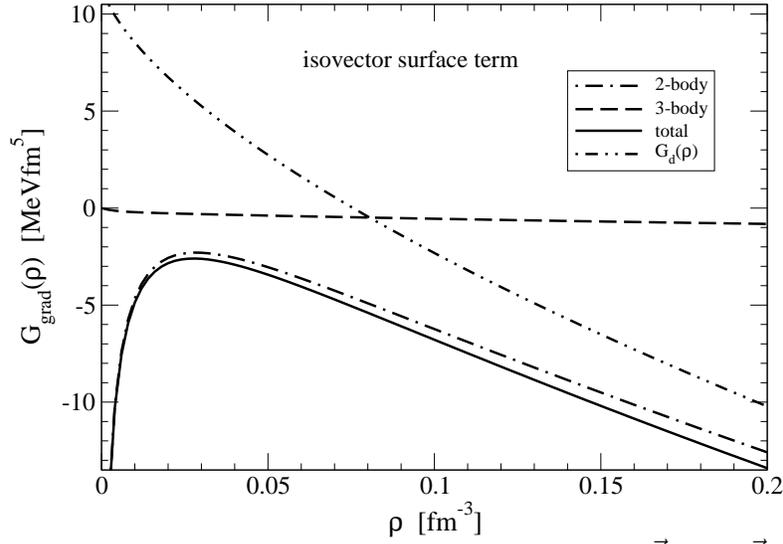}
\end{center}
\vspace{-.9cm}
\caption{Strength function $G_\nabla(\rho)$ of the isovector surface term 
$(\vec \nabla \rho_p-\vec \nabla \rho_n)^2$ versus the nuclear density $\rho$.}\label{ggrad}
\end{figure}  
\begin{figure}
\begin{center}
\includegraphics[scale=0.41,clip]{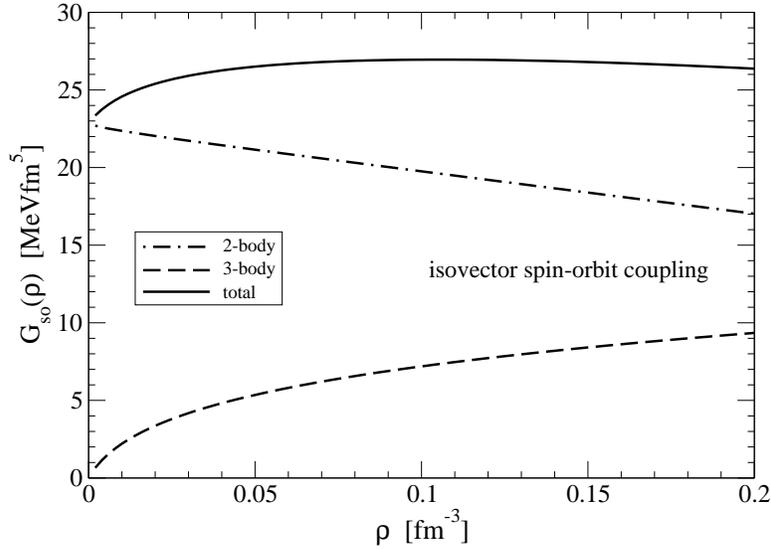}
\end{center}
\vspace{-.9cm}
\caption{Strength function $G_{so}(\rho)$ of the isovector spin-orbit coupling term 
$(\vec \nabla \rho_p-\vec \nabla \rho_n)\cdot (\vec J_p- \vec J_n)$ versus the nuclear 
density $\rho$.}\label{gso}
\end{figure}
\begin{figure}
\begin{center}
\includegraphics[scale=0.38,clip]{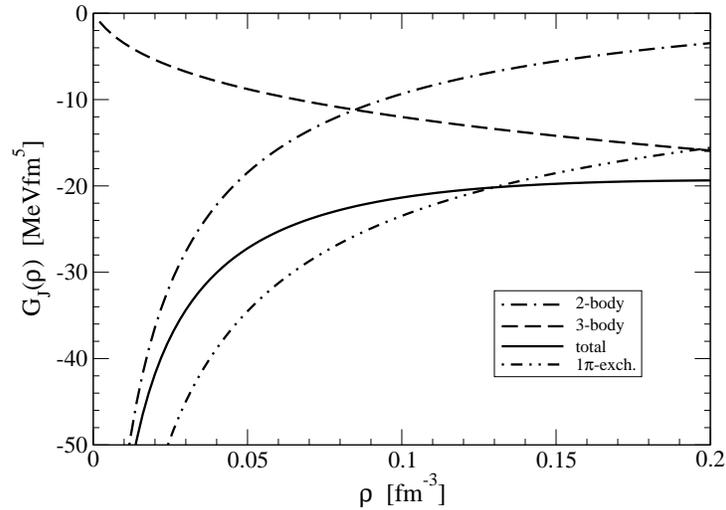}
\end{center}
\vspace{-.9cm}
\caption{Strength function $G_J(\rho)$ multiplying the squared isovector spin-orbit 
density $(\vec J_p-\vec J_n)^2$ versus the nuclear density $\rho$.}\label{gj}
\end{figure}

The results for the isosvector strength functions are shown in 
Figs.\ \ref{asyme}--\ref{gj}. The following features can be observed. 
The Hartree-Fock approximation seems to work better for isovector 
quantities. The asymmetry energy at saturation density comes out as $A(\rho_0) = 26.5\,$MeV, 
which agrees within $20\%$ with the empirical value $(35\pm 2)\,$MeV. In fact, a second-order 
calculation \cite{fermiliq} with two-nucleon forces increased the asymmetry energy by 
approximately $20\%$ and reached the empirical value. The strength functions
$G_\tau(\rho)$ and  $G_\nabla(\rho)$, related to the splitting of the proton and neutron effective
masses and the isosvector surface term $(\vec \nabla \rho_p- \vec \nabla \rho_n)^2$, are typically 
one order of magnitude smaller than their isoscalar analogs $F_\tau(\rho)$ and  $F_\nabla(\rho)$.
Due to the absence of a large three-body term, the isovector spin-orbit coupling strength 
$G_{so}(\rho)$ is now dominated by the short-distance contribution (namely, the low-energy constant 
$C_5/8$) and thus has only a very weak 
density dependence. The quadratic spin-orbit coupling strength $G_J(\rho)$ is again dominated by 
$1\pi$-exchange contribution which induces a strong density dependence. It should be noted that 
within Skyrme phenomenology the isovector part of the energy density functional is 
presently not well determined. For example, no definite choice could be made in 
ref.\ \cite{flocard} between different density dependences ($\sim \rho_p+ \gamma \rho_n, 
\gamma=0,1,2)$ for the neutron spin-orbit potential in Pb isotopes. 

The physics of nuclear structure at large neutron excesses will be explored in the near future 
at rare isotope facilities. The predictions obtained from chiral two- and three-body 
interactions with their definite isospin-dependence can serve as a guideline for the 
theoretical exploration of this field.

\section{Chiral Effective Interactions and Fermi Liquid Theory}
\label{flt}

Landau's theory of normal Fermi liquids \cite{landau57} has long served as a standard 
basis for understanding strongly-interacting Fermi systems at low temperatures. Although 
more than fifty years old, Landau's original insights foreshadowed key ideas in the 
development of effective field theories \cite{polchinski93} and the renormalization 
group \cite{shankar94,chitov95,dupuis96}. In Fermi liquid theory, the relevant degrees 
of freedom associated with low-energy, long-wavelength excitations of a many-body 
Fermi system are quasiparticles interacting in the vicinity of the Fermi surface through 
a residual force. Although the quasiparticle interaction is not necessary weak, 
small perturbations of the system excite relatively few quasiparticles, which then forms 
the basis for an expansion of the theory in terms of the quasiparticle density.

The microscopic foundation for Fermi liquid theory is based on an analysis of the one- 
and two-particle Green's functions \cite{abrikosov59}. In the vicinity of the Fermi surface, 
the one-body Green's function can be decomposed into quasiparticle and background
contributions, and the effects of the background contribution are absorbed into
the effective coupling between quasiparticles. In the original applications of Fermi liquid 
theory to finite nuclei \cite{migdal63,migdal67}, the quasiparticle couplings were adjusted 
to empirical data, resulting in a phenomenological theory capable of linking seemingly 
disconnected phenomena. The perturbative approach to nuclear Fermi liquid theory 
based on microscopic models of the strong nuclear force was initiated by Brown and 
collaborators \cite{brown71} and has in recent years been updated to take advantage
of the improved convergence properties of chiral and low-momentum nuclear two- and
three-body forces \cite{schwenk02,schwenk03,fermiliq,holt12,holt13}.

In the present section we review the Fermi liquid description of nuclear and neutron 
matter in the context of chiral effective field theory. A microscopic approach framed 
in many-body perturbation theory is shown to provide a successful description of many 
nuclear matter observables when the leading-order medium corrections from two- and 
three-body forces are included. The extrapolation to pure neutron matter is considered, 
and applications relevant for neutron star structure and evolution are discussed. We 
focus on the convergence of the perturbative expansion with chiral nuclear forces and 
the role of chiral 
three-nucleon forces. For a more general review of the Fermi liquid approach to nuclear 
many-body systems, we refer the reader to ref.\ \cite{friman12}.

\subsection{\it Symmetric Nuclear Matter}

Numerous properties of isospin-symmetric nuclear matter in the vicinity of the 
saturation density $\rho_0$ are well constrained by empirical data and serve as 
a benchmark for microscopic approaches to nuclear structure. Within the 
framework of Fermi liquid theory, specific nuclear observables are directly 
related to the central components of the quasiparticle interaction, which has
the general form
\begin{equation}
{\cal F}_{\rm cent}({\vec p}_1, {\vec p}_2)=\frac{1}{N_0}\sum_{L=0}^\infty \left [
F_L + F^\prime_L \,\vec \tau_1 \cdot
\vec \tau_2 + (G_L + G^\prime_L \,\vec \tau_1 \cdot \vec \tau_2)\, \vec \sigma_1 
\cdot \vec \sigma_2 \right ]P_L({\rm cos}\, \theta),
\label{ffunction2}
\end{equation}
where $N_0 = 2M^* k_f/\pi^2$ is the density of states at the Fermi surface, and we 
have set $|\vec p_1| = |\vec p_2| =k_f$, which allows one to write the quasiparticle
interaction in terms of $\cos \theta = \hat p_1 \cdot \hat p_2$. The noncentral
parts of the quasiparticle interaction, which are particularly relevant for neutron 
matter, will be discussed in the following section. The relationship between individual 
Fermi liquid parameters and nuclear observables are well known \cite{migdal67}:
\begin{eqnarray}
{\rm Quasiparticle \, \, effective \, \, mass:} \hspace{.2in} \frac{M^*}{M_N} &=& 1+F_1/3, 
\nonumber \\
{\rm Compression \, \, modulus:} \hspace{.2in} {\cal K}&=&\frac{3k_F^2}{M^*} 
\left (1+F_0\right ), \nonumber \\
{\rm Isospin\, \,  asymmetry \, \, energy:} \hspace{.2in} \beta &=& 
\frac{k_F^2}{6M^*}(1+F_0^\prime), \nonumber \\
{\rm Orbital \, \,} g\mbox{-factor:} \hspace{.2in} 
g_l &=& \frac{1+\tau_3}{2} + \frac{F^\prime_1-F_1}{6(1+F_1/3)}\tau_3 , \nonumber \\
{\rm Spin\mbox{-}isospin\, \, response:} \hspace{.2in} g_{NN}^\prime &=& 
\frac{4M_N^2}{g_{\pi N}^2N_0} G_0^\prime.
\label{obs}
\end{eqnarray}

\begin{figure}
\begin{center}
\includegraphics[width=15cm]{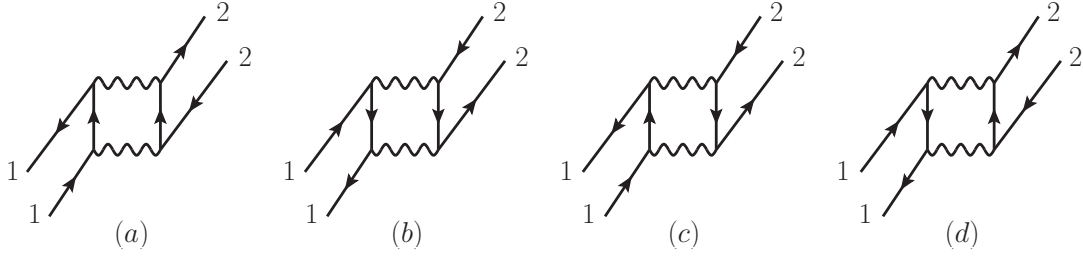}
\end{center}
\vspace{-.5cm}
\caption{Diagrams contributing to the second-order quasiparticle interaction (all 
interactions represented by wavy lines are antisymmetrized): (a) particle-particle diagram, 
(b) hole-hole diagram, and (c)+(d) particle-hole diagrams.}
\label{pphhph}
\end{figure}

Within the framework of many-body perturbation theory, the quasiparticle interaction
is derived by functionally differentiating the total energy density twice with respect
to the particle occupation densities $\delta n_{\vec p, s, t}$:
\be
\delta {\cal E} = \sum_{\vec p s t} \epsilon_{\vec p}\, \delta 
n_{\vec p s t} + \frac{1}{2} \sum_{\stackrel{{\vec p}_1 s_1 t_1}{
{\vec p}_2 s_2 t_2}}{\cal F}({\vec p}_1 s_1 t_1;
{\vec p}_2 s_2 t_2)\, \delta n_{{\vec p}_1 s_1 t_1}
\,\delta n_{{\vec p}_2 s_2 t_2} + \cdots,
\label{deltae}
\ee
where $s_i$ and $t_i$ label the spin and isospin quantum numbers. The first- and
second-order contributions are shown diagrammatically in Fig.\ \ref{pphhph}. As
shown in the first row of Table \ref{n3loc}, the quasiparticle interaction computed
at leading order with chiral two-body forces shares many features with previous 
calculations employing $G$-matrix effective interactions. In particular, the 
quasiparticle effective mass $M^*/M_N = 0.70$ is significantly less than unity,
the nuclear symmetry energy $\beta = 21$\,MeV is too small compared to the empirical 
value $\beta_{emp} \simeq 30-35$\,MeV, and the compression modulus
${\cal K} = -44$\,MeV is negative and implies the instability of nuclear matter to 
density fluctuations. This last feature was quite common in previous calculations 
of the Fermi liquid parameters with $G$-matrix effective interactions
\cite{sjoberg73a,sjoberg73b,backman85} and was remedied with the inclusion
of virtual collective excitations, represented diagrammatically as the 
sum of particle-hole bubble diagrams and encoded in the Babu-Brown induced
interaction \cite{babu73}, which provided sufficient screening in the scalar-isoscalar 
channel to render nuclear matter stable.

\setlength{\tabcolsep}{.07in}
\begin{table}
\begin{center}
\begin{tabular}{|c||c|c|c|c||c|c|c|c|} \hline
\multicolumn{9}{|c|}{Chiral 2N + 3N interactions at $k_F=1.33$ fm$^{-1}$} \\ \hline
 & $f_0$ [fm$^2$] & $g_0$ [fm$^2$] & $f^\prime_0$ [fm$^2$] & $g^\prime_0$ [fm$^2$] & 
$f_1$ [fm$^2$] & $g_1$ [fm$^2$] & $f^\prime_1$ [fm$^2$] & $g^\prime_1$ [fm$^2$] \\ \hline 
$V_{2N}^{(1)}$ & $-$1.274 & 0.298 & 0.200 & 0.955 & $-$1.018 & 0.529 & 0.230 & 0.090   \\ \hline \hline
$V_{2N}^{(2-pp)}$ & $-$1.461 & 0.023 & 0.686 & 0.255 & 0.041 & $-$0.059 & 0.334 & 0.254 \\ \hline
$V_{2N}^{(2-hh)}$ & $-$0.271 & 0.018 & 0.120 & 0.041 & 0.276 & 0.041 & $-$0.144 & $-$0.009 \\ \hline
$V_{2N}^{(2-ph)}$ & 1.642 & $-$0.057 & 0.429 & 0.162 & 0.889 & $-$0.143 & 0.130 & 0.142  \\ \hline \hline
$V_{3N}^{(1)}$ & 1.218 & 0.009 & 0.009 & $-$0.295 & $-$0.073 & $-$0.232 & $-$0.232 & $-$0.179 \\ \hline
\end{tabular}
\caption{Fermi liquid parameters ($L=0,1$) computed at nuclear matter 
saturation density from the Idaho N$^3$LO chiral nucleon-nucleon potential 
as well as from the leading N$^2$LO chiral three-nucleon force.}
\label{n3loc}
\end{center}
\end{table}

Recent chiral effective field theory calculations have systematically studied the
second-order contributions to the quasiparticle interaction, given in rows 3-5 of
Table \ref{n3loc}. The particle-particle term (represented diagrammatically Fig.\ 
\ref{pphhph}(a)) is found to enhance the attraction in the spin- and 
isospin-independent interaction, while the particle-hole terms (Fig.\ \ref{pphhph}(c) 
and (d)) are found to be strongly repulsive. Together these two contributions largely 
cancel in the calculation of the $F_0$ Landau parameter, and since the second-order
particle-hole diagram accounts for much of the full Babu-Brown induced interaction, 
the stabilization mechanism against density fluctuations for chiral nuclear interactions
must be qualitatively different than in the case of the $G$-matrix effective interactions. As
seen in Table \ref{n3loc} both the $f_1$ and $f_0^\prime$ Landau parameters 
receive coherent contributions from all second-order diagrams. This results in a
quasiparticle effective mass at the Fermi surface that is nearly equal to the 
free-space nucleon mass. Taken alone, this would substantially decrease the
isospin asymmetry energy. However, the second-order contributions also play a 
dominant role in enhancing the isospin dependence of the quasiparticle interaction.
At nuclear matter saturation density, the resulting isospin asymmetry energy is 
$\beta = 34$\,MeV, which is in very good agreement with the empirical value.
These results obtained from the Idaho chiral two-nucleon force are qualitatively the
same those from evolved low-momentum NN interactions \cite{fermiliq}.

\begin{figure}[t]
\begin{center}
\includegraphics[height=9cm,angle=270]{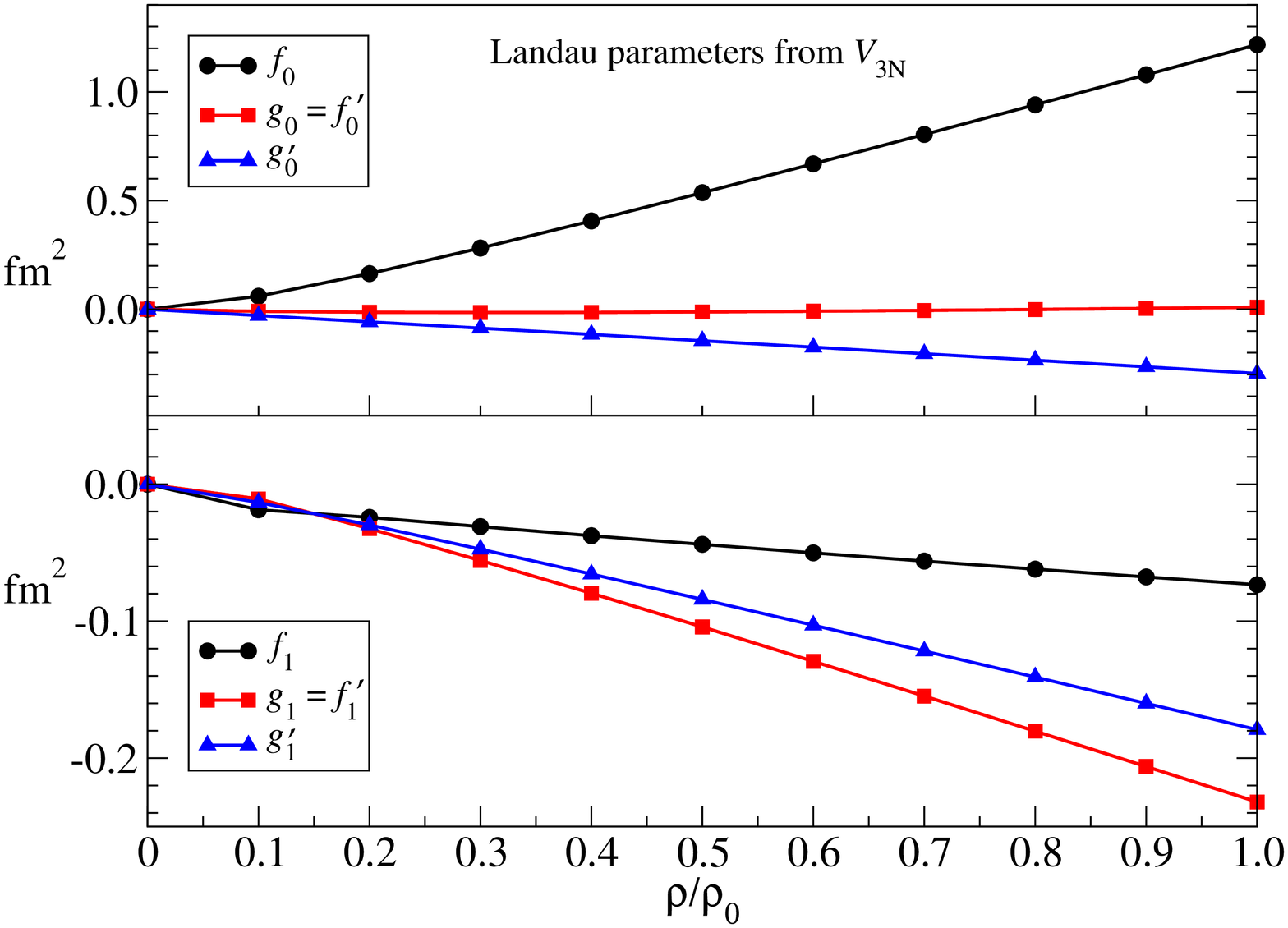}
\includegraphics[height=9cm,angle=270]{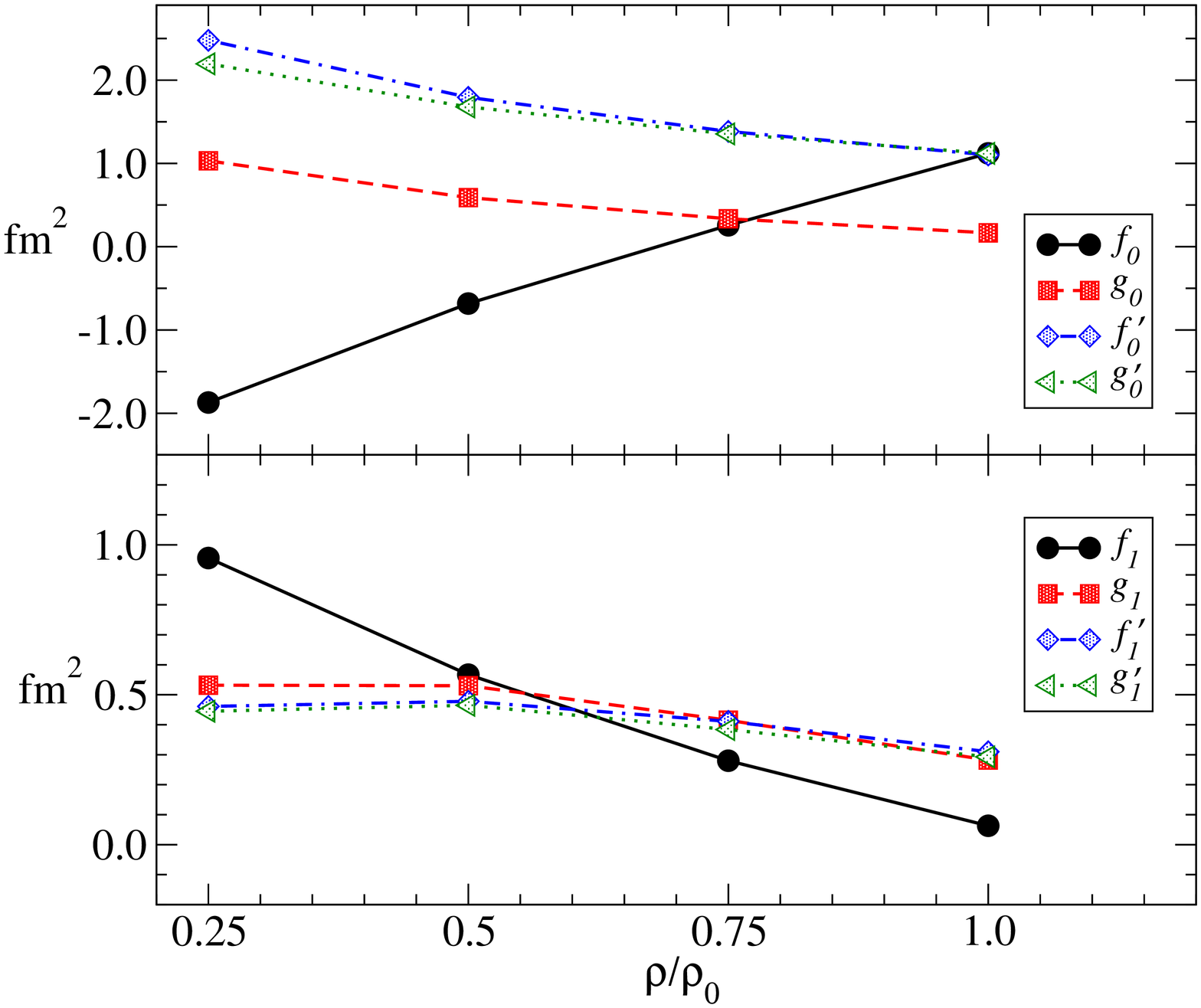}
\end{center}
\vspace{-.5cm}
\caption{Left panel: density-dependent $L=0,1$ Fermi liquid parameters obtained from the 
leading-order chiral three-nucleon force with low-energy constants given in ref.\ \cite{gazit09}. 
Right panel: density-dependent $L=0,1$ Fermi liquid parameters computed from
low-momentum chiral two-nucleon and three-nucleon interactions at the resolution scale 
$\Lambda =2.1$\,fm$^{-1}$.}
\label{DFLPn3lo}
\end{figure}

In the chiral effective field theory description of nuclear interactions, three-body 
forces enter at N$^2$LO in the chiral power counting. The first-order contributions from the 
leading-order three-nucleon force are shown diagrammatically in Fig.\ \ref{mfig1}
and have been computed analytically for the first time in ref.\ \cite{holt12}. The
surprising feature revealed in these calculations is the substantial repulsion 
present in the spin- and isospin-independent quasiparticle interaction resulting
from two-pion exchange dynamics and to a lesser extent the chiral three-nucleon
contact interaction. As shown in the left panel of Fig.\ \ref{DFLPn3lo}, the Landau parameter $f_0$ 
grows strongly with the density and provides more than sufficient repulsion to stabilize 
nuclear matter at the empirical saturation density, resulting in a compression modulus
of ${\cal K} = $\,MeV. This qualitative behavior is foreseen 
already in calculations of the nuclear matter equation of state with chiral and low-momentum
interactions, which require 
three-nucleon forces in order to achieve saturation at the empirical density \cite{vlowkreview}. 
However, the magnitude of $f_0$ relative to the other Landau parameters, as seen in Fig.\ 
\ref{DFLPn3lo} and compiled at nuclear matter density in the last row of Table \ref{n3loc},
is unexpectedly large. The momentum dependence of the quasiparticle interaction in this 
channel is especially weak, producing only a very small decrease in the quasiparticle
effective mass. The scalar-isovector and vector-isoscalar interactions are identical and 
nearly zero when averaged over all angles. Three-nucleon forces therefore play a 
minor role in determining the isospin asymmetry energy of nuclear matter, which was
already well described with two-nucleon forces alone.

In the right panel of Fig.\ \ref{DFLPn3lo} we show the density dependence of the full
quasiparticle interaction for an evolved low-momentum nucleon-nucleon potential
combined with a three-nucleon force whose low-energy constants are
refit at the low-momentum cutoff scale \cite{vlowkreview}. Qualitatively similar results
hold for the unevolved Idaho nucleon-nucleon interaction and associated three-nucleon
force, but the variation in cutoff scale and low-energy constants provides an important 
means for studying the theoretical uncertainties present in the perturbative calculation. 
From Fig.\ \ref{DFLPn3lo} one sees that the density dependence of the spin- and 
isospin-independent quasiparticle interaction is particularly strong. This suggests that 
obtaining a quantitatively correct description of the effective mass and compressibility 
of nuclear matter at the saturation density is nontrivial. In contrast, the remaining
components of the quasiparticle interaction have a weak and strikingly similar density 
dependence. Although most properties of symmetric nuclear matter are well reproduced 
as discussed above, the anomalous orbital $g$-factor $\delta g_l \simeq 0.07 \pm 0.02$
is considerably weaker than the value $\delta g_{l,\rm exp} \simeq 0.23 \pm 0.03$ extracted 
from an experimental analysis of the isovector giant dipole sum rule \cite{brown80}:
\begin{equation}
\int_0^{\omega_{\rm max}} d\omega \, \sigma(E1) = \frac{2\pi \alpha}{M_N} \frac{NZ}{A} (1+2\delta g_l)\, ,
\end{equation}
where $\alpha = 1/137$.
Overall, however, the perturbative calculation including the first-order (free-space)
contribution from two-body forces together with the leading Pauli-blocking and 
core polarization effects from two- and three-body forces provides a successful
description of isospin-symmetric nuclear matter in the vicinity of the empirical
saturation density.



\subsection{\it Neutron Matter}

In contrast to the case of isospin-symmetric nuclear matter, the quasiparticle 
interaction in pure neutron matter is largely unconstrained empirically but is highly 
relevant for neutrino processes in neutron stars \cite{iwamoto82,
navarro99,lykasov08,bacca09} as well as the response of neutron star matter to 
strong magnetic fields \cite{haensel82,olsson04,pethick09,garcia09}. Previous 
calculations with microscopic and phenomenological nuclear potentials have 
given qualitatively different predictions for the magnetic susceptibility of neutron
matter, with some phenomenological potentials even giving rise to a ferromagnetic 
phase transition at several times nuclear matter saturation density \cite{garcia09,
fantoni01,rios05}. Both the magnetic susceptibility of neutron matter as well as 
neutrino absorption, emission, and elastic scattering rates depend on noncentral 
components of the quasiparticle interaction \cite{haensel75,schwenk04} 
as discussed in refs.\ \cite{olsson04,lykasov08,bacca09}.
Recently, these questions have been revisited within the framework of chiral 
effective field theory including three-nucleon forces \cite{holt13}. 

In the long-wavelength limit, the quasiparticle interaction in pure neutron matter
has the general form \cite{schwenk04}:
\begin{eqnarray}
{\cal F}(\vec p_1, \vec p_2\,) &=& f(\vec p_1, \vec p_2\,) + g(\vec p_1, 
\vec p_2\,) \vec \sigma_1 \cdot \vec \sigma_2 + h (\vec p_1, \vec p_2\,) 
S_{12}(\hat q) + k (\vec p_1, \vec p_2\,) S_{12}(\hat P) \nonumber \\ 
&& +l (\vec p_1, \vec p_2\,) (\vec \sigma_1 \times \vec \sigma_2)\cdot 
(\hat q \times \hat P),
\label{qpi}
\end{eqnarray}
where $\vec q = \vec p_1 - \vec p_2$ denotes the momentum transfer in the 
exchange channel, $\vec P = \vec p_1 + \vec p_2$ is the center-of-mass 
momentum of the quasiparticle pair, and $S_{12}(\hat v)$ defines the tensor
operator $S_{12}(\hat v) = 3 \vec \sigma_1 \cdot \hat v\, \vec
\sigma_2 \cdot \hat v -\vec \sigma_1 \cdot\vec \sigma_2$. 
The quasiparticle interaction in eq.\ (\ref{qpi}) is invariant
under parity, time-reversal, and transformations that interchange particle labels. 
The presence of the medium allows for the possibility of terms that break Galilean
invariance and that depend explicitly on the center of mass momentum $\vec P$; namely, 
$S_{12}(\hat P)$ and $A_{12}(\hat q, \hat P) = (\vec \sigma_1 \times \vec \sigma_2)\cdot 
(\hat q \times \hat P)$.
In ref.\ \cite{holt13} a practical method has been developed for extracting the various 
scalar functions from a spin-space decomposition of the quasiparticle interaction. 
Additionally, three-nucleon force contributions to the noncentral quasiparticle interaction
were computed for the first time.

\begin{table}
\begin{center}
\begin{tabular}{|c||c|c|c||c|c|c||c|c|c||}\hline
\multicolumn{1}{|c||}{} & \multicolumn{9}{c||}{Chiral 2NF + 3NF \hspace{.3in}($k_f=1.7$\,fm$^{-1}$)} \\ \hline
\multicolumn{1}{|c||}{$L$} & \multicolumn{3}{c||}{0} & \multicolumn{3}{c||}{1} & 
\multicolumn{3}{c||}{2} \\ \hline 
\multicolumn{1}{|c||}{} & \multicolumn{1}{c|}{$V_{2N}^{(1)}$} & \multicolumn{1}{c|}{$V_{2N}^{(2)}$} & 
\multicolumn{1}{c||}{$V_{3N}^{(1)}$} & \multicolumn{1}{c|}{$V_{2N}^{(1)}$} & 
\multicolumn{1}{c|}{$V_{2N}^{(2)}$} & \multicolumn{1}{c||}{$V_{3N}^{(1)}$} &
\multicolumn{1}{c|}{$V_{2N}^{(1)}$} & \multicolumn{1}{c|}{$V_{2N}^{(2)}$} & 
\multicolumn{1}{c||}{$V_{3N}^{(1)}$} \\ \hline 
$f$ [fm$^2$] & $-$0.700 & 0.069 & 1.319 & $-$1.025 & 1.197 & $-$0.037 & $-$0.230 & $-$0.500 & $-$0.293  \\ \hline 
$g$ [fm$^2$] & 1.053 & 0.293 & $-$0.283 & 0.613 & 0.159 & $-$0.364 &  0.337 & $-$0.089 & 0.043  \\ \hline 
$h$ [fm$^2$] & 0.270 & $-$0.212 & 0.075 & 0.060 & 0.106 & 0.164 & $-$0.040 & $-$0.143 & $-$0.087  \\ \hline
$k$ [fm$^2$] & 0 & $-$0.156 & 0 & 0 & 0.085 & 0 & 0 & 0.063 & 0  \\ \hline
$l$ [fm$^2$] & 0 & 0.135 & $-$0.168 & 0 & $-$0.031 & $-$0.134 & 0 & $-$0.279 & 0.083  \\ \hline
\end{tabular}
\caption{Fermi liquid parameters for the quasiparticle interaction in
neutron matter at a density corresponding to a Fermi momentum of $k_f=1.7$\,fm$^{-1}$. 
The low-energy constants of the N$^2$LO chiral three-nucleon force are chosen to
be $c_1 =-0.81\,$GeV$^{-1}$ and $c_3=-3.2\,$GeV$^{-1}$.}
\label{flpkf}
\end{center}
\end{table}

The central quasiparticle interaction in neutron matter is
qualitatively similar to that described in the previous section for isospin-symmetric nuclear 
matter (with $\vec \tau_1 \cdot \vec \tau_2 \rightarrow 1$). The individual contributions from 
two- and three-nucleon forces to the $L=0,1,2$ Landau parameters at various orders 
in perturbation theory for $k_f = 1.7$\,fm$^{-1}$ are shown in Table \ref{flpkf}. The first-order 
contribution from two-nucleon forces is simply a kinematically constrained version of the 
free-space interaction and therefore must be independent of the total quasiparticle momentum 
$\vec P$. Second-order contributions give rise to all possible noncentral interactions, which 
from Table \ref{flpkf} are seen to be significantly smaller than the central components. 
The exchange tensor interaction present in the free-space nucleon-nucleon interaction is 
largely reduced by Pauli-blocking effects in the nuclear medium. 
The center-of-mass tensor interaction arises solely from Pauli-blocking and core polarization 
effects at second-order in perturbation theory with two-body forces, while the novel
spin-nonconserving cross-vector interaction $A_{12}(\hat q, \hat P)$ is generated both from
two- and three-body forces, whose momentum-averaged contributions are of comparable strength.

\begin{figure}
\begin{center}
\includegraphics[width=9.5cm,angle=270]{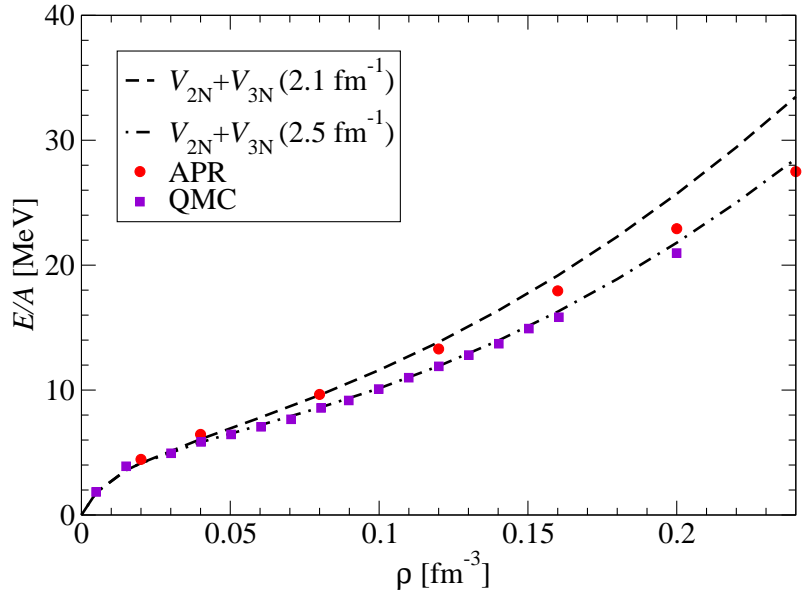}
\end{center}
\vspace{-.5cm}
\caption{Energy per particle of neutron matter from both chiral and low-momentum
two- and three-body interactions. The cutoff scale associated with the low-momentum
interaction is $\Lambda = 2.1$\,fm$^{-1}$. The curve labeled `APR' is taken from the
variational calculations of Akmal {\it et al.} \cite{akmal}, and the curve 
labeled `QMC' is taken from the quantum Monte Carlo calculations in refs.\
\cite{armani11,gandolfi}.}  
\label{ea2n3n}
\end{figure}

\begin{figure}
\begin{center}
\includegraphics[width=10cm,angle=270]{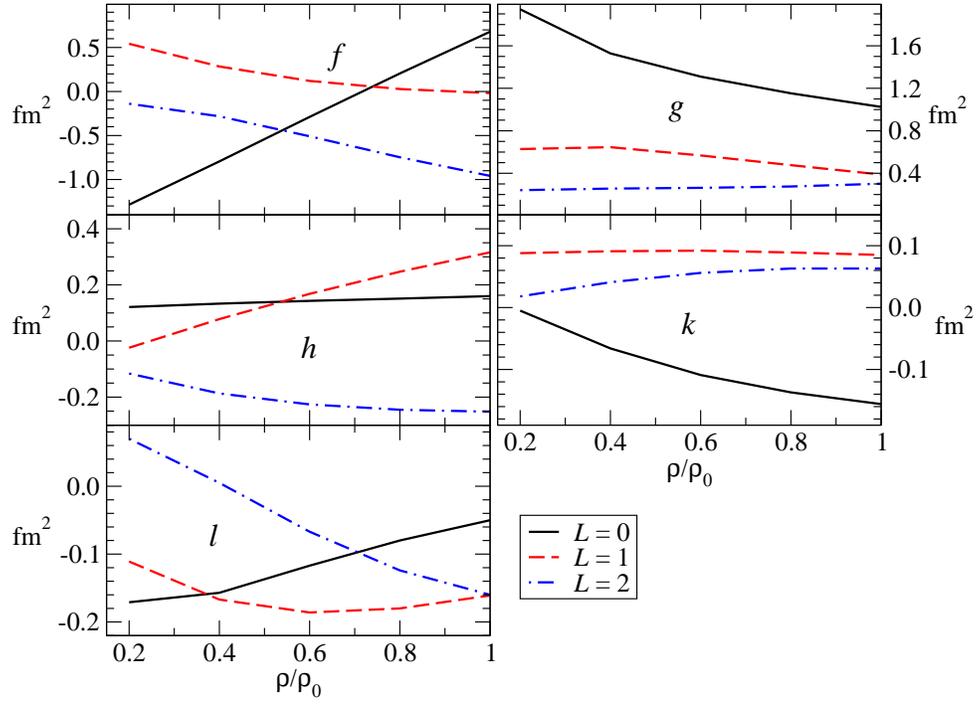}
\end{center}
\vspace{-.5cm}
\caption{(Color online) Density-dependent Fermi liquid parameters inlcuding first- and second-order
contributions from the chiral N$^3$LO nucleon-nucleon potential of ref.\ \cite{entem03}
as well as the N$^2$LO chiral three-nucleon force to leading order.}
\label{ddflp}
\end{figure}

At leading order the compressibility ${\cal K}$ of neutron matter at $\rho_0$ is
unphysically small in comparison to auxiliary-field diffusion Monte Carlo 
calculations with realistic two- and three-nucleon forces which find that at 
$\rho=0.16$\,fm$^{-3}$, ${\cal K} \simeq 520$\,MeV \cite{fantoni01}. Second-order
contributions from chiral two-body forces do little to change this picture, but as in the case 
of symmetric nuclear
matter the leading-order chiral three-nucleon force provides substantial additional
repulsion, yielding a compression modulus ${\cal K} = 550$\,MeV. The $f_1$ Landau
parameter governing the quasiparticle effective mass at the Fermi surface again receives
a nearly negligible contribution from three-body forces, but strong polarization effects at 
second order increases the effective mass from $M^*/M_N = 0.82$ to $M^*/M_N = 1.04$. 
As seen in Fig.\ \ref{ddflp} the Landau parameters $f_L$ associated with the 
spin-independent quasiparticle interaction depend sensitively on the neutron density.

Neglecting effects from noncentral interactions, the magnetic susceptibility is determined
by the Landau parameter $G_0$:
\be
\chi = \mu_n^2 \frac{N_0}{1+G_0},
\label{susc}
\ee
where $\mu_n=-1.913$ is the free-space neutron magnetic moment in units of 
the nuclear magneton, and $G_0 = N_0 g_0$. Noncentral components in the 
quasiparticle interaction result in effective charges (magnetic moments) that are not
scalars under rotations of the quasiparticle momentum. The spin susceptibility then 
involves separately the longitudinal and transverse components of the magnetic 
moment \cite{haensel82,olsson04}. The Landau parameter $g_0$ decreases with the 
nuclear density as shown in Fig.\ \ref{ddflp}, but there is no evidence for a spin-instability 
in the vicinity of nuclear matter saturation density.

As a consistency check on the perturbative calculation of the quasiparticle interaction 
in neutron matter, we reproduce in Fig.\ \ref{ea2n3n} the neutron matter equation of state 
at zero temperature computed to the same order in many-body perturbation theory.  
The results agree well with variational and Monte Carlo calculations of 
neutron matter employing realistic and phenomenological nuclear forces 
\cite{akmal,armani11,gandolfi}. The recent neutron matter calculation 
\cite{tews12} including the N$^3$LO chiral three- and four-neutron forces gives 
results that are very similar to those shown in Fig.\ \ref{ea2n3n}. The dependence on the
resolution scale arises mostly from the different choice of the low-energy constant
$c_3$ in the two calculations. Recently, it has been shown that refitting $c_3$ to 
peripheral partial waves with the running of the cutoff scale yields a much reduced
regulator dependence of the neutron matter equation of state \cite{coraggio13}.

\begin{figure}
\begin{center}
\includegraphics[scale=0.6,clip]{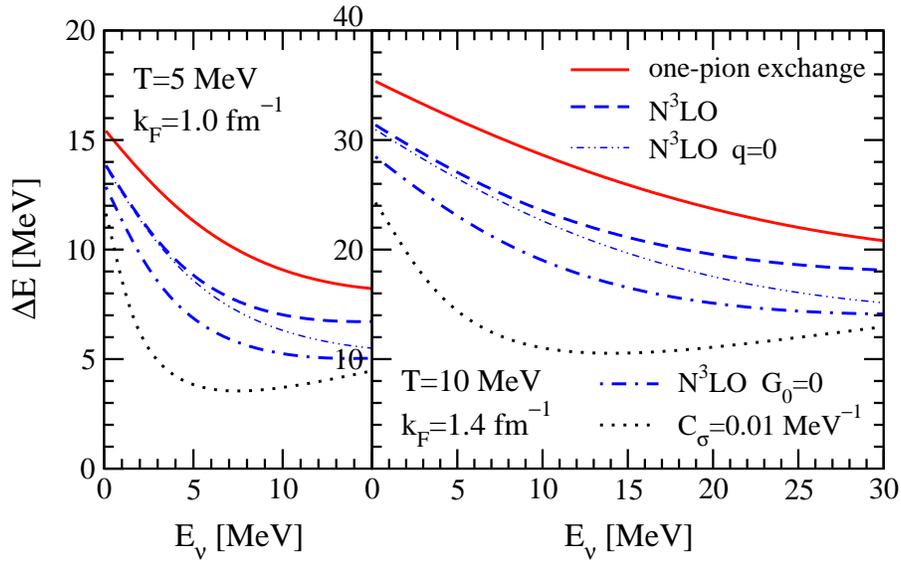}
\end{center}
\vspace{-.5cm}
\caption{The rms energy transfer in neutrino scattering from neutron matter as a function
of the temperature, density, and neutrino energy. Results are computed with chiral two-body
forces at different orders in the chiral power counting. Figure 
reproduced from ref.\ \cite{bacca09}.
}
\label{etrans}
\end{figure}

The dynamics of supernovae explosions are particularly sensitive to neutrino processes
occurring in the neutrinosphere, a warm and low-density ($n \simeq \rho_0/10$) gas of 
neutron-rich matter. At such small densities, three-nucleon correlations are expected to be 
negligible, and an accurate understanding of
neutrino scattering, production, and annihilation rates is accessible from two-nucleon forces 
alone. These processes determine the supernova neutrino spectrum and may be important
for heating effects that re-energize the shock wave. The typical scale for 
neutrino energies $\epsilon_\nu \simeq 30$\,MeV and 
momenta $p_\nu \simeq 0.15$\,fm$^{-1}$ justifies a treatment within linear response theory. 
Supernova simulations generally include neutrino-pair bremsstrahlung and absorption rates 
computed in the one-pion exchange approximation. For degenerate conditions, where the 
temperature is small relative to the neutron Fermi energy, Landau's Fermi liquid theory is 
applicable has been used in recent work \cite{bacca09} to compute spin-spin response function
of neutron matter employing chiral two-body interactions at N$^3$LO.

Within Fermi liquid theory, the density-density and spin-spin response functions are computed 
by solving the transport equation for quasiparticles. The relaxation rate that is used to 
approximate the collision integral in the Landau transport equation involves contributions from 
central and noncentral components of the quasiparticle interaction, the latter being essential
for describing two-nucleon response. Contributions from chiral nuclear interactions beyond 
the leading-order one-pion exchange potential substantially reduce the relaxation rate, 
which consequently reduces the mean-square energy loss in neutrino scattering, as shown in 
Fig.\ \ref{etrans}.

\section{Nuclear Chiral Thermodynamics}
\subsection{\it Nuclear Phase Diagram and Liquid-Gas Transition}

The thermodynamic properties of nuclear and neutron matter play an important role in 
applications to heavy-ion collisions and astrophysics. The plateau observed in the caloric 
curve of the nuclear fragments in nucleus-nucleus collisions can be viewed as the trace of a 
first-order liquid-gas phase transition \cite{gross,pocho,natowitz2}. In the context of nuclear 
astrophysics, the recent observation of a two-solar-mass neutron star \cite{demorest} strongly
constrains the equation of state of hadronic matter and rules out various exotic models which 
tend to produce softer equations of state. 

Previous chapters have introduced and developed the framework of chiral effective field theory and 
its application to the nuclear many-body problem at zero temperature. In the present section we 
review progress that has been achieved in extending this framework to finite temperatures. The previous
ordering scheme for the nuclear energy density in terms of the number of medium insertions is now 
generalized and applied to the free energy density as a function of density $\rho$ and temperature $T$. 
Introducing the free energy per nucleon, 
$\bar{F}(\rho, T)$, the free energy density can be written as the sum of convolution integrals:
\begin{eqnarray}  \label{convolution}
	\rho \, \bar{F}(\rho, T) &=& 4 \int\limits_0^\infty dp \, p \,\mathcal{K}_1(p) \, n(p) 
+ \int\limits_0^\infty dp_1 \int\limits_0^\infty dp_2 \, \mathcal{K}_2(p_1,p_2) \, n(p_1) \, n(p_2) 
\nonumber \\ && + \int\limits_0^\infty dp_1 \int\limits_0^\infty dp_2 \int\limits_0^\infty dp_3 \, 
\mathcal{K}_3(p_1,p_2,p_3) \, n(p_1) \, n(p_2) \, n(p_3) + \rho \, \bar \mathcal{A}(\rho,T) \ ,
\end{eqnarray}        
where
\begin{equation}\label{density}
	n(p) = {p \over 2\pi^2} \bigg[ 1 + \exp{\frac{p^2 /2 M_N -\tilde{\mu}}{T}} \bigg]^{-1}
\end{equation}
is the density of nucleon states in momentum space. It is the product of the temperature-
dependent Fermi-Dirac distribution and a kinematical prefactor $p/2\pi^2$ that has been 
included in $n(p)$ for convenience. Complete expressions of the kernels 
$ \mathcal{K}_j $ can be found in \cite{FKW2012}. The term $ \bar\mathcal{A}(\rho,T)$ is called 
the anomalous contribution. It is associated with a Fock term involving second order 
$ 1\pi$-exchange \cite{nucmatt} and arises from the smoothing of the Fermi 
surface at nonzero temperature. In the range of temperatures considered here the effect of
the anomalous term is small. The one-body effective ``chemical potential''  $ \tilde{\mu} $ 
in the distribution $ n(p)$ is related to the density by
\begin{equation}
	\rho = 4 \int\limits_0^\infty dp \, p \, n(p) \,.
\end{equation}
The kernel $ \mathcal{K}_1$ is the non-interacting nucleon gas contribution, while the 
kernels $ \mathcal{K}_2 $ and $ \mathcal{K}_3 $ encode the effects of the interactions 
among two and three nucleons. The pressure is then computed from the standard thermodynamic 
relation
\begin{equation} \label{pressure}
P(\rho, T) = \rho^2 \, \frac{\partial \bar{F}(\rho, T)}{\partial \rho} \, . 
\end{equation}

In addition to isospin-symmetric nuclear matter, the extension to asymmetric nuclear matter and 
neutron matter has also been explored. The only changes required involve isospin factors and the introduction 
of separate proton and neutron thermal occupation probabilities $d_p(p_j)$ and $d_n(p_j)$.
Note that for pure neutron matter the Pauli principle forbids a (momentum independent) 
three-body contact interaction. The behavior at very low density is governed by the 
large neutron-neutron scattering length, $ a_{nn} \simeq 19 $ fm \cite{chen}, and as discussed in Section
\ref{rsnm} an all-order resummation is required for a realistic 
description at low densities ($ \rho_n < 0.02\,$fm$^{-3}$).

In this section we first concentrate on the equation of  state of isospin-symmetric nuclear matter for 
different temperatures, featuring the liquid-gas phase transition. The calculations are then
extended to isospin-asymmetric matter, focusing on the change of 
thermodynamic properties with varying proton-neutron asymmetry. The topics of interest are
the asymmetry free energy, its dependence on density and temperature  and the validity of 
the (empirically observed) parabolic approximation for the free energy as a function of the 
asymmetry parameter, $ \delta = (\rho_n-\rho_p)/ (\rho_n+\rho_p) $.  

\begin{figure}[tbp] 
 \center
 \includegraphics[scale=1.0,clip]{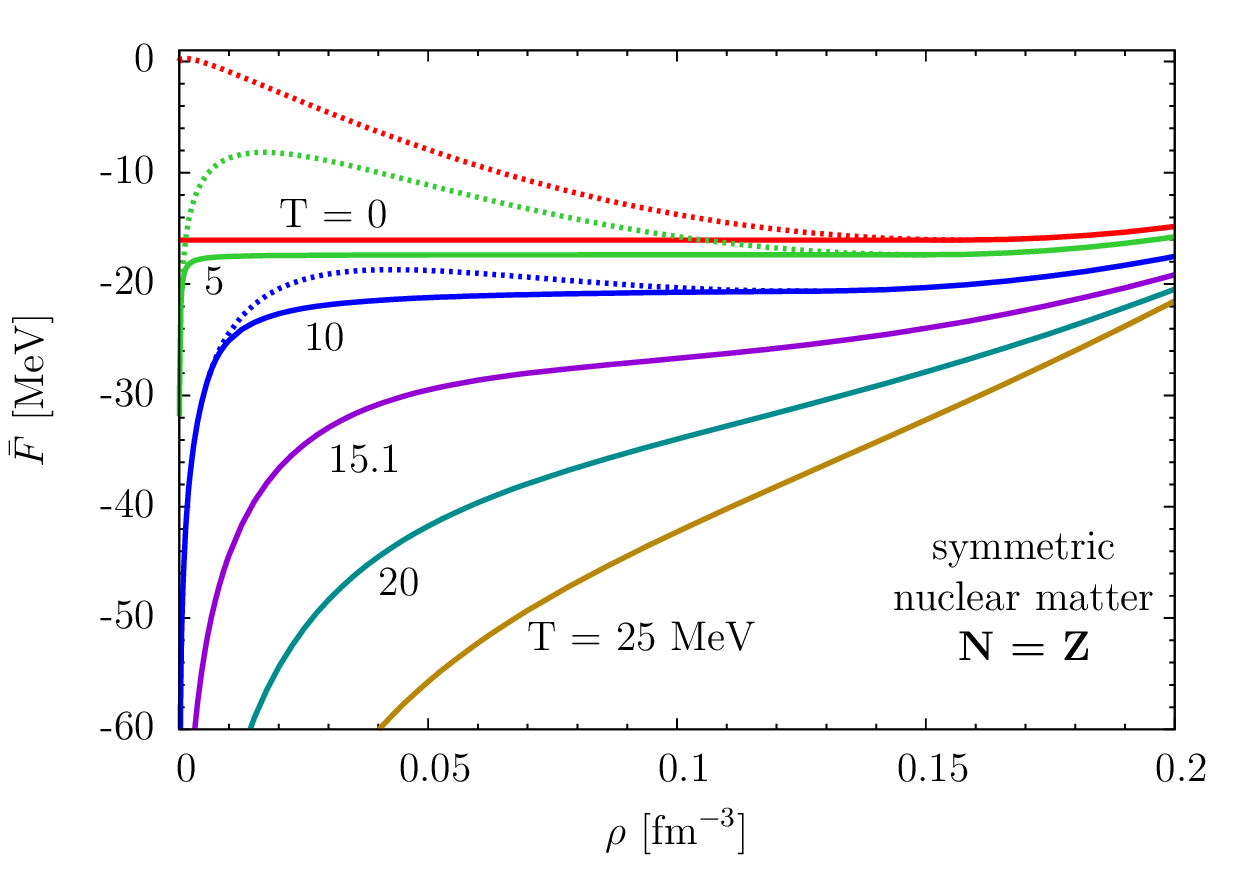}     
\caption{Free energy per particle of isospin-symmetric nuclear matter as a function of the
temperature and baryon density $\rho$. The dotted line denotes the non-physical 
part of the free energy in the liquid-gas coexistence region. The physical free energy (solid line) 
is obtained using the Maxwell construction.}\label{fenergy50}
\end{figure}

The free energy per particle $ \bar{F} (\rho, T) $ is calculated using eq.\ (\ref{convolution})
as a function of density $ \rho $ for a sequence of temperatures up to 25 MeV, with the 
input for the interaction kernels $ \mathcal{K}_n $ specified in ref.\ \cite{FKW2012}. The result 
is shown in Fig.\ \ref{fenergy50}. The dotted lines indicate the non-physical behavior of 
the equation of state in the first-order liquid-gas transition region. This unphysical part 
is then replaced by the physical one (solid lines) obtained from the Maxwell construction. 
At zero temperature the free energy is equal to the internal energy of the system, which has
the well-known saturation point located at $ \bar{E}_0 = -16.0 $ MeV and $ \rho_0 = 0.16\,$ 
fm$^{-3}$, corresponding to a Fermi momentum of $ k_{f0} = 1.33\ \mbox{fm}^{-1} = 262 $ MeV. 
At finite temperatures the free energy develops a singular behavior for $ \rho \rightarrow 0 $,
a feature that is present in numerous other many-body calculations \cite{akmal,horowitz}.

\begin{figure}[tbp]
\center
 \includegraphics[scale=1.0,clip]{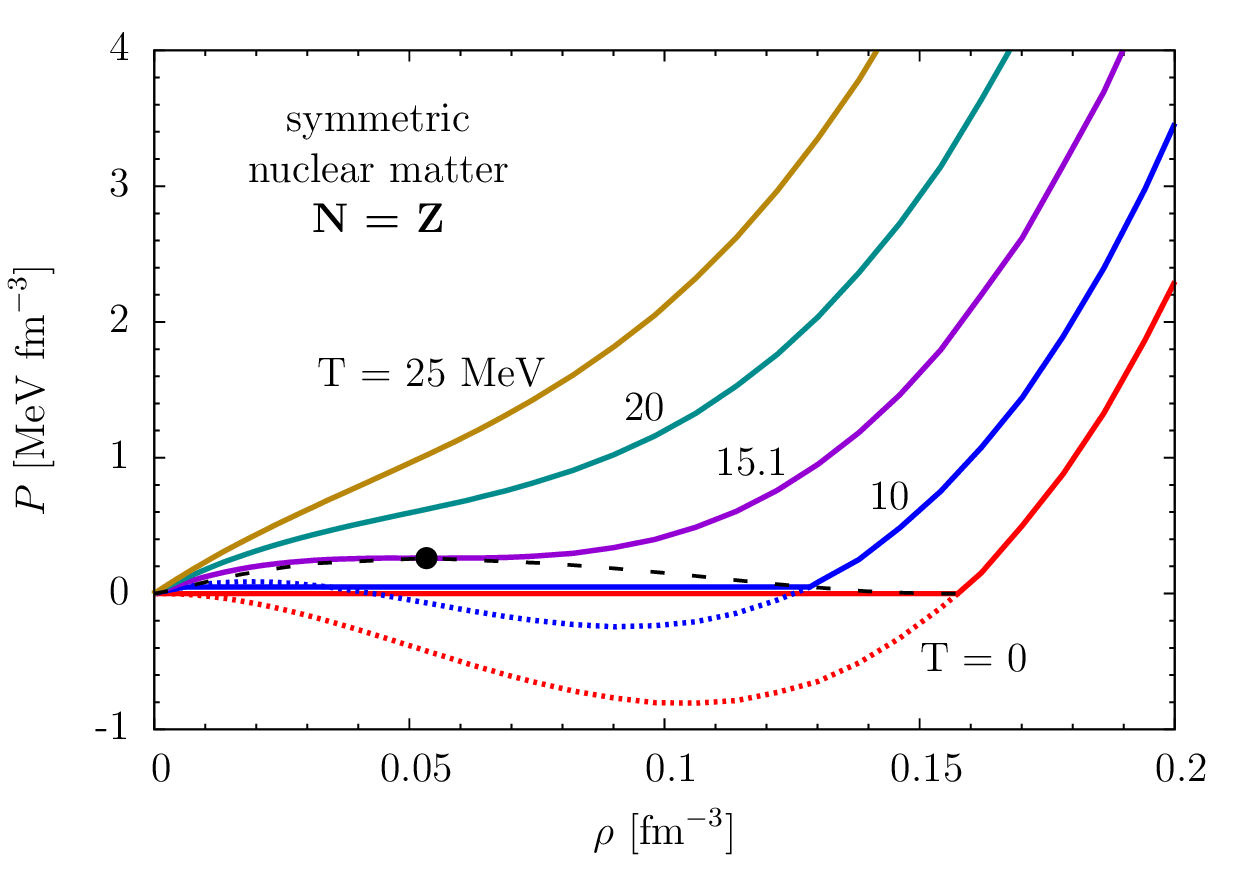}  
 \caption{Pressure isotherms as a function of density and temperature for 
 isospin-symmetric nuclear matter displaying 
a first-order liquid-gas phase transition. The dotted lines at low temperature indicate the 
non-physical behavior of the isotherms in the transition region, and the physical pressure is 
calculated from the Maxwell construction. The dashed line delineates the boundary of the 
coexistence region, and the dot denotes the critical point at $ T_c \simeq 15.1 $ MeV and 
$ \rho_c \simeq \rho_0/3 $.} 
 \label{eos50}
 \end{figure}

In Fig.\ \ref{eos50} we show isotherms of the pressure $P(\rho, T)$ as a function of the temperature
and density. The picture that emerges is reminiscent of a Van der Waals gas with its 
generic first-order liquid-gas phase transition. As outlined in Section 4.3, 
chiral nucleon-nucleon dynamics generates intermediate-range attractive interactions
(e.g., from $2\pi$-exchange with intermediate $\Delta$-excitations) that exhibit a characteristic 
$ e^{-2\,m_\pi r} / r^6 $ behavior at distances on the order of $1-2$\,fm (see ref.\ \cite{gerstendorfer}). 
Such mechanisms provide nearly half of the attraction required to bind nuclear matter at zero
temperature. The 
liquid-gas phase transition then results from a sensitive balance between intermediate 
range attraction and short-range repulsion, the latter encoded in contact terms representing 
dynamics unresolved at the scales relevant for typical nuclear Fermi momenta $k_f$ considered
here.

The critical temperature for the liquid-gas phase transition is observed at $T_c \simeq 15.1\,$ 
MeV. For temperatures less than the critical temperature $T_c $, the usual Maxwell construction is 
applied, where the pressure is kept constant in the liquid-gas coexistence region. 
Empirical estimates of the critical temperature are deduced from fission and multi-fragmentation 
measurements, which indicate a value of $T_c \simeq 15-20$\,MeV \cite{karnaukhov}. Similar values
for the critical temperature are obtained from calculations \cite{sauer} employing phenomenological 
Skyrme interactions.

\begin{figure}[htbp]
\center
\includegraphics[scale=1.0,clip]{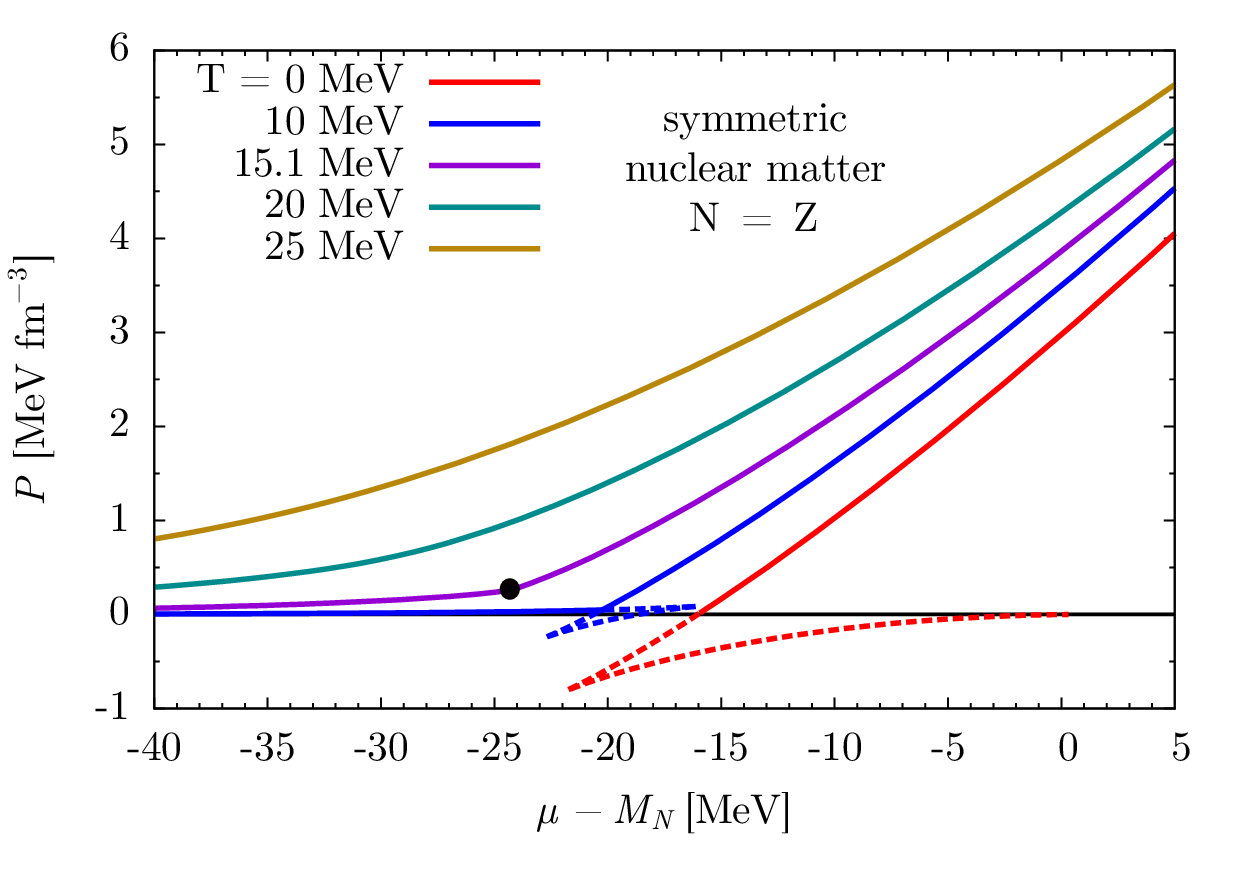}  
 \caption{Dependence of the pressure isotherms on the nucleon chemical potential $\mu$ for 
symmetric nuclear matter. The non-physical behavior of the 
equation of state in the liquid-gas coexistence region for temperatures below $T_c$ is denoted with 
the dotted lines. The physical pressure and chemical potential in the coexistence region are constant 
and determined from the Maxwell construction. The critical point is denoted with the large dot.} 
 \label{pchem50}
 \end{figure}

In Fig.\ \ref{pchem50} we show the dependence of the pressure on the nucleon chemical potential (including the free nucleon mass),
\begin{equation}
\mu = M_N + \left( 1+\rho\, \frac{\partial}{\partial \rho} \right) \bar{F}(\rho, T)  \,.
\end{equation}
The non-physical part of the pressure is shown as the dotted curves in Fig.\ \ref{eos50} and manifested 
in the double-valued behavior of $P$ at temperatures below the critical temperature. In the 
coexistence region, the physical pressure and chemical potential are 
constant and given according to the 
Maxwell construction. The temperature at which the pressure becomes single-valued is the critical 
temperature $T_c$.

\begin{figure}[htbp]
\center
\includegraphics[scale=0.5,clip]{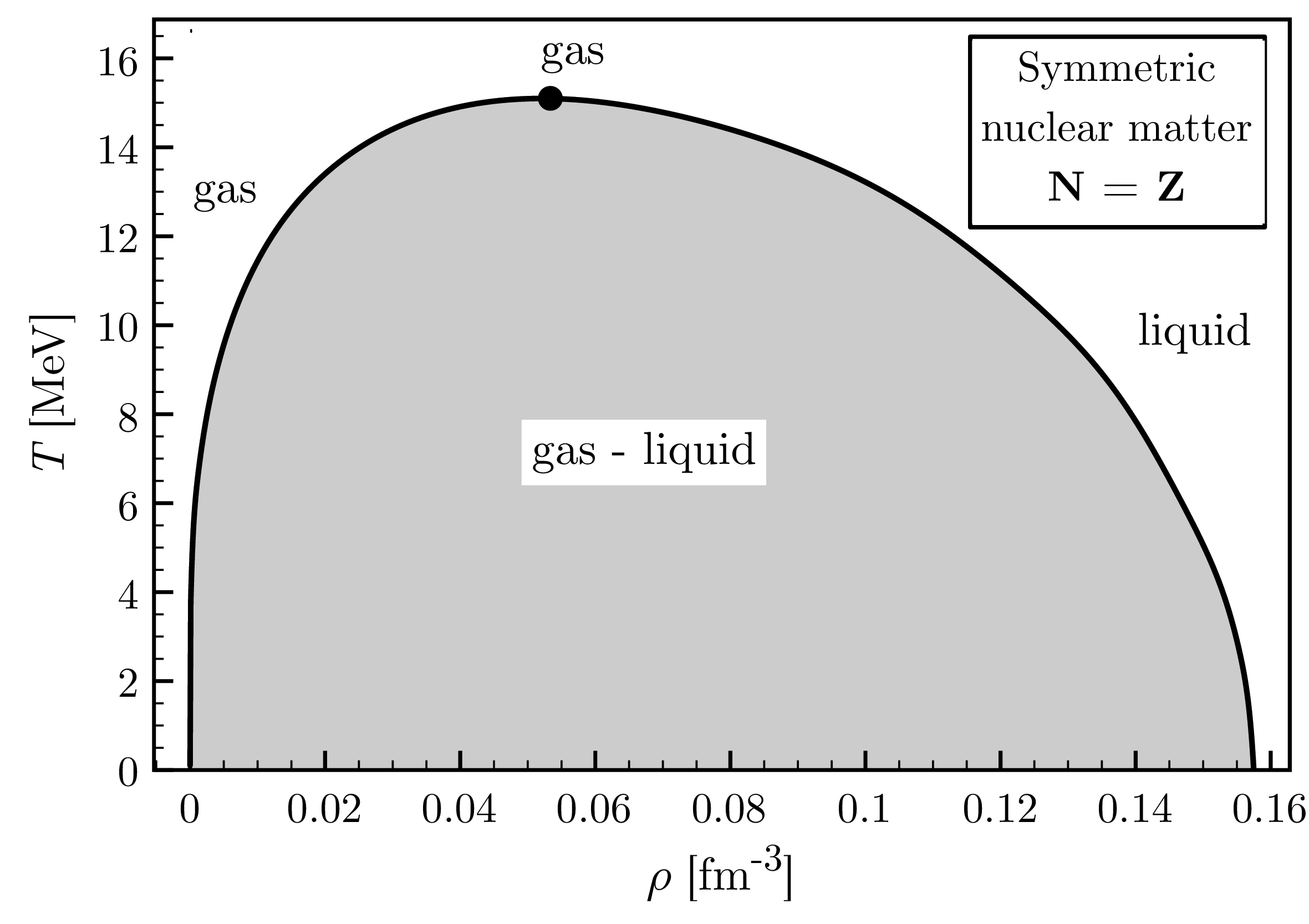} 
\vspace{-.1in}
 \caption{The phase diagram of isospin-symmetric nuclear matter. The critical point is indicated by
 the dot on the boundary of the phase coexistence region.} 
 \label{phasdiag}
 \end{figure}

The $T-\rho$ phase diagram, shown in Fig.\ \ref{phasdiag}, collects the relevant information regarding
the liquid-gas phase transition. The first-order transition region ends at the 
critical point ($T_c=15.1$\,MeV) and is denoted with the dot. The associated critical values of the 
pressure, baryon chemical potential, and density are found to be $ P_c \simeq 0.261\,$ MeV\,fm$^{-3}$, $ \mu_c \simeq 914.7 $ MeV, and $ \rho_c \simeq 0.053\, {\rm fm}^{-3}$. The $T-\rho $ phase diagram
indicates that the liquid-gas coexistence region extends over a wide range of densities
up until the Fermi liquid phase is realized at $\rho_0 = 0.16$\,fm$^{-3}$, the equilibrium density of 
nuclear matter.

At zero temperature the third law of thermodynamics provides important constraints on features of
the $T-\rho$ diagram. In particular, at $T=0$ the boundary of the phase coexistence
region has an infinite slope in the $ T - \rho $ diagram. Additionally, the 
chemical potential at zero temperature is given by the total energy per particle at 
the saturation point, that is, $ \mu = M_N + \bar{E}_0 \simeq 923 $ MeV.

\subsection{\it Isospin-Asymmetric Nuclear Matter} 

It is now of interest to investigate how the liquid and gas phases characteristic of 
isospin-symmetric ($N = Z$) nuclear matter evolve with changing proton fraction, 
$x_p = Z/A$. The properties 
of isospin-asymmetric matter with $0.5 > x_p > 0$ are governed by the detailed isospin
dependence of the chiral NN interaction. It is almost entirely controlled by the isospin 
dependence of one- and two-pion exchange processes once the relevant nucleon-nucleon
contact terms are fit to the empirical isospin asymmetry energy $A(\rho_0,T=0) \simeq 34$\,MeV. 

\begin{figure}[tbp]
\center
\includegraphics[scale=1.0,clip]{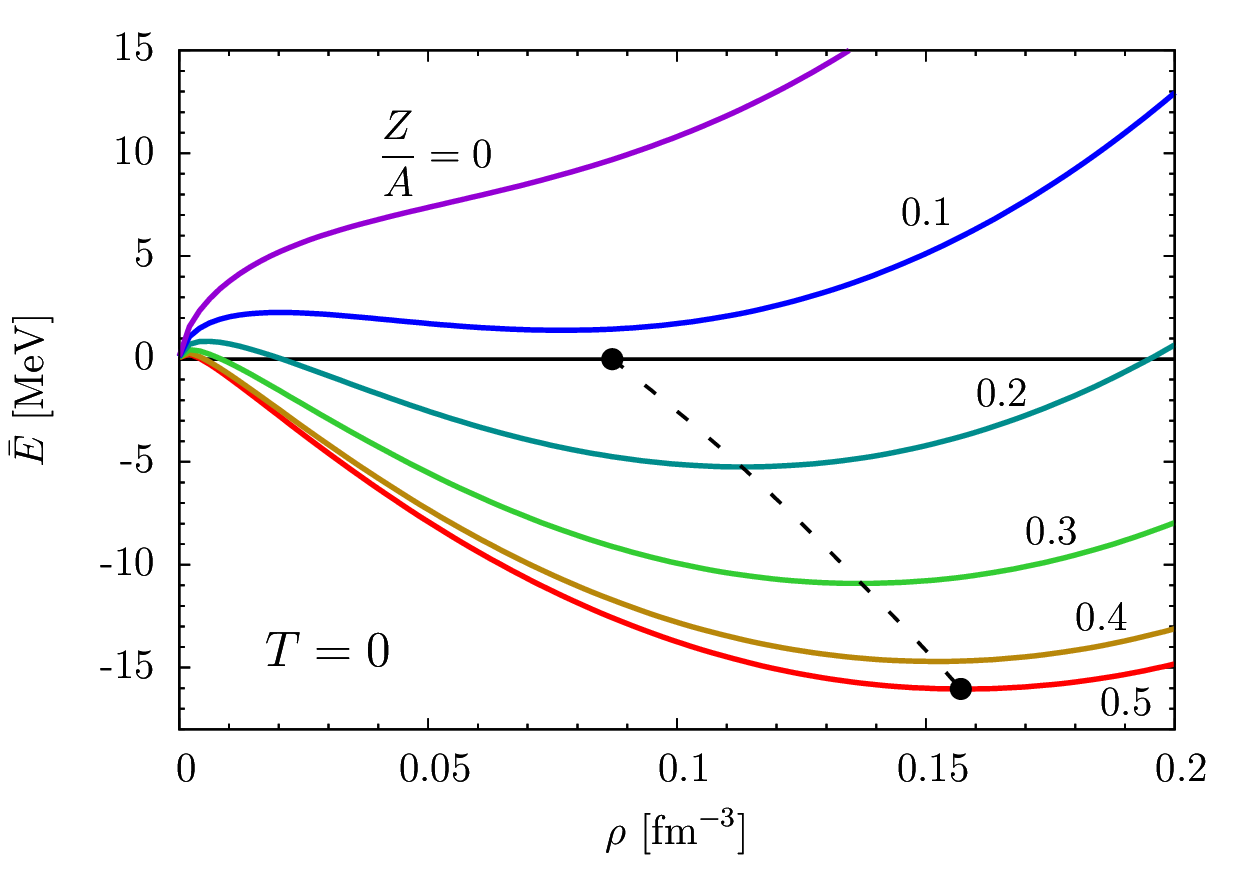} 
\caption{Dependence of the energy per nucleon (and saturation point) of nuclear matter 
on the proton fraction $x_p$ at $T=0$. The dashed line denotes the evolution of the 
saturation point with varying proton fraction. For proton fractions below $ x_p = 0.12$,
the energy remains positive at all densities.}
\label{satpoint}
\end{figure}

Consider the evolution of the saturation point, defined as the 
minimum of the energy per nucleon at $T = 0$, as the proton fraction decreases. 
The result is shown in Fig.\ \ref{satpoint}. Starting from its value
for symmetric nuclear matter, $ \bar{E}_0 \simeq -16 $ MeV at $ \rho_0 \simeq 0.157\,
{\rm fm}^{-3} $, the binding energy per particle continuously reduces in magnitude with
decreasing $ x_p $ until it vanishes for a value of the proton fraction $ x_p \simeq 0.12$,
beyond which the neutron-rich matter is unbound at zero temperature.  

\begin{figure}[tbp]
\center
\includegraphics[scale=0.5,clip]{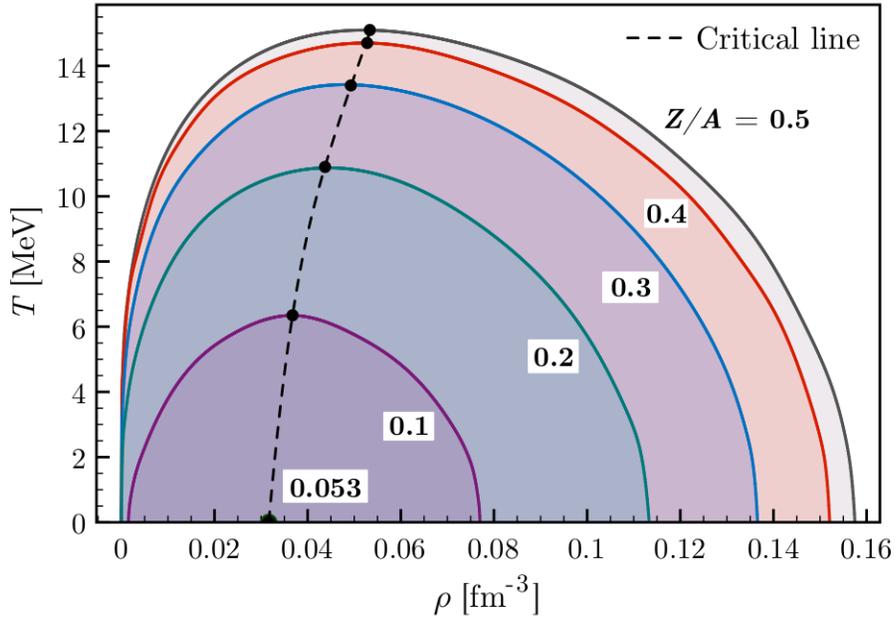} 
\caption{Dependence of the nuclear matter phase diagram on the proton 
fraction $x_p$. The evolution of the critical point is given by the dashed line.}
\label{phaseasy}
\end{figure}

In Fig.\ \ref{phaseasy} we show the phase diagram as a function of the proton fraction $x_p$.
The dashed line shows the trajectory of the critical point and its disappearance for proton
fractions below $ x_p \simeq 0.05 $. At this value of the proton fraction, the coexistence 
region reduces to a single point, indicating the absence of a liquid-gas phase 
transition. Neutron-rich matter with $ x_p \leq 0.05 $ therefore exists only 
in a gaseous phase. We note as well that the critical point disappears at a small but nonzero 
value of the pressure.

From Fig.\ \ref{phaseasy} we see that for proton fractions above $x_p \simeq 0.1$, 
the liquid-gas coexistence region starts at $\rho=0$. At $T=0$ and from $ x_p = 0.1$ to 0.5, 
the boundary between the coexistence and Fermi liquid regimes covers a density range 
between about $ 0.5\, \rho_0 $ and nuclear matter saturation density density $ \rho_0 $. 
The behavior of the coexistence region changes qualitatively for a proton fraction $ x_p = 0.12$, 
which is the point at which the zero-temperature binding energy vanishes. 
From Fig.\ \ref{satpoint} we observe that for $ x_p\leq 0.12$ the absolute minimum of the energy
per particle is located at $ \rho = 0 $ but there may still exist a local minimum in $ \bar{E}(\rho, x_p) $ 
at a finite value of the density. This indicates that neutron-rich nuclear matter with values of $ x_p 
\leq 0.12 $ is in the gaseous phase at very low density and $ T=0 $. As the density increases, the
neutron-rich matter then enters the coexistence region. For values of the proton fraction 
$ 0.05 \leq x_p \leq 0.12$, nuclear matter will not be self-bound but may nevertheless still have a 
liquid-gas phase transition.

The treatment of nuclear matter chiral thermodynamics presented thus far is oversimplified 
at low densities where nuclear clustering occurs (mainly driven by the Coulomb repulsion 
of the protons). Previous work \cite{typel} that combines the appearance of deuteron, triton 
and helium clusters with relativistic mean-field theory suggests, however, that 
only modest changes in the $ T-\mu $ phase diagram arise. When clustering effects are included, 
the position of the critical point changes by less than $ 1 \% $ in $ \mu_c $ and less than 
$ 10 \% $ in $ T_c $.
 
The free energy per particle can be expanded in powers of the nuclear asymmetry parameter 
$\delta = (\rho_n - \rho_p) / \rho = 1-2x_p$ around the free energy at $\delta=0$:
 \begin{equation} 
\bar{F}(\rho_p,\rho_n, T) = \bar{F}(\rho, T) + A(\rho, T)\, \delta^2 + 
\mathcal{O} (\delta^4) \, , \end{equation}
which defines the asymmetry free energy per particle $ A(\rho, T) $. If isospin-symmetry 
breaking effects are ignored, the expansion of the free energy per particle includes only even 
powers of $ \delta$. In this approximation, nuclear matter is invariant under the interchange of 
neutrons and protons.
 
The validity of the parabolic approximation has been shown in ref.\ \cite{FKW2012}, where the 
free energy difference relative to isospin-symmetric nuclear matter as a function of 
$ \delta^2 $ for different nucleon densities $ \rho = \rho_n + \rho_p $ has been analyzed.
At $ T = 0 $ the quadratic dependence on $\delta$ holds very well even up 
to large values of the isospin asymmetry parameter. For higher temperature (e.g., $ T = 20\,$MeV) 
one observes a slight bending, especially at low density.

At nuclear matter saturation density $\rho_0$, the contact terms of the isospin-dependent 
part of the nucleon-nucleon interaction have been fit to the isospin asymmetry 
energy $A(\rho_0, T = 0) \simeq 34.0 $\,MeV. Previous determinations from
fits of nuclear masses \cite{blaizot,seeger} as well as relativistic mean-field models constrained by
the properties of specific nuclei \cite{vretenar} or by giant dipole resonances \cite{cao} give values
of $ A(\rho_0) $ in the range $33 - 37$\,MeV. The asymmetry energy at a lower density, $ \rho = 0.1 $ 
fm$^{-3} $, has also been estimated in the latter paper and yields a value for the isospin
asymmetry energy between 21 and 23 MeV. This range is slightly below the value
$ A(\rho = 0.1\, {\rm fm}^{-3}, T = 0) \simeq 23.9 $ MeV computed within the present chiral 
effective field theory framework.

In the vicinity of the nuclear matter saturation density $\rho_0$, the asymmetry energy at zero
temperature can be expanded in powers of $\rho-\rho_0$:
\begin{equation}
A(\rho) = A(\rho_0) + L\,\frac{\rho - \rho_0}{3\,\rho_0} + \frac{K_{as}}{2} 
\left(  \frac{\rho - \rho_0}{3\,\rho_0} \right)^2 + \dots  \end{equation}
The coefficients of the linear and quadratic terms (in $\rho-\rho_0$) are found in the present
case to be $L \simeq 90$ MeV and $ K_{as} \simeq 153 $ MeV. In particular, this value of 
$L$ is consistent with empirical constraints from isospin diffusion, which give $ L = 88 \pm 25 $ 
MeV \cite{baoanli}. 

For small values of the isospin asymmetry $\delta$, the saturation density is reduced to 
$ \rho_0 (1 - 3L\,\delta^2/K )$. The nuclear compression modulus $ K(\delta) $ is then often 
expressed in terms of an expansion in powers of $ \delta $:
\begin{equation}
K(\delta) = K + K_\tau \delta^2 + \mathcal{O}(\delta^4) \ , \qquad K_\tau= K_{as} - 6\,L \ ,
\end{equation}
where $ K $ is the compression modulus of isosopin-symmetric nuclear matter, and $ K_\tau $ is
referred to as the isobaric compressibility. The value calculated in the present framework is 
$ K_\tau = -388 $ MeV. Recent measurements of giant monopole resonances in even-mass-number 
isotopes yields the result $ K_\tau = -550 \pm 100 $ MeV \cite{li}.

\subsection{\it Comparisons to other Approaches}

In the present section we discuss recent calculations of the thermodynamic properties of nuclear 
matter within the self-consistent Green's function method or mean-field theory 
\cite{rios,vidana,soma,rios2,wu}. As noted in ref.\ \cite{rios}, many previous studies of 
nuclear matter at nonzero temperature carried out in the mean-field approximation treat the 
temperature dependence in a simplistic way, namely, the zero-temperature step-function 
momentum distribution is just replaced by the corresponding Fermi-Dirac distribution. Such an
approach neglects the temperature dependence of the phenomenological interactions that account 
for nuclear correlations. In microscopic many-body methods, on the other hand, Pauli-blocking
effects generated by the nuclear medium are weakened as the temperature increases. 
Hence, nuclear matter properties and correlations depend on the temperature in a non-trivial way. 

In the SCGF formalism, the set of equations relating the in-medium $T$-matrix, the nucleon self-energy,
and the single-particle Green's function are solved self-consistently. The equation of state of 
isospin-symmetric nuclear matter has been calculated in this approach \cite{soma} starting from 
the high-precision CD-Bonn and Nijmegen nucleon-nucleon potentials. The contributions from 
three-nucleon forces are included through an effective (medium-dependent) two-body interaction. 
Three-body correlations are found to strongly affect the pressure $ P(\rho, T, \delta)$ and result
in a liquid-gas coexistence region that is much reduced in size. The critical temperatures associated
with the CD-Bonn nucleon-nucleon potential and the Nijmegen potential are $ T_c = 12.5$\,MeV and 
$ T_c = 11.5$\,MeV, respectively. The critical density is found to be $ \rho_c \simeq 0.09 - 0.11\,
{\rm fm}^{-3} $, while the critical pressure has the value $ 0.15 \, 
{\rm MeV fm}^{-3} $. The results obtained within a chiral effective field theory treatment and described 
in previous sections are therefore significantly different than the results based on the self-consistent 
Green's function approach with one-boson exchange nuclear interactions.

The properties of nuclear matter have also been calculated \cite{wu} within the framework of 
relativistic mean field theory incorporating density-dependent meson-nucleon couplings that account 
for effects of the medium. Such a method has been successful in previous calculations of finite 
nuclei and infinite nuclear matter. The critical temperature $ T_c = 13.2 $ MeV is found for symmetric 
nuclear matter, and for proton fractions below $ x_p \simeq 0.07 $ the liquid-gas coexistence region 
vanishes. The picture that emerges in these calculations is closer to the one reported in the 
present review. The main difference between the two is that mean field theory gives a slightly smaller
phase transition region.

\begin{figure}[tbp]
\center
\includegraphics[scale=0.3,clip]{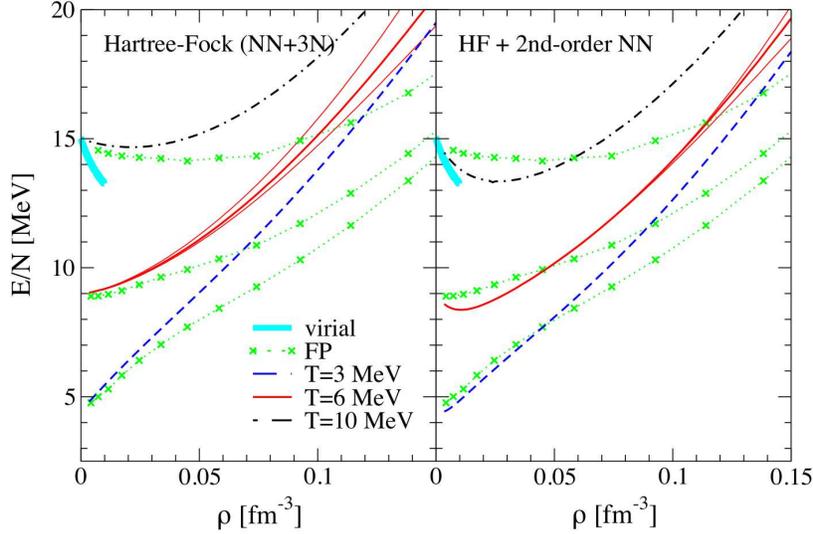}
\vspace{-.1in}
\caption{Energy per particle of neutron matter calculated at temperatures $T = 3, 6$ and $10$ MeV
as function of density, using low momentum NN and 3N interactions. Left: Hartree-Fock results.
Right: results including 2nd order NN interactions. Also shown for comparison are curves from
a many-body calculation (FP) \cite{FP81} and a model-independent virial expansion (virial) \cite{horowitz}
at low density.}
\label{tolosfigure}
\end{figure}

To close this section we mention the work of ref.\ \cite{tfs2008} in which the energy per 
particle of neutron matter has been calculated at several temperatures $T \leq 10$ MeV
as a function of density, using a low-momentum neutron-neutron interaction that is close 
in spirit to the chiral interaction discussed in this review. Two- and three-body interactions
are taken into account. Representative results from these calculations are displayed 
in Fig.\ \ref{tolosfigure}. The left panel shows first order (Hartree-Fock) results, including a comparison 
with a model-independent virial expansion at low density. The right panel demonstrates
significant improvements when the NN interaction is treated in second order.

\section{Thermodynamics of the Chiral Condensate}

The chiral condensate $\langle \bar{q}q \rangle$ is defined as the expectation value 
of the scalar quark density and plays a fundamental role in QCD as an order parameter 
for the spontaneous breaking of chiral symmetry that occurs at low energy scales. The 
dependence of $\langle \bar{q}q \rangle$ on the temperature and baryon 
density is key to locating the chiral restoration transition boundary in the phase diagram of
QCD. The condensate is expected to vanish at high temperatures and/or
densities, signaling the crossover from the Nambu-Goldstone phase, in which chiral symmetry
is spontaneously broken, to the Wigner-Weyl realization of chiral symmetry in QCD.

It is therefore of principal interest to perform a systematic calculation of the 
thermodynamics of the chiral condensate \cite{condpap}. For such a 
calculation it is necessary to understand the dependence of the free energy density on 
the light quark mass (or equivalently, the pion mass). In-medium chiral effective field 
theory provides the appropriate framework for such a study, given that the theory 
provides explicit access to the pion mass dependence through one- and two-pion exchange 
dynamics and the resulting two- and three-body correlations in the nuclear medium. 

The starting point is the free energy density, ${\cal F}(\rho,T) = \rho \bar 
F(\rho,T)$, of spin-saturated isospin-symmetric nuclear matter, given by the series
of convolution integrals in eq.\ (\ref{convolution}). This series is written in terms of one-body, two-body and 
three-body kernels, ${\cal K}_1, {\cal K}_2$ and ${\cal K}_3$, respectively. 

The one-body kernel ${\cal K}_1$ in eq.\ (\ref{convolution}) represents the contribution 
of the non-interacting nucleon gas to the free energy density and it reads
\cite{nucmatt}:
\begin{equation} \label{K1}
{\cal K}_1(p) = M_N +\tilde \mu- {p^2\over 3M_N}- {p^4\over 8M_N^3} \,. 
\end{equation}
The first term in ${\cal K}_1$ gives the leading contribution (density 
times nucleon rest mass) to the free energy density. The 
remaining terms account for (relativistically improved) kinetic energy 
corrections.

The two- and three-body kernels, ${\cal K}_2$ and ${\cal K}_3$, specifying 
all one- and two-pion exchange processes up to three loop order for the
free energy density, are given in their explicit form in 
refs.\ \cite{nucmatt,fkw2,FKW2012}. Let us recall that after fixing a few 
contact terms the free energy density computed from these interaction kernels 
provides a realistic nuclear equation of state up to densities $\rho 
\leq 2\,\rho_0$. What matters in the present context is the dependence 
of the kernels ${\cal K}_2$ and ${\cal K}_3$ on the light quark mass, $m_q$, 
or equivalently, on the pion mass, $m_\pi$, that is introduced by pion 
propagators and by pion loops. 

The Feynman-Hellmann theorem is used to establish an exact relation
between the temperature- and density-dependent quark condensate 
$\langle \bar q q\rangle(\rho,T)$ and the derivative of the free energy 
density of nuclear matter with respect to 
the light quark mass $m_q$. From the Gell-Mann-Oakes-Renner relation $m_\pi^2 
f_\pi^2 = -m_q \langle 0|\bar q q|0\rangle$, the ratio of the 
in-medium to vacuum quark condensate is given by
\begin{equation}  
{\langle \bar q q\rangle(\rho,T)\over  \langle 0|\bar q q|0
\rangle} = 1 - {\rho \over f_\pi^2} {\partial \bar F(\rho,T) \over \partial 
m_\pi^2} \,,\end{equation}
where the derivative of the free energy density with respect to $m_\pi^2$ is 
taken at fixed $T$ and $\rho$. Both the vacuum quark condensate 
$\langle 0|\bar q q|0\rangle$ and the pion decay constant $f_\pi$ are taken in the 
chiral limit. Similarly, $m_\pi^2$ represents the 
leading linear term in the expansion of the squared pion mass in terms of the quark mass.

The quark mass dependence in the one-body kernel ${\cal K}_1$ is implicit through its 
dependence on the nucleon mass $M_N$. The 
condition $\partial \rho/\partial M_N=0$ leads to the 
dependence of the effective one-body chemical potential $\tilde \mu$ 
on the nucleon mass $M_N$:
\begin{equation} {\partial \tilde \mu \over \partial M_N }= {3 \rho \over 2M_N 
\Omega_0''} \,, \qquad   \Omega_0''= -4M_N  \int_0^\infty dp\, 
{n(p) \over p}\,. \end{equation} 
The nucleon sigma term $\sigma_N = \langle N|m_q \bar q q |N\rangle= m_\pi^2 \,
\partial M_N/\partial m_\pi^2$ relates the variation of the nucleon mass to
that of the pion mass. Combining the above two relationships then leads to the 
expression for the derivative of the one-body kernel with respect to the pion mass:
\begin{equation} {\partial {\cal K}_1 \over \partial m_\pi^2} = {\sigma_N 
\over m_\pi^2} \bigg\{ 1+ {3 \rho \over 2M_N \Omega_0''} +{p^2 \over 3M_N^2} 
+{3p^4 \over 8M_N^4} \bigg\}\,. \label{derk1}\end{equation}
In the zero temperature limit, the terms in eq.\ (\ref{derk1}) 
provide the linear decrease of the chiral condensate with the nucleon density. The 
corrections from the kinetic energy account for the small difference between the 
scalar and vector densities. The nucleon sigma term is determined empirically at the 
physical pion mass to be
$\sigma_N =(45\pm 8)\,$MeV \cite{GLS91}. Recently, smaller values of the nucleon
sigma term have been 
suggested based on lattice QCD calculations of the quark mass 
dependence of baryon masses and accurate chiral 
extrapolations \cite{sigmaterm}. The value of $\sigma_N$ determined in this way is
consistent with the empirical value to within the experimental uncertainty. However, a considerably 
larger nucleon sigma term, $\sigma_N \simeq 
60\,$MeV, has recently been discussed in ref.\ \cite{camalich}. In the present discussion
based on ref.\ \cite{condpap}, the central value 
$\sigma_N = 45$\,MeV has been employed.

There are some three-loop contributions that are of special relevance for 
the in-medium quark condensate. The corresponding two-pion exchange mechanisms 
arise from the chiral symmetry breaking $\pi\pi NN$ contact-vertex 
proportional to  $c_1m_\pi^2$. Note that this parameter is equivalent to the 
leading contribution (linear in the quark mass) to the nucleon sigma term  
$\sigma_N = -4 c_1m_\pi^2+{\cal O}(m_\pi^3)$. Concerning the 
free energy density $\rho \bar F(\rho,T)$ or the equation of state of nuclear 
matter the $c_1m_\pi^2$ contributions are almost negligible. However,
when taking the derivative with respect to $m_\pi^2$ as required for the 
calculation of the in-medium condensate, these contributions turn out to be 
of  similar importance as other interaction terms. The pertinent two- and 
three-body kernels ${\cal K}_{2,3}^{(c_1)}$ can be found in ref.\ \cite{condpap}.

It is meaningful to incorporate also effects from thermal pions. The pressure 
(or free energy density) of thermal pions gives rise, through its  
$m_\pi^2$-derivative, to a further reduction of the $T$-dependent 
in-medium condensate. In the two-loop approximation of chiral perturbation 
theory one finds the following shift of the chiral condensate ratio due to the presence 
of a pionic heat bath \cite{gerber,toublan,pipit}: 
\begin{eqnarray}
{\delta\langle \bar q q\rangle(T)\over  \langle 0|\bar q q|0
\rangle} &=&-{3m_\pi^2 \over (2\pi f_\pi)^2} H_3\Big({m_\pi 
\over T}\Big) \bigg\{1+ {m_\pi^2 \over 8\pi^2 f_\pi^2} 
\bigg[H_3\Big({m_\pi \over T}\Big)-  H_1\Big({m_\pi \over T}\Big) + {2-3
\bar{\ell}_3 \over 8} \bigg] \bigg\}\,, \nonumber\\ &&\end{eqnarray}
where the functions $H_{1,3}(m_\pi/T)$ are defined by integrals over the 
Bose distribution of thermal pions:
\begin{eqnarray}
 H_1(y) = \int_y^\infty dx\, {1 \over \sqrt{x^2-y^2}
(e^x-1)}\,,\qquad\qquad H_3(y) = y^{-2} \int_y^\infty dx\, 
{\sqrt{x^2-y^2} \over e^x-1}\,. 
\end{eqnarray}

It is worth emphasizing that in-medium chiral perturbation theory with 
this dynamical input yields a realistic nuclear matter equation of state 
\cite{fkw2,FKW2012}, including a proper description of the liquid-gas phase 
transition with critical temperature $T_c\simeq 15\,$MeV. The three-loop
calculation of the free energy density is expected to be reliable up to about twice nuclear
matter saturation density and for temperatures below about $T\sim 100$ MeV, where the hot and dense 
hadronic matter remains in the spontaneously broken chiral symmetry regime.

\begin{figure}
\begin{center}
\includegraphics[width=12cm]{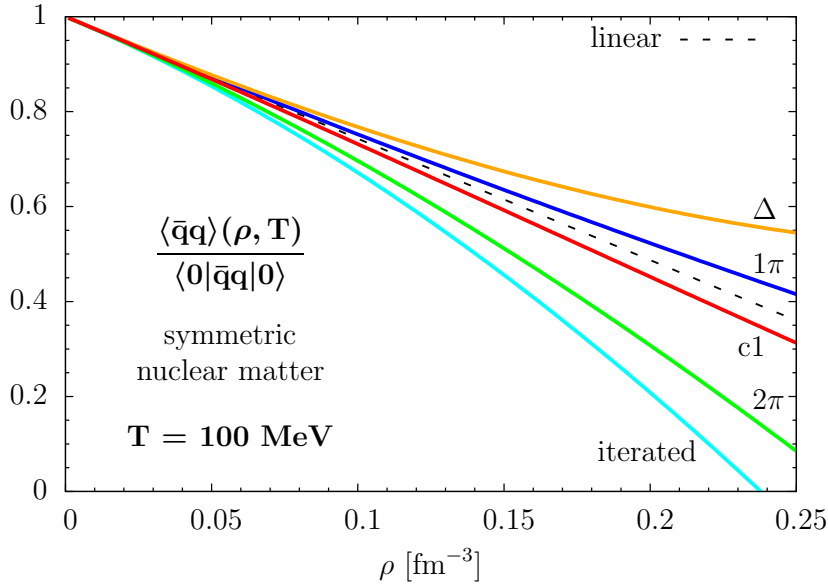}
\end{center}
\vspace{-.3in}
\caption{Density dependence of the chiral condensate at temperature $T = 100$ 
MeV. Starting from the linear density dependence (dashed curve) associated with
the free nucleon Fermi gas, the following interaction contributions are successively 
added: one-pion exchange Fock contribution, iterated one-pion exchange, 
irreducible two-pion exchange, two-pion exchange with intermediate $\Delta$ excitations,
two-pion exchange contribution proportional to the low-energy constant $c_1$. For the 
sake of clarity in presenting the pattern of medium 
modifications, the contribution from thermal pions is omitted here}
\label{condfig2}
\end{figure} 

In Fig.\ \ref{condfig2} we show a typical example ($T = 100$ MeV) 
of the effects of interaction contributions to the density 
dependence of 
$\langle\bar{q}q\rangle(\rho,T)$ arising from the pion mass derivative 
of the chiral two- and three-body kernels ${\cal K}_2$ and ${\cal K}_3$. Of 
particular importance is the pion-mass dependence of the terms 
involving virtual $\Delta(1232)$ excitations, which delay the tendency towards 
chiral symmetry restoration as the 
density increases. Summing together all one- and two-pion exchange processes contributing 
to $\partial {\cal K}_2/\partial m_\pi^2$ and $\partial{\cal K}_3/\partial 
m_\pi^2$, the chiral condensate at $T=100$ MeV is not far from the 
linear density dependence characteristic of a free Fermi gas. However, this behavior
is the result of a subtle balance between attractive and repulsive 
correlations and their detailed pion-mass dependences. Iterated one-pion exchange and 
irreducible two-pion exchange alone would have resulted in the system 
becoming unstable not far above normal nuclear matter saturation density as 
can be seen in Fig.\ \ref{condfig2}. In the chiral limit $(m_\pi \rightarrow 
0)$ this instability would have appeared even at much lower densities. 
This emphasizes once again the importance of including $\Delta(1232)$-excitations, and 
it also highlights the significance of explicit chiral symmetry breaking in governing 
nuclear scales. 

\begin{figure}[htb]
\begin{center}
\includegraphics[width=12cm]{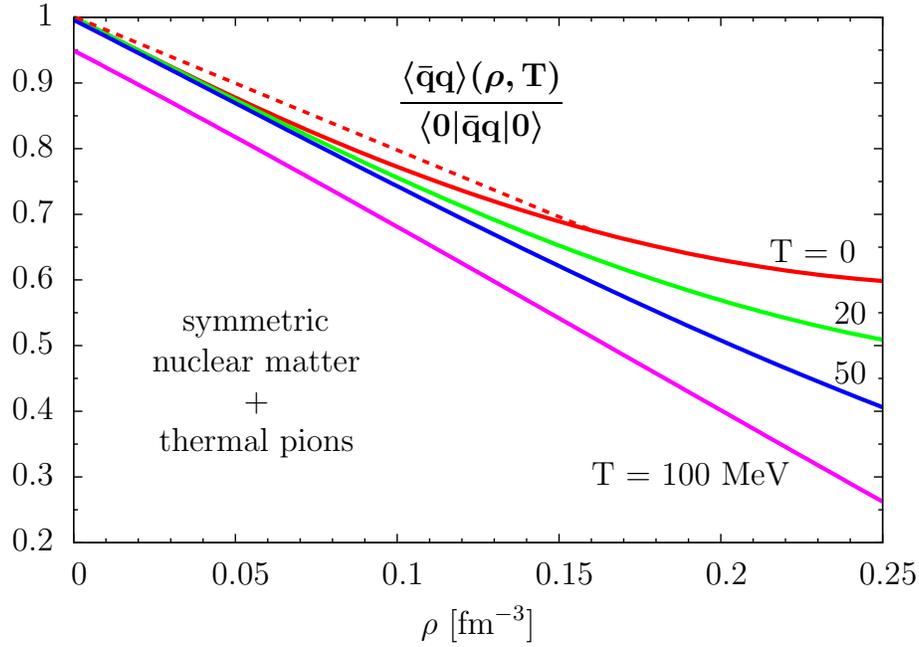}
\end{center}
\vspace{-.2in}
\caption{Ratio of the in-medium chiral condensate to its vacuum value as a function 
of the baryon density $\rho$ and temperature in isospin-symmetric nuclear matter. 
Effects of thermal pions are included, and the dashed line at $T \simeq 0$ results from 
the Maxwell construction in the nuclear liquid-gas coexistence region. }
\label{condfig3}
\end{figure}

In Fig.\ \ref{condfig3} we show the variation of the chiral condensate with both the
baryon density $\rho$ and the temperature $T$. All nuclear correlation effects are included
as well as the small additional shift from thermal pions, which is visible only at the highest 
temperature considered here ($T = 100$ MeV). The actual crossover 
transition can be extracted from lattice QCD simulations and has the value
$T \sim 170$ MeV \cite{lattice}.

At zero temperature, the decrease in the chiral condensate beyond nuclear matter
saturation density is hindered as a result of three-body correlations 
encoded in ${\cal K}_3$ which grow faster than ${\cal K}_2$ as the 
density increases. As the temperature rises, the influence of 
${\cal K}_3$ relative to ${\cal K}_2$ is reduced, 
so that at $T=100$ MeV only a small net effect remains in 
comparison to the free Fermi gas.

At $T = 0$, the solid line in Fig.\ \ref{condfig3} does not yet
account for the coexistence region of the nuclear liquid and gas 
phases \cite{FKW2012}. Any first-order phase transition is expected 
to be visible also in other order parameters, and the dashed 
line in Fig.\ \ref{condfig3} based on the usual Maxwell construction indicates this effect. 
This feature becomes much more pronounced when the chiral condensate is plotted as a 
function of the baryon chemical potential at low temperatures:
\begin{equation} \mu = M_N +\bigg( 1 + \rho {\partial \over\partial \rho}\bigg) 
\bar F(\rho,T) \,.\end{equation} 

\begin{figure}[htb]
\begin{center}
\includegraphics[width=12cm]{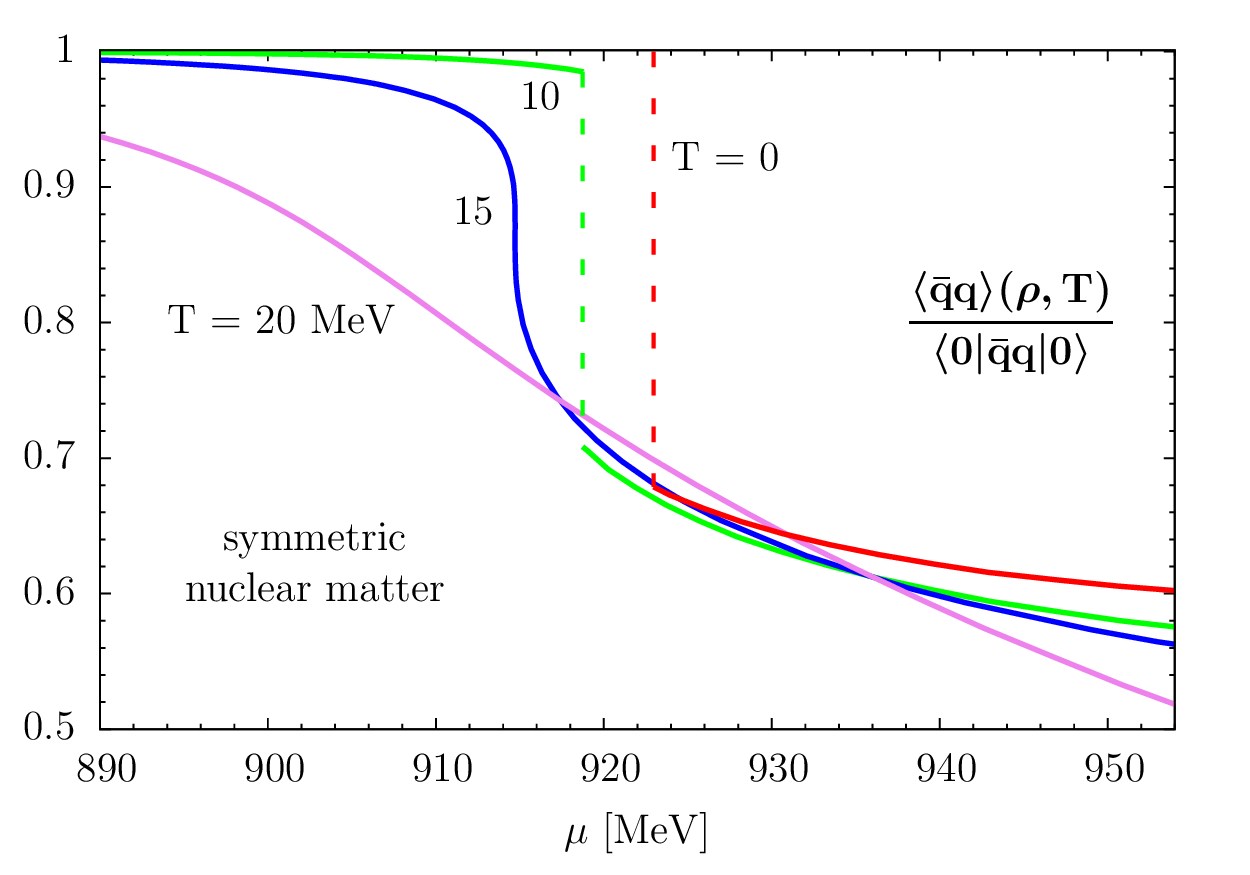}
\end{center}
\vspace{-.2in}
\caption{Ratio of chiral condensate relative to its vacuum value as function 
of baryon chemical potential in isospin-symmetric nuclear matter at low
temperatures within the liquid-gas phase coexistence region.}
\label{condfig4}
\end{figure}

The discontinuity associated with the first-order liquid-gas transition
at $T$ smaller $T_c \simeq 15$\,MeV is clearly visible in Fig.\ \ref{condfig4}. 
Another effect generated by the first-order liquid-gas phase transition is 
that the frequently advocated ``low-density theorem'' requires modification:
\begin{equation}  
{\langle \bar q q\rangle(\rho)\over  \langle 0|\bar q q|0
\rangle} = 1 - {\widetilde\sigma_N \over m_\pi^2f_\pi^2} \rho  \,.\label{ldtmod}
\end{equation}
In this equation the effective nucleon sigma term $\widetilde\sigma_N\simeq 36\,$MeV 
represents the quark mass dependence of the sum $M_N+\bar E_0$, where $\bar E_0 
\simeq -16\,$MeV is the binding energy per particle of nuclear matter at the saturation density. 
The standard version of eq.\ (\ref{ldtmod}) contains the nucleon sigma term $\sigma_N$ 
in vacuum and assumes that at sufficiently low densities nuclear matter can be 
treated as a non-interacting Fermi gas.

The results discussed here set important nuclear physics constraints for 
the QCD equation of state at baryon densities and temperatures that are of 
interest in relativistic heavy-ion collisions. In particular, there is  
no indication for chiral symmetry restoration at temperatures $T\leq  
100\,$MeV and baryon densities at least up to about twice nuclear matter saturation
density. The effects of intermediate $\Delta(1232)$-isobar 
excitations (i.e.\ the strong spin-isospin polarizability of the nucleon) 
together with Pauli blocking in the nuclear medium play a crucial role in stabilizing the 
scalar quark condensate at and beyond nuclear matter saturation density.

\section{Concluding Remarks and Outlook}

Chiral Effective Field Theory, as a low-energy realization of QCD in its sector with light ($u$ and $d$) quarks,
is not only a successful framework for the description of pion-pion and pion-nucleon interactions.
With its systematic counting scheme in powers of ``small" $Q/\Lambda_\chi$ (where $Q$ stands generically 
for momentum, energy or the pion mass, and $\Lambda_\chi \sim 1$ GeV is the spontaneous chiral symmetry 
breaking scale), chiral EFT has also become the basis for the modern theory of nuclear forces. The present review
goes still one step further and reports on recent developments concerning the application of chiral EFT to 
nuclear many-body problems, both at zero and finite temperatures.

The additional ``small" scale introduced by the many-body system is the Fermi momentum. Its typical magnitude
in nuclear physics, $k_f \sim 2 m_\pi$, implies that the chiral dynamics of pions in a low-energy, low-temperature
nuclear environment should also be controlled by an expansion in powers of $k_f/\Lambda_\chi$. Indeed, the typical
average distance between two nucleons in equilibrium nuclear matter, $d \simeq 1.8$ fm, is again comparable to
a ``pionic" scale, just slightly larger than the Compton wavelength of the pion. Furthermore, another characteristic 
parameter of nuclear matter, its compression modulus, happens to be of a magnitude comparable to twice the pion
mass. The only quantity that falls out of this pionic scales scenario is the binding energy per nucleon. Its value 
$B/A \simeq 16$ MeV is abnormally small and indicates a quantitative fine-tuning: the well-known 
subtle balance between intermediate-range attraction and strong short-range repulsion in the nucleon-nucleon
interaction. 

Details of the short-range NN interaction remain unresolved at momentum scales $p \leq k_f$. Chiral EFT
assigns a limited number of NN contact terms to the physics at short distance ($ r < 1$ fm), with prefactors
(a set of low-energy constants) to be determined by experiment. While the strengths of these contact terms
generally require non-perturbative resummation techniques to be performed at low densities, the intermediate and 
long range forces governed by one- and two-pion exchange processes in the medium are supposed to be
accessible to perturbative methods. This is the essence of in-medium chiral perturbation theory. In this perturbative
hierarchy, three-nucleon forces enter naturally at a well-defined order of the low-momentum expansion.

Such a conceptual framework turns out to be remarkably successful. Much of the well established phenomenology
associated with the nuclear many-body problem can now be given more systematic foundation that relies 
substantially on the well-established spontaneous chiral symmetry breaking pattern of low-energy QCD. 
The nuclear mean field experienced by nucleons as quasiparticles, including its extension to the optical potential for scattering, the density functional approach to finite systems, Landau-Migdal Fermi liquid theory, thermodynamics
and the phase diagram of nuclear matter featuring the liquid-gas phase transition - all these important aspects of
nuclear theory are covered in this report. As a bonus of this framework that permits investigating the energy density
and related quantities systematically as a function of the pion mass, reliable statements about the behavior
of the quark condensate at temperatures up to about 100 MeV and densities reaching twice the density of normal
nuclear matter can be made, indicating stability of this chiral order parameter considerably beyond previous 
expectations that were based merely on the Fermi gas limit.

Of course, many further reaching questions still need to be addressed. So far, the calculations of the energy density
(and of the free energy density at finite temperature) have been carried out up to three-loop order. This includes
important three-body forces but does not (yet) cover issues of convergence beyond this order, e.g. concerning the
role of four-nucleon correlations. At this point Monte Carlo studies of the kind performed in ref.\ \cite{MC2013} may 
prove useful in the near future. As a further important topic, the quest for a sufficiently stiff equation-of-state 
satisfying the constraints imposed by the existence of a two-solar-mass neutron star \cite{demorest} is under 
investigation.


\end{document}